\newcommand*\figpath{./}

\documentclass[11pt,a4]{article}

\usepackage{lmodern,sansmath}
\usepackage{textcomp}
\usepackage{microtype}
\usepackage{amsmath, amssymb, amsthm, dsfont, bm}         %% math formulae
\usepackage{amsfonts}         %% math fonts
\usepackage{array}

\usepackage{natbib}
\usepackage[colorlinks=true,linkcolor=red,citecolor=blue]{hyperref}

\usepackage{mathtools}
\newcolumntype{M}[1]{>{\raggedright}m{#1}}
\newcolumntype{C}[1]{>{\centering\arraybackslash}m{#1}}

\usepackage{color}
\usepackage[dvipsnames,svgnames]{xcolor}

\usepackage{booktabs}
\usepackage{graphicx}

\usepackage{caption}
\usepackage{subcaption}
\usepackage{multirow}

\renewcommand{\vec}[1]{\boldsymbol{#1}}

\newcommand {\R} {{\mathds R}}

\newcommand{\ie}{{\em i.\thinspace{}e. }}

\newcommand{\eg}{{\em e.\thinspace{}g. }}

\newcommand{\ds}{\displaystyle}
\newcommand{\pl}{\partial}

\newcommand{\TF} {{\hbox{\tiny TF}}}

\newcommand{\omegap}{\omega_{\perp}}

\newcommand{\trap}{{\rm trap}}

\newcommand{\eff}{{\rm eff}}

\newcommand{\ff}{FreeFem{\small +$\!$+}  }

\newcommand{\ii}{\mathrm{i}}

\newcommand{\C}{\ensuremath{\mathbb C}}
\usepackage{braket}

\usepackage{xcolor}
\usepackage{textcomp}
\usepackage{listings}
\makeatletter
\renewcommand*{\eqref}[1]{%
	\hyperref[{#1}]{\textup{\tagform@{\ref*{#1}}}}%
}
\makeatother

\begin{document}

%\begin{frontmatter}

%% Title, authors and addresses

%% use the tnoteref command within \title for footnotes;
%% use the tnotetext command for theassociated footnote;
%% use the fnref command within \author or \address for footnotes;
%% use the fntext command for theassociated footnote;
%% use the corref command within \author for corresponding author footnotes;
%% use the cortext command for theassociated footnote;
%% use the ead command for the email address,
%% and the form \ead[url] for the home page:
%% \title{Title\tnoteref{label1}}
%% \tnotetext[label1]{}
%% \author{Name\corref{cor1}\fnref{label2}}
%% \ead{email address}
%% \ead[url]{home page}
%% \fntext[label2]{}
%% \cortext[cor1]{}
%% \address{Address\fnref{label3}}
%% \fntext[label3]{}

\title{Parallel finite-element codes for the Bogoliubov-de Gennes
stability analysis of Bose-Einstein condensates}

%% use optional labels to link authors explicitly to addresses:
%% \author[label1,label2]{}
%% \address[label1]{}
%% \address[label2]{}

%\author[lmrs]{Georges Sadaka}
%\ead{georges.sadaka@univ-rouen.fr}
%
%\author[LIP6]{Pierre Jolivet}
%\ead{pierre@joliv.et}
%
%\author[lmrs,CPSU]{Efstathios G. Charalampidis}
%\ead{echarala@calpoly.edu}
%
%\author[lmrs]{Ionut Danaila\corref{io}}
%\ead{ionut.danaila@univ-rouen.fr}
%
%\address[lmrs]{Univ Rouen Normandie, CNRS, Laboratoire de Math\'ematiques
%Rapha\"el Salem, UMR 6085, F-76000 Rouen, France}
%
%\address[LIP6]{Sorbonne Universit\'e - CNRS, LIP6, 4 place Jussieu, 75252 Paris, France}
%
%\address[CPSU]{Mathematics Department, California Polytechnic State University, %
%San Luis Obispo, CA 93407-0403, USA}
%
%\cortext[io]{Corresponding author. Tel.: (+33) 2 32 95 52 50}

\author{
	Georges Sadaka$^1$, Pierre Jolivet$^2$, Efstathios G. Charalampidis$^{1, 3}$ and Ionut Danaila$^{1, *}$ 
\\ \\
$^1$Univ Rouen Normandie, CNRS,  \\
Laboratoire de Math{\'e}matiques Rapha{\"e}l Salem,\\
UMR 6085, F-76000 Rouen, France\\
$^2$Sorbonne Universit{\'e}, CNRS,
LIP6,\\
4 place Jussieu, 75252 Paris, France.	\\
$^3$ Mathematics Department, California Polytechnic State University,\\
San Luis Obispo, CA 93407-0403, USA\\
$^*$ Corresponding author: ionut.danaila@univ-rouen.fr
}

\date{\today}
\maketitle

\begin{abstract}
We present and distribute a parallel finite-element toolbox
written in the free software \ff for computing the Bogoliubov-de Gennes (BdG)
spectrum of stationary solutions to one- and two-component Gross-Pitaevskii
(GP) equations, in two or three spatial dimensions. The parallelization of
the toolbox relies exclusively upon the recent interfacing of \ff with the
\texttt{PETSc} library. The latter contains itself a wide palette of
state-of-the-art linear algebra libraries, graph partitioners, mesh generation
and domain decomposition tools, as well as a suite of eigenvalue solvers that
are embodied in the \texttt{SLEPc} library. Within the present toolbox,
stationary states of the GP equations are computed by a Newton method.
Branches of solutions are constructed using an adaptive
step-size continuation algorithm. The combination of mesh adaptivity tools from \ff with
the parallelization features from \texttt{PETSc} makes  the toolbox
efficient and reliable for the computation of stationary states. Their BdG spectrum
is computed using the \texttt{SLEPc} eigenvalue
solver. We perform extensive tests and validate our programs by comparing the
toolbox's results with known theoretical and numerical findings that have been
reported in the literature.
\end{abstract}

%\begin{keyword}
%% keywords here, in the form: keyword \sep keyword
%Bose-Einstein condensate \sep Gross-Pitaevskii equation \sep %
%Bogoliubov-de Gennes equation \sep finite elements \sep %
%FreeFEM \sep PETSc \sep SLEPc.
%% PACS codes here, in the form: \PACS code \sep code

%% MSC codes here, in the form: \MSC code \sep code
%% or \MSC[2008] code \sep code (2000 is the default)

%\end{keyword}
%
%
%
%\end{frontmatter}

\begin{small}
\noindent
{\bf Programm summary}\\
{\em Program Title:}  {FFEM\_BdG\_ddm\_toolbox.zip}\\
{\em CPC Library link to program files:}  \\ % Leave blank, supplied by Technical Editor. \\
{\em Developer's repository link:}        \\ % EGC: We may want to add this to our websites?
{\em Code Ocean capsule:} %Leave blank, supplied by Technical Editor. \\
{\em Licensing provisions:} GPLv3 \\
% EGC: I picked one out of the many options below; feel free to change:
% CC0 1.0/CC By 4.0/MIT/Apache-2.0/BSD 3-clause/BSD 2-clause/GPLv3/GPLv2/LGPL/CC BY NC 3.0/MPL-2.0  \\
{\em Programming language:} \ff (v 4.12) free software (www.freefem.org)\\
%
%
% EGC: Not sure if the following lines below are needed; please check/see the guidelines:
% https://www.elsevier.com/journals/computer-physics-communications/0010-4655/guide-for-authors
%
%
 {\em Catalogue identifier:}\\ %Leave blank, supplied by Elsevier.\\
 {\em Program summary URL:}\\
 {\em Program obtainable from:}\\
 {\em No. of lines in distributed program, including test data, etc.: }  4,054 \\
 {\em No. of bytes in distributed program, including test data, etc.: } 200Ko\\
 {\em Distribution format:} .zip\\
 {\em Computer:} PC, Mac, Super-computer.\\
 {\em Operating system:} Mac OS, Linux, Windows.\\
%{\em RAM:}  Megabytes\\
%{\em PACS:} 02.60.-x, 02.70.-c, 31.15.-p, 31.15.xf.\\
%{\em Classification:} \\

% 50-250 words (per journal's instructions).

\noindent
{\em Nature of problem:} Among the plethora of configurations that
may exist in Gross-Pitaevskii (GP) equations modeling  one or two-component Bose-Einstein
condensates, only the ones that are deemed spectrally stable (or even,
in some cases, weakly unstable) have high probability to be observed
in realistic ultracold atoms experiments. To investigate the spectral
stability of solutions requires the numerical study of the linearization
of GP equations, the latter commonly known as the Bogoliubov-de Gennes
(BdG) spectral problem. 
The present software offers an efficient and reliable tool for the computation of 
eigenvalues (or modes) of the BdG problem for a given two- or three-dimensional GP
configuration. Then, the spectral stability (or instability) can be inferred
from its spectrum, thus predicting (or not) its observability in experiments.\\
% 50-250 words (per journal's instructions).

\noindent{\em Solution method:}
The present toolbox in \ff consists of
the following steps. At first, the GP equations in two (2D) and three (3D) spatial
dimensions are discretized
by using P2 (piece-wise quadratic) Galerkin triangular (in 2D) or tetrahedral
(in 3D) finite elements. For a given configuration of interest, mesh adaptivity
in \ff is deployed in order to reduce the size of the problem, thus reducing
the toolbox's execution time. Then, stationary states of the GP equations are
obtained by a Newton method whose backbone involves the use  of a reliable and efficient linear solver 
 judiciously selected
from the \texttt{PETSc}%
\footnote{\texttt{https://petsc.org/}} library. Upon identifying
stationary configurations, to trace branches of such solutions a parameter continuation method over the chemical
potential in the GP equations (effectively controlling the number of atoms in a
BEC) is employed with step-size adaptivity of the continuation parameter. Finally, the computation of the stability
of branches of solutions (\ie the BdG spectrum), is carried out by accurately solving, at each point in the parameter space,  the underlying
eigenvalue problem by using the
\texttt{SLEPc}\footnote{\texttt{https://slepc.upv.es}} library. Three-dimensional computations are made affordable in 
the present toolbox by using the domain decomposition method (DDM). In the course of the
computation, the toolbox stores not only the solutions but also the eigenvalues and
respective eigenvectors emanating from the solution to the BdG problem. We offer examples
for computing stationary configurations and their BdG spectrum in one- and two-component
GP equations.\\
{\em Running time:} From minutes to hours depending on the mesh resolution and space dimension.\\
\end{small}
%=================================================================================
%\pagebreak
%%% 1 %%%%%%%%%%%%%%%%%%%%%%%%%%%%%%%%%%%%%%%%%%%%%%%%%%%%%%%%%%%%%%%%%%%%%%%%%

%{
%\tableofcontents
%}

%\pagebreak

\section{Introduction}
% =================================================================================================
%\tableofcontents
The study of Bose-Einstein condensates (BECs) has admittedly enjoyed a substantial
interest for more than two decades since their first observation in ultracold atoms
experiments \citep{anderson1995observation,davis1995bose}. Both theoretical and
experimental developments have been summarized in \citep{pethick_book2011,pitaevskii2015bose}.
These studies revealed the emergence of interesting wave configurations including vortices
and vortex structures \citep{fetter_prl2001,engels_2004,pgk_mod_2004,fetter_mod_2009},
and the quest for experimentally creating and studying new states has been an exciting
and active area of research. Indeed, a plethora of experimental techniques have been
developed including imprinting techniques \citep{PhysRevLett.83.2498,BEC-physV-2002-imprint,becker2008oscillations},
stirring the condensate above a certain critical angular
speed \citep{madison_prl_2000,BEC-physV-2001-haljan}, counterflow techniques \citep{yan2011multiple},
the use of anisotropic potentials \citep{theocharis2010multiple}, nonlinear interference
between different condensate fragments \citep{anderson_prl_2007} as well as the
so-called Kibble-Zurek mechanism \citep{anderson_nature_2008}, among many others.
The variety of configurations that have emerged through these studies is vast, and has
sparked theoretical and computational investigations over the years. Some basic examples
of such structures are dark solitons, single vortex lines (with I-, U- or S-shaped ones in
rotated BECs) \citep{dan-2003-aft}, as well as vortex rings \citep{pgk_siam_book} (see
also the review \citep{malomed_physicaD_2019} and references therein). More complex states,
such as multiple vortex lines and rings, vortex stars, and hopfions have also been reported
in the literature  (see for example \citep{dan-2004-cras,bisset2015robust,wang2017single}),
together with recent computational techniques for vortex identification  \cite{dan-2023-CPC-postproc}.
Alongside prototypical bound modes in multi-component BECs that can identified (\eg
dark-bright \citep{PhysRevE.91.012924}, vortex-bright \citep{kody_prl_2011,PhysRevE.94.022207}
and dark-antidark \citep{dan-2016-PRA}, as well as vortex-ring-bright and vortex-line-bright
solitons \citep{wang2017two}), more exotic configurations have been reported, including
skyrmions \citep{anglin_prl_2001,sutcliffe_prl_2002}, monopoles and Alice
rings \citep{anglin_prl_2006,PhysRevA.105.053303}. Even more, with the introduction of
state-of-the-art bifurcation techniques for partial differential equations (PDEs), more
and more multi-component solutions were identified \citep{charalampidis2020bifurcation,panos-boulle-2022}.

The principal model for the above theoretical and computational studies has
been the Gross-Pitaevskii (GP) equation \citep{pitaevskii2015bose} (and variants,
including multi-component settings) which is a PDE known to describe the
properties of a BEC in the mean-field approximation. Note  that the GP model is a nonlinear Schr\"odinger (NLS) equation that
incorporates an external potential to confine the atoms in the condensate \citep{pitaevskii2015bose}.
One of the key steps in these studies, however, is concerned with the response
of the pertinent waveforms under the presence of a perturbation induced,  \eg
by imperfections in the initial state preparation in the BEC. This crucial
step involves the study of the spectral stability \citep{keith_book} of the
solution to the GP equation at the theoretical/computational level, and it is
a two-fold process. At first, a stationary solution to the GP equation is identified by means of (spatial) discretization methods and
root-finding, \ie fixed-point techniques. Then, the GP equation is linearized
about this (stationary) solution, thus resulting into a spectral (eigenvalue)
problem, known as the Bogoliubov-de Gennes (BdG) problem \citep{Bogolyubov-1947,deGennes-1966}.
The numerical  solution of the BdG problem provides important information
about the spectral characteristics of waveforms that may have high probability
to be observed numerically if they are deemed stable (or even weakly unstable,
depending on the growth rates of the unstable eigenvalues).

Up until now, there has been a wide variety of publicly available programs
(written in \texttt{C}, \texttt{Fortran}, \texttt{MATLAB}, and \ff$\!\!$)
dedicated exclusively to the computation of stationary states to the GP
equation that employ spectral methods%
 \citep{BEC-CPC-2007-dion-cances,BEC-CPC-2013-Caliari,BEC-CPC-2014-antoine-duboscq}, %
finite elements \citep{BEC-CPC-2016-FEM,dan-2016-CPC} (see, also \citep{uecker_wetzel_rademacher_2014}
which includes a working example on GP equations),
and finite differences~%
\citep{BEC-CPC-2009-Muruganandam,BEC-CPC-2012-Vudragovic,BEC-CPC-2014-simplectic,%
BEC-CPC-2014-Hohenester,BEC-CPC-2019-rotating}. In almost all of these studies,
stationary solutions are computed when the L2-norm of the wave function is fixed
(this is accomplished by imposing a Lagrange multiplier constraint for the
number of atoms). Alternative approaches for the computation of solutions to
the GP equation involves the re-formulation of the problem as a bifurcation
one where the chemical potential (controlling the number of atoms) is varied
by using numerical continuation \citep{Allgower-1990} coupled with Newton's
method \citep{kelley_book_2003}. This approach has been adopted in a series of
studies that employ finite-element%
 \citep{dan-2016-PRA,carreterogonzalez2016vortex,boulle2020deflation,charalampidis2020bifurcation},
finite-difference,%
 \citep{bisset2015robust,wang2017single,charalampidis2020bifurcation},
as well as spectral (spatial) discretization methods \citep{wang2017single}.
However, to the best of our knowledge, a limited number of publicly available codes for studying the BdG
spectrum of configurations to the GP equation exist.
One such a code (written in \texttt{Fortran}) is the FACt toolbox \citep{arko2020fact}
which computes thermal fluctuations in BECs by solving the associated BdG
equations. Recent efforts in that same vein involve the publicly available toolbox
in \ff that was developed by a subset of the present authors \citep{sadaka_2023}.
It utilizes mesh adaptation techniques (that are built-in in \ff) and employs
the ARPACK eigenvalue solver \citep{hecht-2012-JNM} (which is interfaced with
\ff$\!\!$) for solving the BdG equations although the calculations therein
are carried out in sequential mode.

Building upon the recent work in \citep{sadaka_2023}, we present and distribute
herein a \textit{parallel} finite-element toolbox written in \ff for computing
the Bogoliubov-de Gennes (BdG) spectrum of stationary solutions to one- and
two-component Gross-Pitaevskii (GP) equations in 2D or 3D. The parallelization
of the toolbox relies exclusively on the recent interfacing of \ff with the
\texttt{PETSc} library  \cite{petsc} (see also \cite{pierre_interfacing}).
The combination of mesh adaptivity and the simplification in the use of parallel
linear solvers in \ff (such as distributed direct solvers and domain decomposition methods%
  \cite{pierre_book_2015,FFD:Tournier:2019}) renders the present toolbox an ideal
framework for computing configurations in one- and two-component BECs in 2D or
3D. This further paves the path for the efficient and reliable computation of
the BdG spectrum  by using the
\texttt{SLEPc} \citep{hernandez2005slepc} library. 

Our ultimate goal with the
present toolbox is to offer a versatile and reliable tool to the BEC community
which can perform parallel computations for exploring the existence and BdG
spectrum of 2D and 3D (one- or two-component) configurations of interest within
reasonable computational time. Finally,  the advantage of \ff
in hiding all technicalities of the finite-element
method and using a syntax close to the mathematical formulation of the problem
allows the user to focus on the mathematical and physical aspects of the problem
and easily make changes in the codes to simulate new configurations.

The structure of the paper is as follows. In Sec.  \ref{sec-theory}, we introduce
the one- and two-component GP equations together with the associated BdG models.
In Secs.  \ref{sec-num-meth-gp} and  \ref{sec-num-meth-bdg}, we describe the numerical
methods for computing stationary states to the GP equations and
their respective BdG spectra. We illustrate the validation of our programs in Secs.~%
\ref{sec-valid1c} and \ref{sec-valid2c}, whereas the architecture of the programs
and the description of parameter and output files is discussed in Sec.  \ref{sec-desc-prog}.
Finally, the main features of the toolbox are summarized in Sec.  \ref{sec-conclusions},
where we additionally offer some of its potential extensions.

% \pagebreak

\section{The Gross-Pitaevskii model and Bogoliubov-de Gennes equations}\label{sec-theory}
% ======================================================================
In this section, we present the theoretical setup of  the toolbox. We introduce the one- and two-component
Gross-Pitaevskii (GP) and Bogoliubov-de Gennes equations in Secs. \ref{sec-GP1}
and \ref{sec-GP2}, respectively. We would like to stress out that the model
equations below are expressed in adimensionalized form, and further details
about the physical units of the model equations together with their scaling
can be found in our recent contribution \cite{sadaka_2023} (and references
therein). 
%Moreover, the present toolbox follows and incorporates exactly the
%same scaling and formalism that was used in \cite{sadaka_2023}. 
For the user's
convenience, we include with this toolbox the example scripts (see files \texttt{phys\_to\_adim\_1comp.edp}
and \texttt{phys\_to\_adim\_2comp.edp} in the \texttt{Tools\_scaling} subdirectory)
 that compute non-dimensional parameters from physical values
corresponding to several experimental studies published in the literature.
These programs could guide the user in linking parameters of existing
experiments with non-dimensional parameters used in this contribution (and,
more generally, in theoretical studies).

\subsection{The one-component case: Gross-Pitaevskii and Bogoliubov-de
Gennes equations} \label{sec-GP1}
% ----------------------------------------------------------------------
The spatio-temporal behavior of a Bose-Einstein condensate (BEC) is
described by a complex-valued wave function
$\psi(\vec x,t):{\cal D}\times \R^{+} \rightarrow \C$, where ${\cal D} \in \R^d$  and $d$ is the spatial
dimension, \ie $d=1,2,3$. In the mean-field
approximation for the interparticle interactions in a BEC,
the wave function $\psi(\vec x,t)$ is a solution of the Gross-Pitaevskii
(GP) equation  \cite{pitaevskii2015bose}:
\begin{align}
\label{eq-scal-GP}
\ii \frac{\partial \psi}{\partial t}=-\frac{1}{2} \nabla^2  \psi %
+ C_\trap  \psi + \beta |\psi|^2  \psi,
\end{align}
where $\nabla^2$ stands for the Laplacian,
$\beta$ is the nonlinearity strength corresponding to repulsive ($\beta>0$) or
attractive ($\beta<0$) interactions
and $\ii$ is the imaginary unit ($\ii^2=-1$). The external potential $C_\trap(\vec{x})$ confining the atoms in the condensate is taken to be the harmonic
oscillator potential 
\begin{align}
\label{eq-scal-GP-V}
C_\trap(x,y,z)=\frac{1}{2}\left(
\omega_{x}^2 {x}^2 + \omega_y^2 {y}^2 + \omega_z^2 {z}^2 \right), \,\, %
\end{align}
where $\omega_{x}, \omega_{y}, \omega_{z}$ are the trapping frequencies.
Note that the interplay of different aspect ratios of the
trapping frequencies can lead to difference BEC scenarios  \cite{pgk_siam_book}.
For example, the case $\omega_{x}\equiv\omega_{y}\gg \omega_{z}$ leads to
1D BEC configurations (ideal to study bright/dark solitons), whereas 
$\omega_{x}\equiv\omega_{y}\ll \omega_{z}$ leads to 2D configurations (\eg
ideal to study vortices). For 3D configurations (\eg where vortex lines and vortex rings could be observed),
usually $\omega_{x}\equiv\omega_{y}\sim \omega_{z}$ (the interested
reader may also want to see the so-called dimension-reduction approach in
 \citep{frantzeskakis2010dark,bao2013mathematical}).

Stationary solutions to Eq.  \eqref{eq-scal-GP} are sought by using the standing wave ansatz
\begin{align}
\label{eq-GP-phi}
\psi({\vec x},t) = \phi({\vec x})e^{-\ii\mu t},
\end{align}
where $\mu$ is the chemical potential. The stationary GP equation is then obtained:
\begin{align}
\label{eq-scal-GP-stat}
 -\frac{1}{2} \nabla^2  \phi + C_\trap   \phi + %
 \beta |\phi|^2  = \mu  \phi.
\end{align}
The presence of the external potential $C_\trap$ in \eqref{eq-scal-GP-stat}
makes the atomic density $n({\vec x})=|\phi({\vec x})|^2$ vanish rapidly
outside the condensate.
%(specified by the Thomas-Fermi radius, see Eqs.~(8)
%and (9) in  \cite{sadaka_2023}, as well as  \cite{pgk_siam_book}). 
This
necessitates the use of homogeneous Dirichlet boundary conditions for the
stationary wave function, \ie we impose $\phi=0$ on $\partial \cal D$.

In this work, we compute stationary solutions for fixed $\mu$. Branches
of such solutions are obtained by performing numerical continuation  \cite{Allgower-1990}
over $\mu$, which corresponds here  to a bifurcation parameter. At each
step in the continuation process, we monitor the energy
\begin{align}\label{eq-NRJ}
{\mathcal E}(\phi) = \int_{\cal D}
\left(\frac{1}{2}|\nabla\phi({\vec x})|^2 + %
V_\trap({\vec x})|\phi({\vec x})|^2 + \frac{\beta}{2}|\phi({\vec x})|^4 \right)d{\vec x},
\end{align}
and the L2-norm of the solution $\phi$
\begin{align}
\label{eq-GP-N}
N(\phi)=\int_{\cal D} \phi\overline{ \phi}\, d{\vec x}=\int_{\cal D} |\phi|^2 d{\vec x},
\end{align}
the latter representing the total number of atoms in the condensate
(the overbar stands for complex conjugation). Note that both the
energy and number of atoms
are conserved quantities for the time-dependent GP equation   \eqref{eq-scal-GP}.
%

%Having presented the theoretical setup for finding stationary solutions
%$\phi$ (and branches thereof) to the GP equation [cf.~Eq.  \eqref{eq-scal-GP-stat}],
We proceed with the setup of the spectral stability analysis problem.
This is obtained by linearizing Eq.  \eqref{eq-scal-GP}, and this process
gives the Bogoliubov-de Gennes (BdG) problem we aim to solve. 
We first consider the ansatz
\begin{align}
\psi(\vec{x},t) =\left[\phi(\vec{x}) + %
\delta\left(A(\vec{x}) e^{-\ii\omega t} + %
\overline{B}(\vec{x}) e^{\ii\overline{\omega}t}\right)\right]%
e^{-\ii\mu t}, \quad \delta\ll 1,
\label{eq-BdG-psi}
\end{align}
where $\phi(\vec{x})$ is a stationary state (that we linearize
the GP equation about), $A$ and $B$ are complex-valued functions,
and $\omega$ is a complex number. Upon inserting \eqref{eq-BdG-psi}
into  \eqref{eq-scal-GP}, we obtain at order $\mathcal{O}(\delta)$
the linear eigenvalue problem called the  BdG equation:
\begin{align}
\label{eq-scal-BdG}
\begin{pmatrix}
{\mathcal H} -\mu + 2 \beta|\phi|^2 & \beta\phi^2\\
-\beta{\overline{\phi}}^2 & -({\mathcal H} -\mu + 2 \beta|\phi|^2)\\
\end{pmatrix}
\begin{pmatrix} A \\ B \end{pmatrix} =
\omega \begin{pmatrix} A \\ B \end{pmatrix},
\end{align}
where 
\begin{equation}
	\label{eq-scal-BdG-H}
{\mathcal H} \equiv  -\frac{1}{2} \nabla^2  + C_\trap.
\end{equation}
 
 The present toolbox
computes the eigenvalue-eigenvector pair $(\omega, A, B)$ for a given
stationary solution $\phi \in \C$. Note that $A$ and $B$ represent
the components of the eigenvector $(A, B)$ in Eq.  \eqref{eq-scal-BdG}.
We direct the reader to  \cite{sadaka_2023} for a detailed discussion
on the properties of the BdG problem mentioned above.
We recall the definition of the Krein signature 
	\begin{equation}
K= sign(\omega \int_{\cal D} \left(|A|^2 - |B|^2\right) d\vec{x}).
	\label{eq-BdG-Krein}
\end{equation}
If $ \int_{\cal D} \left(|A|^2 - |B|^2\right) d\vec{x}  \neq 0$, only real eigenvalues $\omega$ are possible  and $K$ becomes an important diagnostic tool assessing on the energetic stability of a solution $\phi$: if $K>0$ for all modes, then $\phi$ is the global minimum of the energy, \ie the ground state; if there exists a mode with $K < 0$, then the excitation reduces the energy of the system and the stationary state is thus energetically unstable, \ie excited state (or local minimum of the energy). If $\int_{\cal D} \left(|A|^2 - |B|^2\right) d\vec{x}  =0$, complex eigenvalues $\omega=\omega_r+i \omega_i$ are possible; moreover,  if $\omega_i \ne 0$, then the BdG mode is dynamically unstable.

\subsection{The two-component case: Gross-Pitaevskii and Bogoliubov-de %
Gennes equations} \label{sec-GP2}
% ----------------------------------------------------------------------
In the mean-field approximation, a mixture of two BECs  (\eg different hyperfine states of the same species) is
described by a coupled system of GP equations  \cite{pethick_book2011,pitaevskii2015bose}.
We consider here two wave functions $\psi_1$ and
$\psi_2$ accounting for the two components  and satisfying the dimensionless coupled system of GP equations  \cite{pgk_siam_book}:
\begin{align}
\label{eq-scal-GP2c}
\begin{dcases}
\ii\frac{\partial \psi_1}{\partial t} = \left(-\frac{1}{2} %
\nabla^2 + C_\trap + \beta_{11} |\psi_1|^2  %
+ \beta_{12} |\psi_2|^2\right) \psi_1,\\
\ii\frac{\partial \psi_2}{\partial t} = \left(-\frac{1}{2} %
\nabla^2 + C_\trap + \beta_{21} |\psi_1|^2 %
+ \beta_{22} |\psi_2|^2\right) \psi_2.\\
\end{dcases}
\end{align}
Coefficients $\beta_{11}$ and $\beta_{22}$ in  \eqref{eq-scal-GP2c}
represent the interaction strengths between atoms of same species (or
spin states) whereas the $\beta_{12}$ and $\beta_{21}$ represent the
ones between different species. Note that Eqs.  \eqref{eq-scal-GP2c} consider (for the sake of
simplicity) the same trapping potential $C_\trap$ (given by Eq.  \eqref{eq-scal-GP-V})
across the two components. If necessary, the user of the present toolbox
can easily implement different potentials in the provided scripts (see
the next section).

Similarly to the one-component case, we compute stationary states
to Eqs.  \eqref{eq-scal-GP2c} by using the separation of variables
Ans\"atze
\begin{align}
\label{eq-GP2c-phi}
\psi_j({\vec x},t) = \phi_j({\vec x}) e^{-\ii\mu_j t}, \quad j=1,2,
\end{align}
with chemical potentials $\mu_1$ and $\mu_2$. Indeed, if we plug
 \eqref{eq-GP2c-phi} into \eqref{eq-scal-GP2c}, we obtain
the following coupled system of (stationary) GP equations
\begin{align}
\label{eq-scal-GP2c-stat}
\begin{dcases}
\mu_1\phi_1= \left(-\frac{1}{2} \nabla^2 + C_\trap %
+ \beta_{11} |\phi_1|^2  + \beta_{12} |\phi_2|^2\right) \phi_1,\\
\mu_2\phi_2 = \left(-\frac{1}{2} \nabla^2 + C_\trap %
+ \beta_{21} |\phi_1|^2+ \beta_{22} |\phi_2|^2\right)  \phi_2,
\end{dcases}
\end{align}
with homogeneous Dirichlet boundary conditions,
$\phi_{j}=0$  on $\partial \cal D$ for $j=1,2$. 
The system \eqref{eq-scal-GP2c-stat} is solved (for fixed values of $\mu_1$ and $\mu_2$) using a Newton method that we discuss in the next section.
The characterization  of a stationary solution (and branches of
solutions thereof) to the GP system \eqref{eq-scal-GP2c-stat}
is based on the total energy
\begin{align}
\label{eq-GP2c-energ}
\ds {\cal E}(\phi_1,\phi_2) = \ds \int_{\cal D} %
\sum_{i=1}^2 \left(\frac{1}{2} |\nabla \phi_i|^2 + C_\trap\, |\phi_i|^2 +
\frac{1}{2}\sum_{j=1}^2  \beta_{ij}  |\phi_i|^2 |\phi_j|^2 \right)  d{\vec x},
\end{align}
as well as the total number of atoms
\begin{align}
\label{eq-GP2c-N}
N(\phi_{1},\phi_{2})=N(\phi_{1})+N(\phi_{2}),
\end{align}
where $N(\cdot)$ is given by Eq.  \eqref{eq-GP-N}.

To study the spectral stability of the
solutions we proceed as follows.
We consider the perturbation Ans\"atze
\begin{subequations}
\begin{align}
\label{eq-BdG-psi_1}
\psi_{1}(\vec{x},t)&=\left[\phi_{1}(\vec{x})+\delta%
\left(A(\vec{x}) e^{-\ii\omega t} + \overline{B}(\vec{x}) e^{\ii\overline{\omega}t}\right)%
\right]e^{-\ii\mu_{1} t},\\
\label{eq-BdG-psi_2}
\psi_{2}(\vec{x},t)&=\left[\phi_{2}(\vec{x})+\delta%
\left(C(\vec{x}) e^{-\ii\omega t} + \overline{D}(\vec{x}) e^{\ii\overline{\omega}t}\right)%
\right]e^{-\ii\mu_{2} t},
\end{align}
\end{subequations}
where $A,B,C,D,\omega\in \C$. After plugging Eqs.  \eqref{eq-BdG-psi_1}-\eqref{eq-BdG-psi_2}
into Eqs.  \eqref{eq-scal-GP2c}, we obtain the BdG equations at order $\mathcal{O}(\delta)$
for the two-component case, conveniently written as
\begin{align}
\label{eq-BdG2c-M}
M\begin{pmatrix} A \\ B \\C \\D \end{pmatrix} =
\omega \begin{pmatrix} A \\ B \\C \\D \end{pmatrix},
\end{align}
where the matrix $M$ is:
\begin{align}
M =
\begin{pmatrix}
M_{11} & \beta_{11}\phi_1^2 & \beta_{12}\phi_1\overline{\phi_2} & \beta_{12}\phi_1\phi_2\\
-\beta_{11}\overline{\phi_1}^2 & M_{22} & -\beta_{12}\overline{\phi_1}\overline{\phi_2} %
& -\beta_{12}\overline{\phi_1}\phi_2\\
\beta_{21}\overline{\phi_1}\phi_2 & \beta_{21}\phi_1\phi_2 & M_{33} & \beta_{22}\phi_2^2\\
-\beta_{21}\overline{\phi_1}\overline{\phi_2} & -\beta_{21}\phi_1\overline{\phi_2} %
& -\beta_{22}\overline{\phi_2}^2 & M_{44}
\end{pmatrix},
\end{align}
with matrix elements
\begin{align}
\label{eq-BdG2c-M1}
\begin{cases}
M_{11} &= {\mathcal H} - \mu_1 + 2\beta_{11}|\phi_1|^2 + \beta_{12}|\phi_2|^2,\\
M_{22} &=  -M_{11},\\
M_{33} &= {\mathcal H} - \mu_2 + \beta_{21}|\phi_1|^2 + 2\beta_{22}|\phi_2|^2,\\
M_{44} &=  -M_{33},
\end{cases}
\end{align}
and ${\mathcal H}$ is given by Eq.  \eqref{eq-scal-BdG-H}. 
%Having presented
%the theoretical setup for the existence and BdG analysis of stationary
%solutions for the one- and two-component GP equations, we discuss next
%their computational setup and implementation in \ff.

%%%%%%%%%%%%%%%%%%%%%%%%%%%%%%%%%%%%%%%%%%%%%%%%%%%%%%%%%%%%%%%%
\section{The computation of stationary solutions to the GP equations}\label{sec-num-meth-gp}
%%%%%%%%%%%%%%%%%%%%%%%%%%%%%%%%%%%%%%%%%%%%%%%%%%%%%%%%%%%%%%%%
In this section we discuss the computational methods we employed
in \ff in order to obtain stationary solutions to the GP equations.
We begin our discussion by considering the existence problem
for the one-component setting in Sec.  \ref{subsec-num-meth-GP} first, and
then move on with the two-component case in Sec.  \ref{subsec-num-meth-GP2c}.
Both implementations in \ff are discussed in Sec.  \ref{ff-impl}.

\subsection{Newton's method for a single-component BEC}\label{subsec-num-meth-GP}
% ----------------------------------------------------------------------

For the computation of stationary solutions to Eq.  \eqref{eq-scal-GP-stat},
we use Newton's method  \cite{kelley_book_2003}. We first
split the complex-valued wave function $\phi$ into real and imaginary parts
via $\phi = \phi_r+\ii\,\phi_i$ and obtain from  \eqref{eq-scal-GP-stat}
the following coupled system of nonlinear equations
\begin{align}
\label{eq-num-GP-stat}
\begin{dcases}
-\frac{1}{2} \nabla^2 \phi_r + C_\trap \phi_r + %
\beta f(\phi_r,\phi_i)\phi_r -\mu\phi_r \!\!\!\!\!&=0,\\
-\frac{1}{2} \nabla^2 \phi_i + C_\trap \phi_i + %
\beta f(\phi_r,\phi_i)\phi_i - \mu\phi_i \!\!\!\!\!&=0.
\end{dcases}
\end{align}
We introduced in \eqref{eq-num-GP-stat} the (scalar) function 
$f(\phi_r,\phi_i)=|\phi|^2=\phi_r^2+\phi_i^2$ $f$ that corresponds to the cubic nonlinearity in the GP  equation. 
Note that the  expression
of $f$ has been programmed in the toolbox in a general way;
other types of expressions (corresponding to the GP equation with different nonlinearity than cubic) can be used and easily adopted in the
toolbox.

The homogeneous Dirichlet conditions for the complex-valued
wave function $\phi$ translate into imposing the same boundary
conditions for $\phi_r$ and $\phi_i$, that is $\phi_r=\phi_i=0$ on $\pl \mathcal{D}$.
After setting the classical Sobolev spaces  \cite{adams_sobolev_book}
$V=H^1_0(\mathcal{D})$ for $\phi_r$ and $\phi_i$, we define the weak formulation (mandatory for the finite-element implementation)
of Eq.  \eqref{eq-num-GP-stat} as:
find $(\phi_r,\phi_i) \in V\times V=V^2$, such that for all test functions
$(v_r,v_i) \in V^2$
\begin{align}
\label{eq-num-GP-stat-weak}
\begin{dcases}
\begin{aligned}
\mathcal F_r(\phi_r,\phi_i,v_r) &= \int_{\mathcal D} (C_\trap - \mu)\phi_r v_r %
+\int_{\mathcal D}\frac{1}{2}\nabla \phi_r\cdot\nabla v_r + %
\int_{\mathcal D} \beta f(\phi_r,\phi_i)\phi_r v_r\!\!\!\!\!&=0,\\
\mathcal F_i(\phi_r,\phi_i, v_i) &= \int_{\mathcal D}(C_\trap - \mu)\phi_i v_i %
+\int_{\mathcal D}\frac{1}{2}\nabla \phi_i\cdot\nabla v_i + %
\int_{\mathcal D}\beta f(\phi_r,\phi_i)\phi_i v_i\!\!\!\!\!&=0.
\end{aligned}
\end{dcases}
\end{align}

The above coupled system of nonlinear equations
is discretized using finite elements in \ff (see Sec.  \ref{ff-impl}),
and solved by means of Newton's method which requires a sufficiently
good initial guess. For a given value of  $\mu$ and an initial guess $(\phi_r^{0},\phi_i^{0})$,
Newton's method computes corrections to the solution components
$(\phi_r,\phi_i)$ iteratively via
\begin{align}
q = \phi_r^k-\phi_r^{k+1},\quad s = \phi_i^k-\phi_i^{k+1}, \quad k \geq 0,
\end{align}
where $q$ and $s$ are solutions of the linearized equations
\begin{eqnarray}
\label{eq-num-GP-stat-diff-sys}
\begin{pmatrix}
\left(\dfrac{\pl \mathcal F_r}{\pl \phi_r}\right)_{\phi_r=%
\phi_r^k,\phi_i = \phi_i^k} & \left(\dfrac{\pl \mathcal F_r}{\pl \phi_i}\right)_{\phi_r%
=\phi_r^k,\phi_i = \phi_i^k} \\ \left(\dfrac{\pl \mathcal F_i}{\pl \phi_r}\right)_{\phi_r%
=\phi_r^k,\phi_i = \phi_i^k} & \left(\dfrac{\pl \mathcal F_i}{\pl \phi_i}\right)_{\phi_r%
=\phi_r^k,\phi_i = \phi_i^k}
\end{pmatrix}
\begin{pmatrix}
q \\ s
\end{pmatrix}
= \begin{pmatrix}
\mathcal F_r(\phi_r^k,\phi_i^k, v_r)\\
\mathcal F_i(\phi_r^k,\phi_i^k, v_i)
\end{pmatrix},
\end{eqnarray}
with the corresponding weak formulation
{\small
\begin{align}\label{eq-num-GP-stat-Newton}
\begin{dcases}
\begin{aligned}
&\int_{\mathcal D}(C_\trap - \mu) q v_r+%
\int_{\mathcal D}\frac{1}{2}\nabla  q\cdot %
\nabla v_r + \int_{\mathcal D} \beta\left(\dfrac{\pl f}{\pl\phi_r}(\phi_r^k,\phi_i^k)\,\phi_r^k q %
+ \dfrac{\pl f}{\pl\phi_i}(\phi_r^k,\phi_i^k)\, \phi_r^k s + f(\phi_r^k,\phi_i^k) q\right) v_r\\
&\hspace{1cm}=\int_{\mathcal D}(C_\trap - \mu)\phi_r^k v_r %
+ \int_{\mathcal D}\frac{1}{2}\nabla \phi_{r}^k\cdot \nabla v_r %
+ \int_{\mathcal D} \beta f(\phi_r^k,\phi_i^k)\phi_r^k v_r,\\
&\int_{\mathcal D}(C_\trap - \mu) s v_i%
+\int_{\mathcal D}\frac{1}{2}\nabla  s\cdot\nabla v_i %
+ \int_{\mathcal D} \beta\left(\dfrac{\pl f}{\pl\phi_r}(\phi_r^k,\phi_i^k)\,\phi_i^k q %
+ \dfrac{\pl f}{\pl\phi_i}(\phi_r^k,\phi_i^k)\,  \phi_i^k s %
+ f(\phi_r^k,\phi_i^k) s\right) v_i\\
&\hspace{1cm} =\int_{\mathcal D}(C_\trap-\mu) \phi_i^k v_i
+\int_{\mathcal D}\frac{1}{2}\nabla \phi_i^k\cdot\nabla v_i %
+ \int_{\mathcal D} \beta f(\phi_r^k,\phi_i^k)\phi_i^k v_i.
\end{aligned}
\end{dcases}
\end{align}
}

Note that the implementation of Eqs.  \eqref{eq-num-GP-stat-Newton}
in \ff takes a form very similar to the mathematical formulation
of the problem due to its versatile metalanguage used therein.
This is an advantage for the user who can thus build bug-free
numerical codes when cumbersome mathematical expressions are coded.

\subsection{Newton method for the two-component BEC}\label{subsec-num-meth-GP2c}
% ----------------------------------------------------------------------

The two-component GP system  \eqref{eq-scal-GP2c-stat}
is solved similarly by means of Newton's method, after splitting $\phi_{1}$ and $\phi_{2}$
into real and imaginary parts via $\phi_1 = \phi_{1r}+\ii\,\phi_{1i}$ and
$\phi_{2} = \phi_{2r} + \ii\,\phi_{2i}$. Equations  \eqref{eq-scal-GP2c-stat}
are thus converted into a system consisting of four real-valued (nonlinear) equations:
\begin{align}
	\label{eq-num-GP2c-stat}
\begin{dcases}
- \frac{1}{2} \nabla^2 \phi_{1r} + (C_\trap-\mu_1) \phi_{1r}
+ \beta_{11} f(\phi_{1r},\phi_{1i})\phi_{1r} + \beta_{12} f(\phi_{2r},\phi_{2i})\phi_{1r}\!\!\!\!\!&= 0,\\
- \frac{1}{2} \nabla^2 \phi_{1i} + (C_\trap-\mu_1) \phi_{1i} + \beta_{11}f(\phi_{1r},\phi_{1i})\phi_{1i} + %
\beta_{12} f(\phi_{2r},\phi_{2i})\phi_{1i}\!\!\!\!\!&=0,\\
- \frac{1}{2} \nabla^2 \phi_{2r} + (C_\trap-\mu_2) \phi_{2r}
+ \beta_{21} f(\phi_{1r},\phi_{1i})\phi_{2r} + \beta_{22} f(\phi_{2r},\phi_{2i})\phi_{2r} \!\!\!\!\!&=0,\\
- \frac{1}{2} \nabla^2 \phi_{2i} + (C_\trap-\mu_2) \phi_{2i} + \beta_{21} f(\phi_{1r},\phi_{1i})\phi_{2i}%
+ \beta_{22} f(\phi_{2r},\phi_{2i})\phi_{2i}\!\!\!\!\! &=0.
\end{dcases}
\end{align}
Again,  homogeneous Dirichlet boundary conditions on $\phi_{1}$
and $\phi_{2}$ are imposed:
$\phi_{1r}=\phi_{1i} = \phi_{2r} = \phi_{2i}=0$ on $\pl \mathcal{D}$.
The weak formulation of Eqs.  \eqref{eq-num-GP2c-stat}
can be written as follows: find $(\phi_{1r},\phi_{1i},\phi_{2r},\phi_{2i}) \in V^4$,
such that for all test functions $ (v_{1r}, v_{1i}, v_{2r}, v_{2i}) \in V^4$
{\small
\begin{align}
		\label{eq-num-GP2c-stat-weak}
\begin{dcases}
\begin{aligned}
\mathcal F_{1r} = &\int_{\mathcal D}(C_\trap - \mu_1)\phi_{1r} v_{1r} +\int_{\mathcal D}\frac{1}{2} \nabla \phi_{1r}\cdot\nabla v_{1r} + \int_{\mathcal D}(\beta_{11} f(\phi_{1r},\phi_{1i})+\beta_{12} f(\phi_{2r},\phi_{2i}))\phi_{1r} v_{1r} \!\!\!\!\!&= 0,\\
\mathcal F_{1i} = &\int_{\mathcal D}(C_\trap - \mu_1)\phi_{1i} v_{1i} + \int_{\mathcal D}\frac{1}{2} \nabla \phi_{1i}\cdot\nabla v_{1i} + \int_{\mathcal D}(\beta_{11}f(\phi_{1r},\phi_{1i})+\beta_{12} f(\phi_{2r},\phi_{2i}))\phi_{1i} v_{1i} \!\!\!\!\!&= 0,\\
\mathcal F_{2r} = &\int_{\mathcal D}(C_\trap -\mu_2)\phi_{2r} v_{2r} + \int_{\mathcal D}\frac{1}{2} \nabla \phi_{2r}\cdot\nabla v_{2r} + \int_{\mathcal D}(\beta_{21} f(\phi_{1r},\phi_{1i}) + \beta_{22} f(\phi_{2r},\phi_{2i}))\phi_{2r} v_{2r} \!\!\!\!\!&= 0,\\
\mathcal F_{2i} = &\int_{\mathcal D}(C_\trap-\mu_2)\phi_{2i} v_{2i} + \int_{\mathcal D}\frac{1}{2} \nabla \phi_{2i}\cdot\nabla v_{2i} + \int_{\mathcal D}(\beta_{21} f(\phi_{1r},\phi_{1i}) + \beta_{22} f(\phi_{2r},\phi_{2i}))\phi_{2i} v_{2i}\!\!\!\!\! &= 0.\\
\end{aligned}
\end{dcases}
\end{align}
}
Newton's method computes, for fixed chemical potentials $\mu_{1}$ and $\mu_{2}$ and given
initial guess $\left(\phi_{1r}^{0},\phi_{1i}^{0},\phi_{2r}^{0},\phi_{2i}^{0}\right)$,
 the corrections 
\begin{align}
q_1 = \phi_{1r}^k-\phi_{1r}^{k+1}, \quad s_1 = \phi_{1i}^k-\phi_{1i}^{k+1}, %
\quad q_2 = \phi_{2r}^k-\phi_{2r}^{k+1}, \quad s_2 = \phi_{2i}^k-\phi_{2i}^{k+1},
\end{align}
which are solutions to the following system of linear
equations:
{\small
\begin{equation}
\label{eq-num-GP2c-stat-Newton-1}
\begin{aligned}
& \int_{\mathcal D}(C_\trap-\mu_1)q_1  v_{1r} + %
\int_{\mathcal D} \frac{1}{2} \nabla q_1\cdot \nabla  v_{1r} %
+ \int_{\mathcal D} (\beta_{11} f(\phi_{1r}^k,\phi_{1i}^k) %
+ \beta_{12} f(\phi_{2r}^k,\phi_{2i}^k))q_1v_{1r}\\
&\hspace{1cm}+ \int_{\mathcal D} \beta_{11}\left(\dfrac{\pl f}{\pl\phi_r}(\phi_{1r}^k,\phi_{1i}^k)%
\phi_{1r}^k q_1 + \dfrac{\pl f}{\pl\phi_i}(\phi_{1r}^k,\phi_{1i}^k)\phi_{1r}^k s_1\right) v_{1r}\\
&\hspace{1cm}+ \int_{\mathcal D} \beta_{12}\left(\dfrac{\pl f}{\pl\phi_r}(\phi_{2r}^k,\phi_{2i}^k)%
\phi_{1r}^k q_2 + \dfrac{\pl f}{\pl\phi_i}(\phi_{2r}^k,\phi_{2i}^k)\phi_{1r}^k s_2\right) v_{1r}\\
&\hspace{.5cm}= \int_{\mathcal D}(C_\trap - \mu_1)\phi_{1r}^k v_{1r} +\int_{\mathcal D}\frac{1}{2} %
\nabla \phi_{1r}^k\cdot\nabla v_{1r} + \int_{\mathcal D}(\beta_{11} f(\phi_{1r}^k,\phi_{1i}^k)%
+\beta_{12} f(\phi_{2r}^k,\phi_{2i}^k))\phi_{1r}^k v_{1r},
\end{aligned}
\end{equation}
\begin{equation}
\label{eq-num-GP2c-stat-Newton-2}
\begin{aligned}
& \int_{\mathcal D}(C_\trap-\mu_1)s_1 v_{1i} %
+ \int_{\mathcal D}\frac{1}{2} \nabla s_1\cdot\nabla v_{1i} %
+ \int_{\mathcal D} (\beta_{11} f(\phi_{1r}^k,\phi_{1i}^k) %
+ \beta_{12} f(\phi_{2r}^k,\phi_{2i}^k))s_1v_{1i}\\
&\hspace{1cm}+ \int_{\mathcal D} \beta_{11}\left(\dfrac{\pl f}{\pl\phi_r}(\phi_{1r}^k,\phi_{1i}^k)%
\phi_{1i}^k q_1 + \dfrac{\pl f}{\pl\phi_i}(\phi_{1r}^k,\phi_{1i}^k)\phi_{1i}^k s_1\right) v_{1i}\\
&\hspace{1cm}+ \int_{\mathcal D} \beta_{12}\left(\dfrac{\pl f}{\pl\phi_r}(\phi_{2r}^k,\phi_{2i}^k)%
\phi_{1i}^k q_2 + \dfrac{\pl f}{\pl\phi_i}(\phi_{2r}^k,\phi_{2i}^k)\phi_{1i}^k s_2\right) v_{1i}\\
&\hspace{.5cm} = \int_{\mathcal D}(C_\trap - \mu_1)\phi_{1i}^k v_{1i} + \int_{\mathcal D}\frac{1}{2} %
\nabla \phi_{1i}^k\cdot\nabla v_{1i} + \int_{\mathcal D}(\beta_{11}f(\phi_{1r}^k,\phi_{1i}^k)%
+\beta_{12} f(\phi_{2r}^k,\phi_{2i}^k))\phi_{1i}^k v_{1i},
\end{aligned}
\end{equation}
\begin{equation}
\label{eq-num-GP2c-stat-Newton-3}
\begin{aligned}
& \int_{\mathcal D}(C_\trap-\mu_2)q_2 v_{2r} %
+\int_{\mathcal D} \frac{1}{2} \nabla q_2\cdot\nabla v_{2r} %
+ \int_{\mathcal D} (\beta_{22} f(\phi_{2r}^k,\phi_{2i}^k) %
+ \beta_{21} f(\phi_{1r}^k,\phi_{1i}^k))q_2v_{2r}\\
&\hspace{1cm}+ \int_{\mathcal D} \beta_{21}\left(\dfrac{\pl f}{\pl\phi_r}(\phi_{1r}^k,\phi_{1i}^k)%
\phi_{2r}^k q_1 + \dfrac{\pl f}{\pl\phi_i}(\phi_{1r}^k,\phi_{1i}^k)\phi_{2r}^k s_1\right) v_{2r}\\
&\hspace{1cm}+ \int_{\mathcal D} \beta_{22}\left(\dfrac{\pl f}{\pl\phi_r}(\phi_{2r}^k,\phi_{2i}^k)%
\phi_{2r}^k q_2 + \dfrac{\pl f}{\pl\phi_i}(\phi_{2r}^k,\phi_{2i}^k)\phi_{2r}^k s_2\right) v_{2r}\\
&\hspace{.5cm} = \int_{\mathcal D}(C_\trap -\mu_2)\phi_{2r}^k v_{2r} + \int_{\mathcal D}\frac{1}{2} %
\nabla \phi_{2r}^k\cdot\nabla v_{2r} + \int_{\mathcal D}(\beta_{21} f(\phi_{1r}^k,\phi_{1i}^k) %
+ \beta_{22} f(\phi_{2r}^k,\phi_{2i}^k))\phi_{2r}^k v_{2r},
\end{aligned}
\end{equation}
\begin{equation}
\label{eq-num-GP2c-stat-Newton-4}
\begin{aligned}
& \int_{\mathcal D}(C_\trap-\mu_2)s_2 v_{2i} %
+\int_{\mathcal D} \frac{1}{2} \nabla s_2\cdot\nabla v_{2i} %
+ \int_{\mathcal D} (\beta_{22} f(\phi_{2r}^k,\phi_{2i}^k) %
+ \beta_{21} f(\phi_{1r}^k,\phi_{1i}^k))s_2v_{2i}\\
&\hspace{1cm}+ \int_{\mathcal D} \beta_{21}\left(\dfrac{\pl f}{\pl\phi_r}(\phi_{1r}^k,\phi_{1i}^k)%
\phi_{2i}^k q_1 + \dfrac{\pl f}{\pl\phi_i}(\phi_{1r}^k,\phi_{1i}^k)\phi_{2i}^k s_1\right) v_{2i}\\
&\hspace{1cm}+ \int_{\mathcal D} \beta_{22}\left(\dfrac{\pl f}{\pl\phi_r}(\phi_{2r}^k,\phi_{2i}^k)%
\phi_{2i}^k q_2 + \dfrac{\pl f}{\pl\phi_i}(\phi_{2r}^k,\phi_{2i}^k)\phi_{2i}^k s_2\right) v_{2i}\\
&\hspace{.5cm} = \int_{\mathcal D}(C_\trap -\mu_2)\phi_{2i}^k v_{2i} + \int_{\mathcal D}\frac{1}{2} %
\nabla \phi_{2i}^k\cdot\nabla v_{2i} + \int_{\mathcal D}(\beta_{21} f(\phi_{1r}^k,\phi_{1i}^k) %
+\beta_{22} f(\phi_{2r}^k,\phi_{2i}^k))\phi_{2i}^k v_{2i}.
\end{aligned}
\end{equation}
}Again, the implementation of Eqs.  \eqref{eq-num-GP2c-stat-Newton-1}-\eqref{eq-num-GP2c-stat-Newton-4}
with \ff is very similar to the mathematical formulation.

\subsection{Finite-element implementation with \ff}\label{ff-impl}
% ----------------------------------------------------------------------

We now present the finite-element implementation in the free software \ff \citep{hecht-2012-JNM} of the  weak formulations 
for the one- and two-component GP equations solved with Newton's method.
Note that the main principles of programming and numerical settings presented herein are shared with
the implementation of the BdG problem, see Sec.  \ref{sec-num-meth-bdg}.

One of the main advantages while programming in \ff is that cumbersome
formulas are coded in a compact form, and close to their mathematical
formulation. For example, the weak form of the system of linear
equations given by Eqs.  \eqref{eq-num-GP-stat-Newton} and used in Newton's
method is conveniently implemented as a list of rules and expressions
embodied in a \texttt{Macro} (see, in particular, \texttt{BdG\_1comp\_ddm/A\_macro/Macro\_problem.edp})
in which integral terms are easy to identify:

{\small\begin{lstlisting}[firstnumber=last]
NewMacro problemGP
	macro f(ur,ui) (ur^2 + ui^2)//
	macro dfdur(ur,ui) (2.*ur)//
	macro dfdui(ur,ui) (2.*ui)//
	
	varf vGP([q,s],[vr,vi]) = 
	intN(Th,qforder=ord)((Ctrap - mu)*q*vr + .5*grad(q)'*grad(vr)
	+ (Ctrap - mu)*s*vi + .5*grad(s)'*grad(vi)
	+ beta * (f(phir,phii)*q*vr + f(phir,phii)*s*vi)
	+ beta * phir*vr*(dfdur(phir,phii)*q + dfdui(phir,phii)*s)
	+ beta * phii*vi*(dfdur(phir,phii)*q + dfdui(phir,phii)*s))
	+ intN(Th,qforder=ord)((Ctrap - mu)*phir*vr + .5*grad(phir)'*grad(vr)
	+ (Ctrap - mu)*phii*vi + .5*grad(phii)'*grad(vi)
	+ beta * f(phir,phii) * (phir*vr + phii*vi))
	BCGP;
EndMacro
\end{lstlisting}}
\noindent
Another advantage of this formulation  in \ff
is that it can be invariantly used in any (spatial) dimension ($d=2$
or $d=3$), and for any available type of finite elements. This is
accomplished by simply declaring respective values in the files
defining the computational case. Indicatively, for the computation
of the 2D ground state
% in the large density limit case, \ie Thomas-Fermi
%(TF) limit  \cite{pitaevskii2015bose} 
using a $P2$ finite-element space,
the user can declare (see, in particular, the file \texttt{BdG\_1comp\_ddm/INIT/2D\_TF.inc}):
{\small\begin{lstlisting}[firstnumber=last]
macro dimension 2//
macro FEchoice P2//	
\end{lstlisting}}
\noindent
These choices are transmitted in the main programs, see,  \eg
\texttt{FFEM\_GP\_1c\_2D\_3D\_ddm.edp}:
{\small\begin{lstlisting}[firstnumber=last]
func Pk = [FEchoice,FEchoice];
...
meshN Th; // Local mesh
meshN ThBackup; // Global mesh
fespace Wh(Th,FEchoice);
fespace Whk(Th,Pk);
fespace WhBackup(ThBackup,FEchoice);
fespace WhkBackup(ThBackup,Pk);
...
Wh<complex> phi, phitemp; // Wavefunction
Whk [q,s], [phir,phii];
WhBackup<complex> phiBackup, phitempBackup; // Wavefunction
\end{lstlisting}}

Similarly, for the two-component case, the macro formulation for the
linear system given by Eqs.  \eqref{eq-num-GP2c-stat-Newton-1}-\eqref{eq-num-GP2c-stat-Newton-2}
can be found in the file  \texttt{BdG\_2comp\_ddm/A\_macro/Macro\_problem.edp},
and reads
\pagebreak

{\small\begin{lstlisting}[firstnumber=last]
NewMacro problemGP
	macro f(ur,ui) (ur^2 + ui^2)//
	macro dfdur(ur,ui) (2.*ur)//
	macro dfdui(ur,ui) (2.*ui)//
	
	varf vGP([q1,s1,q2,s2],[v1r,v1i,v2r,v2i])=
	intN(Th,qforder=ord)(
	1./2.*grad(q1)'*grad(v1r) + (Ctrap - mu1)*q1*v1r + (beta11*f(phi1r,phi1i) + beta12*f(phi2r,phi2i))*q1*v1r
	+ beta11*(dfdur(phi1r,phi1i)*phi1r*q1 + dfdui(phi1r,phi1i)*phi1r*s1)*v1r
	+ beta12*(dfdur(phi2r,phi2i)*phi1r*q2 + dfdui(phi2r,phi2i)*phi1r*s2)*v1r
	+1./2.*grad(s1)'*grad(v1i) + (Ctrap - mu1)*s1*v1i + (beta11*f(phi1r,phi1i) + beta12*f(phi2r,phi2i))*s1*v1i
	+ beta11*(dfdur(phi1r,phi1i)*phi1i*q1 + dfdui(phi1r,phi1i)*phi1i*s1)*v1i
	+ beta12*(dfdur(phi2r,phi2i)*phi1i*q2 + dfdui(phi2r,phi2i)*phi1i*s2)*v1i
	+1./2.*grad(q2)'*grad(v2r) + (Ctrap - mu2)*q2*v2r + (beta22*f(phi2r,phi2i) + beta21*f(phi1r,phi1i))*q2*v2r
	+ beta22*(dfdur(phi2r,phi2i)*phi2r*q2 + dfdui(phi2r,phi2i)*phi2r*s2)*v2r
	+ beta21*(dfdur(phi1r,phi1i)*phi2r*q1 + dfdui(phi1r,phi1i)*phi2r*s1)*v2r
	+1./2.*grad(s2)'*grad(v2i) + (Ctrap - mu2)*s2*v2i + (beta22*f(phi2r,phi2i) + beta21*f(phi1r,phi1i))*s2*v2i
	+ beta22*(dfdur(phi2r,phi2i)*phi2i*q2 + dfdui(phi2r,phi2i)*phi2i*s2)*v2i
	+ beta21*(dfdur(phi1r,phi1i)*phi2i*q1 + dfdui(phi1r,phi1i)*phi2i*s1)*v2i
	)
	+ intN(Th,qforder=ord)(
	1./2.*grad(phi1r)'*grad(v1r) + (Ctrap - mu1)*phi1r*v1r + (beta11*f(phi1r,phi1i) + beta12*f(phi2r,phi2i))*phi1r*v1r
	+ 1./2.*grad(phi1i)'*grad(v1i) + (Ctrap - mu1)*phi1i*v1i + (beta11*f(phi1r,phi1i) + beta12*f(phi2r,phi2i))*phi1i*v1i
	+ 1./2.*grad(phi2r)'*grad(v2r) + (Ctrap - mu2)*phi2r*v2r + (beta22*f(phi2r,phi2i) + beta21*f(phi1r,phi1i))*phi2r*v2r
	+ 1./2.*grad(phi2i)'*grad(v2i) + (Ctrap - mu2)*phi2i*v2i + (beta22*f(phi2r,phi2i) + beta21*f(phi1r,phi1i))*phi2i*v2i
	)
	BCGP;
EndMacro
\end{lstlisting}}
\noindent
We highlight here that the user has the flexibility to consider different
trapping in the two-component case if necessary. This can be accomplished
by modifying the \texttt{.inc} files located in the \texttt{INIT} subdirectories, and consider, for example,
\texttt{Ctrap1} and \texttt{Ctrap2} for the first and second components,
respectively.

The programs that we deliver with this toolbox consider $P2$
(piece-wise quadratic) finite elements. 
The mesh
in \ff (generically identified as  \texttt{Th})  is made
of triangles in 2D and tetrahedra in 3D. A fast mesh generator with a simple syntax is built in \ff\!.
A striking feature
of \ff is the ability to perform adaptive mesh refinement:
the grid is refined in regions of large gradients and coarsened
in low gradients ones. This is of paramount importance, especially
for high-dimensional problems where a sufficiently good resolution
of the solution is required. Using a very fine mesh
(with no mesh adaptation) for the entire domain would lead to a
large memory consumption and an excessively long computational
time. With the implementation of adaptive mesh refinement in the
present toolbox in \ff, we maintain reasonable problem sizes, and
thus computational time, while keeping a high degree of accuracy.
We briefly discuss next the key points of the mesh adaptation
techniques the toolbox employs (further details can be found in our
recent contribution  \citep{sadaka_2023}).

For solutions to the 2D GP equations at hand, the mesh is adapted by
using the built-in  \texttt{adaptmesh} command in \ff. In short, the underlying algorithm
modifies the inner products (which consider the Hessian of the solution by default) in the mesh generator to evaluate distance
and volume \citep{hecht-1996-missi,frey-george-1999,moham-piron-2000}.
This way, quasi-equilateral elements are constructed, accordingly to the new
metric. On the other hand, and for 3D configurations,
adaptive mesh refinement in \ff is performed through the use of the
libraries \texttt{mshmet} and \texttt{mmg} \citep{dapogny2014three} where
similar algorithms are employed. Note that the user can adjust
the values of \texttt{hmax} and \texttt{hmin} representing the maximum
and minimum edge sizes of the triangular mesh, respectively during the mesh
adaptation process. This offers the possibility to control the size of the
mesh, and thus find a trade-off between accuracy and computational cost.
While performing mesh adaptation, however, one faces with the important
question about what is the right choice of variables that is suitable for
controlling mesh adaptivity. In the present implementation for computing stationary
2D and 3D configurations to the GP equations, we use adaptive mesh refinement
based on the density of the solution as well as its real and imaginary parts.
This approach has been considered in  \cite{dan-2010-JCP}, and has been proven
quite effective in computing complicated vortex solutions.

The underlying nonlinear equations are solved by means of Newton's method
which is fed by an initial guess (with fixed chemical potential(s)), see
Secs.  \ref{sec-valid1c} and  \ref{sec-valid2c} for example cases. Newton's
iterations are stopped when one of the following criteria is satisfied:
\begin{align}\label{eq-bdg-err1}
\left\lVert
\begin{pmatrix}
q\\
s
\end{pmatrix}
\right\rVert_\infty < \epsilon_{\scriptscriptstyle q},
\quad
\left\lVert
\begin{pmatrix}
\mathcal{F}_r\\
\mathcal{F}_i
\end{pmatrix}
\right\rVert_2 < \epsilon_{\scriptscriptstyle F},
\end{align}
The former controls the convergence (in the infinity norm) in
Newton's method whereas the latter checks the accuracy of the solution
(the residual in the $L2$ norm). In practice, we use $\epsilon_{\scriptscriptstyle q} = 10^{-8}$
and $\epsilon_{\scriptscriptstyle F} = 10^{-16}$ but we found that
both criteria are satisfied simultaneously in all the cases we have
considered in this paper. Moreover, we note that convergence in Newton's method
depends crucially on the choice of the linear solver we employ.
Specifically, in 2D, we use an exact LU decomposition, as computed (within
the \texttt{SLEPc} library) by the MUMPS solver with options:\\
\texttt{"-pc\_type lu -ksp\_type preonly"}\\
The computational cost in 2D  is thus manageable.
For 3D cases, we switch to a more economical preconditioner, and in particular,
the algebraic multigrid method being available in HYPRE with options:\\
\texttt{"-pc\_type hypre -ksp\_type gmres -ksp\_atol 1e-12 -ksp\_rtol 1e-6 -ksp\_gmres\_restart 50 -ksp\_max\_it 500 -ksp\_pc\_side right -sub\_pc\_type lu }%
\newline
\texttt{-sub\_pc\_factor\_mat\_solver\_type mumps"}.
% For 2D problems  we solve the systems of Eqs.  \eqref{eq-num-GP-stat-Newton}
%(single-component case) and Eqs.  \eqref{eq-num-GP2c-stat-Newton-1}-\eqref{eq-num-GP2c-stat-Newton-4}
%(two-component case) with a direct LU method using the parameters
%in \texttt{PETSc}: \texttt{"-pc\_type lu -ksp\_type preonly"}. On the
%other hand, and for 3D problems, we use the following set of parameters:
%\texttt{"-pc\_type hypre -ksp\_type gmres -ksp\_atol 1e-12 -ksp\_rtol 1e-6 -ksp\_gmres\_restart 50 -ksp\_max\_it 500 -ksp\_pc\_side right -sub\_pc\_type lu}%
%\newline
%\texttt{-sub\_pc\_factor\_mat\_solver\_type mumps"}}.

The toolbox can trace branches of stationary configurations to the GP equations by
performing numerical continuation \citep{Allgower-1990} over the parameters of the
model. For the one-component case, we consider the chemical potential $\mu$ as our
principal continuation parameter. In particular, we start from a value of the chemical
potential $\mu_0$ for which the initial guess  is sufficiently close to the
stationary state of interest. Upon convergence in Newton's method (discussed above),
we use the resulting converged state as an initial guess for the next step in the
continuation process with chemical potential $\mu_0 + \delta\mu$. We highlight the
fact that we include a simple adaptive strategy for the selection of the increment
$\delta\mu$ in the toolbox, and it is described briefly next.
Initially, the $\delta\mu$
is fixed to $10^{-3}$ when $\mu_{0}$ is close to the respective state's linear limit.
It then gets doubled, \ie $\delta\mu=2\delta\mu$, at every 10 steps in the continuation
process until it reaches $\delta\mu_{max}=0.015$ whereupon it remains fixed, and the
continuation stops when the final value $\mu_f$  specified by the  user is reached.

Finally, for the two-component setting, we follow different continuation strategies
that involve relevant principal continuation parameters in order to match the toolbox's
results with ones that exist in the literature. For example, the 2D ring-antidark branch
is traced by performing continuation over $\mu_1$ and $\mu_2$ first, and then over the
inter-component interactions $\beta_{12}$ and $\beta_{21}$ (with fixed $\mu_{1}$ and $\mu_{2}$).%
The 2D vortex-antidark branch is traced by fixing $\mu_1$ and $\mu_2$ first,
and continuation over the inter-component interactions $\beta_{12}$ and $\beta_{21}$
(as for the 2D) is performed afterwards. Ultimately, various continuation strategies can
be conveniently designed and implemented in the toolbox by the user involving different
principal continuation parameters.

\pagebreak

\section{Solving the BdG equations}\label{sec-num-meth-bdg}
% ======================================================================
We solve the BdG problems for the one- and two-component cases of
using the \texttt{SLEPc} library \citep{hernandez2005slepc}
which is now interfaced with \ff. First, we need to write the
weak form of the BdG problems that will be supplied to the solver.
Indicatively, and for the one-component case, the weak formulation
of the BdG problem associated with Eq.  \eqref{eq-scal-BdG} reads:
\begin{align}
\label{eq-num-BdG-weak}
\begin{dcases}
\phantom{-}\int_{\mathcal D} \dfrac{1}{2}\nabla A\cdot\nabla v_1 %
+ \int_{\mathcal D} (C_\trap-\mu)A v_1 + \int_{\mathcal D}2\beta |\phi|^2 A v_1 %
+ \int_{\mathcal D}\beta \phi^2B v_1 = \omega \int_{\mathcal D}A v_1,\\
-\int_{\mathcal D}\dfrac{1}{2}\nabla B\cdot\nabla v_2 %
- \int_{\mathcal D}(C_\trap+\mu)B v_2  - \int_{\mathcal D} 2\beta |\phi|^2 B v_2 %
- \int_{\mathcal D} \beta \overline{\phi}^2 A v_2 = \omega \int_{\mathcal D}B v_2.
\end{dcases}
\end{align}
The bilinear terms in the left hand side of this equation form the finite-element
matrix $M$ that is fed to \texttt{SLEPc} library. The implementation of the BdG
problem of Eq.  \eqref{eq-num-BdG-weak} can be straightforwardly made now in \ff:
{\small\begin{lstlisting}[firstnumber=last]
NewMacro problemBdG
	varf vBdGMat([A,B],[v1,v2]) =
	intN(Th,qforder=ord)(.5*grad(v1)'*grad(A) +(Ctrap-mu)*A*v1'
	+ 2.*beta*abs(phi)^2*A*v1' + beta*phi^2*B*v1'
	- .5*grad(v2)'*grad(B) - (Ctrap-mu)*B*v2'
	- 2.*beta*abs(phi)^2*B*v2' - beta*(phi')^2*A*v2')
	BCBdG;

	varf vBdGVec([A,B],[v1,v2]) = intN(Th,qforder=ord)(A*v1' + B*v2');
EndMacro
\end{lstlisting}}
\noindent
It is easy to see the correspondence between the weak formulation of
Eq.  \eqref{eq-num-BdG-weak} and its implementation in the above macro
(see, also, the file \texttt{BdG\_1comp\_ddm/A\_macro/Macro\_problem.edp}).
For the computation of the BdG spectrum, we apply a small shift,  \eg
$\sigma = 10^{-4}$ or $\sigma = 10^{-2}$ that is implemented in the
\texttt{EPSSolve} function of \texttt{SLEPc} by using the parameters:\\
\texttt{"-st\_type sinvert -eps\_target sigma"} \\to slightly regularize
the eigenproblem.
%if we are close to the linear limit and  if we are far from it, this shift is implemented inside the \\
%
% \egc{Georges, are we sure about this? Yuji Nakatsukasa said this (verbatim):
% ``I currently use a  Krylov-Schur algorithm with a shift and invert technique
% to obtain eigenvalues near the target, which is set to 0 (to obtain the smallest
% magnitude eigenvalues) and yields a shift of 0. However, if the system has an
% eigenvalue close to the shift (0 is an eigenvalue in our problem), then the shifted
% linear system will give spurious eigenvalues. Therefore, I now set the shift to a
% positive value like 0.1 as it's quite unlikely to have an eigenvalue of this exact
% magnitude.''
%
% Recall that the BdG problem does have 0 eigenvalues!
% }

Upon computing the eigenvalues and eigenvectors of the BdG problem
in \texttt{SLEPc}, we further check their accuracy by displaying the
residual or Eq. \eqref{eq-scal-BdG}:
\begin{align}\label{eq-bdg-residual}
\left\lVert M \begin{pmatrix} A\\B \end{pmatrix}
- \omega \begin{pmatrix} A\\B \end{pmatrix}\right\rVert_\infty
\end{align}
by using the \texttt{SLEPc} parameters: \texttt{"-eps\_error\_backward ::ascii\_info\_detail"}.

\pagebreak

Finally, we present the weak formulation in the two-component case
emanating from Eqs.  \eqref{eq-BdG2c-M}-\eqref{eq-BdG2c-M1}:
\begin{equation}\label{eq-num-BdG2c-weak}
\begin{dcases}
\begin{aligned}
\int_{\mathcal D} \frac{1}{2}\nabla A\cdot \nabla v_1 &%
+ \int_{\mathcal D}(C_\trap - \mu_1) A v_1 %
+ \int_{\mathcal D} \left(2\beta_{11}|\phi_1|^2 %
+ \beta_{12}|\phi_2|^2\right)A v_1\\
& + \int_{\mathcal D} \beta_{11}\phi_1^2B v_1 %
+ \int_{\mathcal D} \beta_{12}\phi_1\overline{\phi_2}C v_1 %
+ \int_{\mathcal D} \beta_{12}\phi_1\phi_2 D v_1 = \omega \int_{\mathcal D} A v_1,\\
\end{aligned}\\
\begin{aligned}
-\int_{\mathcal D}\frac{1}{2}\nabla B\cdot\nabla v_2 %
&- \int_{\mathcal D}(C_\trap-\mu)B v_2 %
- \int_{\mathcal D}\left(2\beta_{11}|\phi_1|^2 + \beta_{12}|\phi_2|^2\right)B v_2 \\&%
- \int_{\mathcal D}\beta_{11}\overline{\phi_1}^2 A v_2  %
- \int_{\mathcal D}\beta_{12}\overline{\phi_1}\overline{\phi_2}C v_2 %
-\int_{\mathcal D}\beta_{12}\overline{\phi_1}\phi_2 D v_2 = \omega \int_{\mathcal D}B v_2,\\
\end{aligned}\\
\begin{aligned}
\int_{\mathcal D}\frac{1}{2}\nabla C\cdot \nabla v_3 &%
+ \int_{\mathcal D}(C_\trap-\mu)C v_3 + \int_{\mathcal D}\left(2\beta_{22}|\phi_2|^2 %
+ \beta_{21}|\phi_1|^2 \right)C v_3 \\&%
+ \int_{\mathcal D}\beta_{21}\overline{\phi_1}\phi_2 A v_3 %
+ \int_{\mathcal D}\beta_{21}\phi_1\phi_2 B v_3 %
+ \int_{\mathcal D}\beta_{22}\phi_2^2 D v_3 = \omega \int_{\mathcal D}C v_3,\\
\end{aligned}\\
\begin{aligned}
-\int_{\mathcal D}\frac{1}{2}\nabla D\cdot \nabla v_4 &%
- \int_{\mathcal D}(C_\trap-\mu)D v_4 - \int_{\mathcal D}\left(2\beta_{22}|\phi_2|^2 %
+ \beta_{21}|\phi_1|^2\right)D v_4 \\&%
-\int_{\mathcal D}\beta_{21}\overline{\phi_1}\overline{\phi_2}A v_4 %
- \int_{\mathcal D}\beta_{21}\phi_1\overline{\phi_2}B v_4 %
- \int_{\mathcal D}\beta_{22}\overline{\phi_2}^2C v_4 = \omega \int_{\mathcal D}D v_4.\\
\end{aligned}\\
\end{dcases}
\end{equation}
%
% \GS{Stathis Its implementation in \ff is similar to the single-component case,
% and very close to the mathematical formulation above, see
% \texttt{BdG\_2comp\_ddm/A\_macro/Macro\_problem.edp}).

Again, the implementation of the BdG problem of Eq.  \eqref{eq-num-BdG2c-weak}
is easy in \ff (see \texttt{BdG\_2comp\_ddm/A\_macro/Macro\_problem.edp}):
{\small\begin{lstlisting}[firstnumber=last]
NewMacro problemBdG
   varf vBdGMat([A,B,C,D],[v1,v2,v3,v4]) = 
      intN(Th,qforder=ord)(.5*grad(v1)'*grad(A) + (Ctrap - mu1)*A*v1' + (2.*beta11*un2(phi1,phi1) + beta12*un2(phi2,phi2))*A*v1'
         + beta11*phi1*phi1*B*v1' + beta12*phi1*phi2'*C*v1' + beta12*phi1*phi2*D*v1'
         -.5*grad(v2)'*grad(B) - (Ctrap - mu1)*B*v2' - (2.*beta11*un2(phi1,phi1) + beta12*un2(phi2,phi2))*B*v2'
         - beta11*phi1'*phi1'*A*v2' - beta12*phi1'*phi2'*C*v2' - beta12*phi1'*phi2*D*v2'
         +.5*grad(v3)'*grad(C) + (Ctrap - mu2)*C*v3' + (2.*beta22*un2(phi2,phi2) + beta21*un2(phi1,phi1))*C*v3'
         + beta22*phi2*phi2*D*v3' + beta21*phi1'*phi2*A*v3' + beta21*phi1*phi2*B*v3'
         -.5*grad(v4)'*grad(D) - (Ctrap - mu2)*D*v4' - (2.*beta22*un2(phi2,phi2) + beta21*un2(phi1,phi1))*D*v4'
         - beta22*phi2'*phi2'*C*v4' - beta21*phi1'*phi2'*A*v4' - beta21*phi1*phi2'*B*v4'
      )
      BCBdG;

   varf vBdGVec([A,B,C,D],[v1,v2,v3,v4]) = intN(Th,qforder=ord)(A*v1' + B*v2' + C*v3' + D*v4');
EndMacro
\end{lstlisting}}%

The validation of the new toolbox is a necessary task for assessing its
performance, fidelity and reliability. In Secs.  \ref{sec-valid1c}
and  \ref{sec-valid2c}, we perform a series of validation tests that
exist in the literature. Note that the test cases we
present next for the one-component case are the same as the ones
considered in \citep{sadaka_2023}.

\section{Validation test cases for the one-component BEC}\label{sec-valid1c}
% ======================================================================

We begin our discussion on validation test cases of our toolbox by
considering first the one-component GP model in 2D and 3D.~We note
that all the cases we present below consider repulsive interactions (we fix $\beta=1$) and an isotropic trap $C_\trap=\frac{1}{2} \omega_\perp^2 r^2$,
with $r^2=x^2+y^2+z^2$.

To ease our discussion, we further present a summary of the considered
cases, together with typical computational times and mesh sizes (\ie
the number of elements), in Table  \ref{tab-cputime-1c}.
Table  \ref{tab-cputime-1c-3D}
contains the number of unknowns (ndof), the number of tetrahedra (nt), and
the number of non-zero elements (nnz) of the matrix used for the computation
of the BdG spectra.
The toolbox initially builds a mesh by taking into
account the topology of the solution. For example, a disk-shaped mesh with
smaller triangles and minimum edge size $h_{min}=h_{max}/45$ in its center
is used for studying a 2D vortex configuration. The mesh is refined at
each iteration in Newton's method in regions of large gradients (\eg around
solitons or vortices) and de-refined otherwise (zones of constant density) when
adaptive mesh refinement is chosen, see Sec.  \ref{ff-impl}.
For a given
case, we draw comparisons in Table  \ref{tab-cputime-1c} between results
that were obtained with 4 MPI processors and without MPI using adaptive mesh
refinement. We draw also comparisons in Table  \ref{tab-cputime-1c-3D} for more
complex 3D cases. We note too, however, that it is safer to use adaptive mesh
refinement while exploring branches of solutions for which their topology is
unknown.

\begin{table}[h!]
\resizebox{\textwidth}{!}{%
%\begin{tabular}{l|ccc|ccc|}
%\cline{2-7}
%& \multicolumn{3}{c|}{1 MPI processor} & \multicolumn{3}{c|}{4 MPI processors}\\
%& CPU time GP  & CPU time BdG & Mesh size & CPU time GP & CPU time BdG & Mesh size \\ \hline
%\multicolumn{1}{|l|}{2D ground state}          & 00:00:05    & 00:00:26          & 10 900     & 00:00:03          & 00:00:09          & 10 853   \\
%\multicolumn{1}{|l|}{2D dark soliton}          &   00:12:40        & 00:58:18        & 20 956     & 00:06:58        & 00:26:01        & 19 859     \\
%%\multicolumn{1}{|l|}{2D central vortex}        & 00:09:16        & 00:53:20        & 13 300     & 00:09:16        & 00:53:20        & 13 300     \\
%\multicolumn{1}{|l|}{3D ground state}          & 00:31:53        & 06:12:47        & 46 786     & 00:14:44        & 04:58:57        & 47 097    \\
%\hline
%\end{tabular}%
%}
\begin{tabular}{l|cccc|ccc|}
\cline{2-8}
& \multicolumn{4}{c|}{without MPI} & \multicolumn{3}{c|}{4 MPI processors}\\
\cline{2-8}
&  & CPU time  & CPU time & & CPU time & CPU time & \\
& niter &  GP  & BdG & mesh size &GP & BdG & mesh size \\ \hline
\multicolumn{1}{|l|}{2D ground state}      & 1    & 00:00:05    & 00:00:26          & 10,900     & 00:00:03          & 00:00:12          & 10 866   \\
\multicolumn{1}{|l|}{2D dark soliton}        & 208  &   00:19:12        & 00:58:18        & 20,912     & 00:06:58        & 00:26:01        & 19 859     \\
%\multicolumn{1}{|l|}{2D central vortex}    &    & 00:09:16        & 00:53:20        & 13 300     & 00:09:16        & 00:53:20        & 13 300     \\
\multicolumn{1}{|l|}{3D ground state}      & 133    & 01:09:16        & 05:51:27        & 46,681     & 00:14:44        & 04:58:57        & 47 097    \\
\hline
\end{tabular}%
}
\caption{Results on test cases for the one-component GP and BdG problems
with mesh adaptivity. Results are presented with (4 MPI processors)
and without MPI. The computational time, the mesh size (number
of elements) and the number of continuation steps (niter) performed for each
case are shown. When using mesh adaptivity, the size of the mesh for the last
step of the continuation is depicted in the mesh size column. For 2D cases we compute 100 eigenvalues whereas for the 3D ground state, we compute
40 eigenvalues only. the user can compute
more eigenvalues if more memory is available. 
The BdG spectrum is computed every other two (continuation) steps in $\mu$.
The computations were performed on a Macbook pro M1, 16GB of DDR4 2400 MHz RAM.}
\label{tab-cputime-1c}
\end{table}

\begin{table}[h!]
\resizebox{\textwidth}{!}{%
\begin{tabular}{l|c|c|c|c|c|c|}
\cline{2-7}
\qquad GP test cases & Processors  &  CPU time  & niter & ndof & nt & maxRSS\\ \hline
\multicolumn{1}{|l|}{3D dark soliton}          & 28 & 00:01:01 & 168 & 205,822 & 76,455 & 0.78 Gb \\
\multicolumn{1}{|l|}{3D vortex line (1VL)}          & 28 & 00:04:02 & 168 & 329,988 & 122,969 & 0.79 Gb \\
\multicolumn{1}{|l|}{3D vortex ring + 2VLs}          & 56 & 00:05:19 & 201 & 654,802 & 244,597 & 1.10 Gb \\
\hline
\end{tabular}%
}
\vspace{.5cm}
%\end{table}

%\begin{table}[h]
\resizebox{\textwidth}{!}{%
\begin{tabular}{l|c|c|c|c|c|c|}
\cline{2-7}
\qquad BdG test cases & Processors  &  CPU time  & niter  & ndof & nnz & maxRSS\\ \hline
\multicolumn{1}{|l|}{3D dark soliton}          & 28 & 00:01:30 & 56 & 103,116 & 11,782,505 & 3.07 Gb \\
\multicolumn{1}{|l|}{3D vortex line (1VL)}          & 28 & 00:03:41 & 56 & 165,362 & 18,934,023 & 7.09 Gb \\
\multicolumn{1}{|l|}{3D vortex ring + 2VLs}          & 56 & 00:08:02 & 67  & 327,887 & 37,600,455 & 8.57 Gb \\
\hline
\end{tabular}%
}
\caption{Summary of results on 3D test cases for the one-component GP
and BdG problems with mesh adaptivity. The number of
processors, the mean CPU time per each continuation step, the total
number of continuation steps performed (for tracing the respective
branches) are shown. Moreover, the table contains the number of times
the BdG problem was solved (we computed the eigenvalues at every  3
continuation steps in $\mu$), the number of unknowns (ndof), the number of
tetrahedra (nt), the number of non-zero elements (nnz) of the matrix used
for the computation of the BdG spectra, the estimated memory used for
each processor maxRSS. For all test cases, 80 eigenvalues
were computed in the BdG problem only. Again, the user can compute
more eigenvalues if more memory is available. The present computations were performed
on the CRIANN Computing Center and MATRICS platform utilizing an Intel
Broadwell E5-2680 v4 @ 2.40GHz (14 cores per socket) architecture with two
sockets per node and 128 GB of DDR4 2400 MHz RAM.~An Intel Omnipath 100Gb/s low
latency network was used for communications.}
\label{tab-cputime-1c-3D}
\end{table}

%\begin{table}[h]
%\resizebox{\textwidth}{!}{%
%\begin{tabular}{l|c|c|c|c|c|}
%\cline{2-6}
%%& \multicolumn{3}{c|}{1 MPI processor} & \multicolumn{3}{c|}{4 MPI processors}   \\
%& Processes  & CPU time GP & CPU time BdG & Memory GP & Memory BdG\\ \hline
%\multicolumn{1}{|l|}{3D dark soliton}          & 28 & 03:12:20 & 02:01:40 & 20.88 Gb & 82.09 Gb \\
%\multicolumn{1}{|l|}{3D vortex line (1VL)}          & 28 & 11:49:01 & 04:22:17 & 21.05 Gb & 189.55 Gb  \\
%\multicolumn{1}{|l|}{3D vortex ring + 2VL}          & 56 & 18:42:31 & 10:20:49 & 58.84 Gb & 457.57 Gb  \\
%\hline
%\end{tabular}%
%}
%\caption{\GS{Results on test cases for the one-component GP and BdG equation
%with mesh adaptivity. In particular, the number of processors,
%the computational time for GP and for BdG, the used memory estimated for GP and for BdG for each
%case is shown. We note that all test cases for the BdG was done using 80 eigenvalues, we can use more 
%value but we will need to use more memory. All computation were performed using a parallel computer 
%(CRIANN Computing Center and MATRICS platform) based on Intel Broadwell E5-2680 v4 @ 2.40GHz 
%(14 cores per socket) architecture with two sockets per node and 128 GB of DDR4 
%2400 MHz RAM. An Intel Omnipath 100Gb/s low latency network was used for communications.}}
%\label{tab-cputime-1c-3D}
%\end{table}

\clearpage

\subsection{2D case: Ground state}
% ----------------------------------------------------------------------
The distribution of the BdG modes for oscillations of the ground
state in the TF limit for repulsive BECs was derived in \citep{kevrekidis2010distribution},
and is given by
\begin{align}
\label{eq-2DS-pred}
\omega_{m,k}^\TF = \omega_\perp \sqrt{m + 2k^2 + 2k(1+m)},
\end{align}
where $m, k \geq 0$ are integers.

In Table  \ref{tab-2D-TF}, we present
the first $20$ BdG modes our toolbox computed for $\mu=6$ and
$\omega_\perp = 0.2$. Moreover, we compare results that were obtained
with 4 MPI processors and without MPI against the ones from Eq.  \eqref{eq-2DS-pred}.
The Krein signatures  (see, Eq. \eqref{eq-BdG-Krein})
were computed too by the toolbox, and found all to be $1$, thus suggesting
the absence of negative energy modes. Note the perfect match between
numerical and theoretical results together with the fact that mesh adaptation
provides the same results as computations with a refined fixed mesh.
\begin{table}[h!]
\resizebox{\textwidth}{!}{%
\begin{tabular}{c|rrr|rrr|l}
\cline{2-7}
& \multicolumn{3}{c|}{without MPI}             & \multicolumn{3}{c|}{with 4 MPI processors } & \\ \cline{2-8}
& $Re(\omega)$ & $Im(\omega)$ & K & $Re(\omega)$ & $Im(\omega)$ & K & \multicolumn{1}{l|}{$\omega_{m,k}$ from \eqref{eq-2DS-pred}}                                       \\ \hline
\multicolumn{1}{|l|}{$\omega_1$}    & -5.86493e-07 & 8.69311e-14                & 1 & -6.15872e-07 & 8.07869e-14                & 1 & \multicolumn{1}{l|}{\multirow{2}{*}{$\omega_{0,0}^\TF = 0$}}                  \\
\multicolumn{1}{|l|}{$\omega_2$}    & 5.86493e-07  & -8.69062e-14                & 1 & 6.15872e-07  & -8.07646e-14                  & 1 & \multicolumn{1}{c|}{} \\ \hline
\multicolumn{1}{|l|}{$\omega_3$}    & -0.200005    & 4.07202e-15                 & 1 & -0.200005    & 8.1375e-15                 & 1 & \multicolumn{1}{l|}{\multirow{4}{*}{$\omega_{1,0}^\TF = 0.2$}}                \\
\multicolumn{1}{|l|}{$\omega_4$}    & 0.200005     & 7.56706e-16                & 1 & 0.200005     & 7.12452e-15                 & 1 & \multicolumn{1}{l|}{} \\
\multicolumn{1}{|l|}{$\omega_5$}    & -0.200005    & 2.10494e-15                & 1 & -0.200005 & -5.09397e-15                & 1 & \multicolumn{1}{c|}{} \\
\multicolumn{1}{|l|}{$\omega_6$}    & 0.200005     & -8.97438e-18                  & 1 & 0.200005 & 4.42765e-15                & 1 & \multicolumn{1}{c|}{} \\ \hline
\multicolumn{1}{|l|}{$\omega_7$}    & -0.283448    & -1.87346e-14                & 1 & -0.283448 & -3.8324e-16                & 1 & \multicolumn{1}{l|}{\multirow{4}{*}{$\omega_{2,0}^\TF = 0.28284271$}}         \\
\multicolumn{1}{|l|}{$\omega_8$}    & 0.283448 & 2.15388e-15                  & 1 & 0.283448 & 7.56616e-16                 & 1 & \multicolumn{1}{c|}{} \\
\multicolumn{1}{|l|}{$\omega_9$}    & -0.283467 & 1.286e-14                 & 1 & -0.283467 & -1.40959e-15                & 1 & \multicolumn{1}{c|}{} \\
\multicolumn{1}{|l|}{$\omega_{10}$} & 0.283467 & 8.67883e-15                & 1 & 0.283467 & -2.90574e-15                   & 1 & \multicolumn{1}{c|}{} \\ \hline
\multicolumn{1}{|l|}{$\omega_{11}$} &-0.348769    & -5.37074e-16                 & 1 & -0.348769 & 1.69936e-16                 & 1 & \multicolumn{1}{l|}{\multirow{4}{*}{$\omega_{3,0}^\TF = 0.34641016$}}         \\
\multicolumn{1}{|l|}{$\omega_{12}$} & 0.348769     & -6.03743e-15                 & 1 & 0.348769 & -1.13813e-15                 & 1 & \multicolumn{1}{c|}{} \\
\multicolumn{1}{|l|}{$\omega_{13}$} & -0.348769    & -4.43208e-15                & 1 & -0.348769 & -9.10584e-17                 & 1 & \multicolumn{1}{c|}{} \\
\multicolumn{1}{|l|}{$\omega_{14}$} & 0.348769     & 2.56399e-15               & 1 & 0.348769 & -3.7267e-16                 & 1 & \multicolumn{1}{c|}{} \\ \hline
\multicolumn{1}{|l|}{$\omega_{15}$} &-0.400018    & -5.01133e-15                & 1 & -0.400018 & -1.45788e-14                & 1 & \multicolumn{1}{l|}{\multirow{6}{*}{$\omega_{4,0}^\TF = \omega_{0,1}^\TF = 0.4$}} \\
\multicolumn{1}{|l|}{$\omega_{16}$} & 0.400018     & -5.34787e-15                & 1 & 0.400018 & -3.09639e-14                 & 1 & \multicolumn{1}{c|}{} \\
\multicolumn{1}{|l|}{$\omega_{17}$} & -0.405642    & -4.75383e-15                & 1 & -0.405642 & 3.36622e-16                 & 1 & \multicolumn{1}{c|}{} \\
\multicolumn{1}{|l|}{$\omega_{18}$} & 0.405642     & -3.58917e-15                 & 1 & 0.405642 & 2.00132e-16                & 1 & \multicolumn{1}{c|}{} \\
\multicolumn{1}{|l|}{$\omega_{19}$} &-0.405679    & 1.04733e-15                & 1 & -0.405679 & 1.19496e-17                & 1 & \multicolumn{1}{c|}{} \\
\multicolumn{1}{|l|}{$\omega_{20}$} & 0.405679     & 7.04663e-15                 & 1 & 0.405679 & -2.51016e-16               & 1 & \multicolumn{1}{c|}{} \\ \hline
\end{tabular}%
}
\caption{Numerical results on the BdG modes that were obtained
with and without parallelization for the 2D ground state and with the
same shift $\sigma = 0.01$. In addition, the Krein signatures shown as
$K$ are included in the table together with the theoretical prediction
of Eq.  \eqref{eq-2DS-pred}.
}
\label{tab-2D-TF}
\end{table}

\subsection{2D case: Dark soliton}
% ----------------------------------------------------------------------
We test next the 2D dark soliton known also as
the dark-soliton stripe (see  \cite{deflation_2018} and references therein).
At the linear limit, this state is constructed as
\begin{align}
\phi_{DS} = \sqrt{\frac{\omega_\perp}{2\pi}} %
H_0(\sqrt{\omega_\perp}x)H_1(\sqrt{\omega_\perp}y)e^{-\frac{1}{2}\omega_\perp(x^2+y^2)},
\end{align}
where $H_n$ are Hermite polynomials of degree $n$. Similarly as before,
we set $\omega_\perp = 0.2$, and perform a numerical continuation over
$\mu$ all the way up to $\mu = 3$ in order to trace the entire branch.

\begin{figure}[!h]
\begin{subfigure}[t]{0.5\textwidth}
	\centering
	\includegraphics[width=\textwidth]{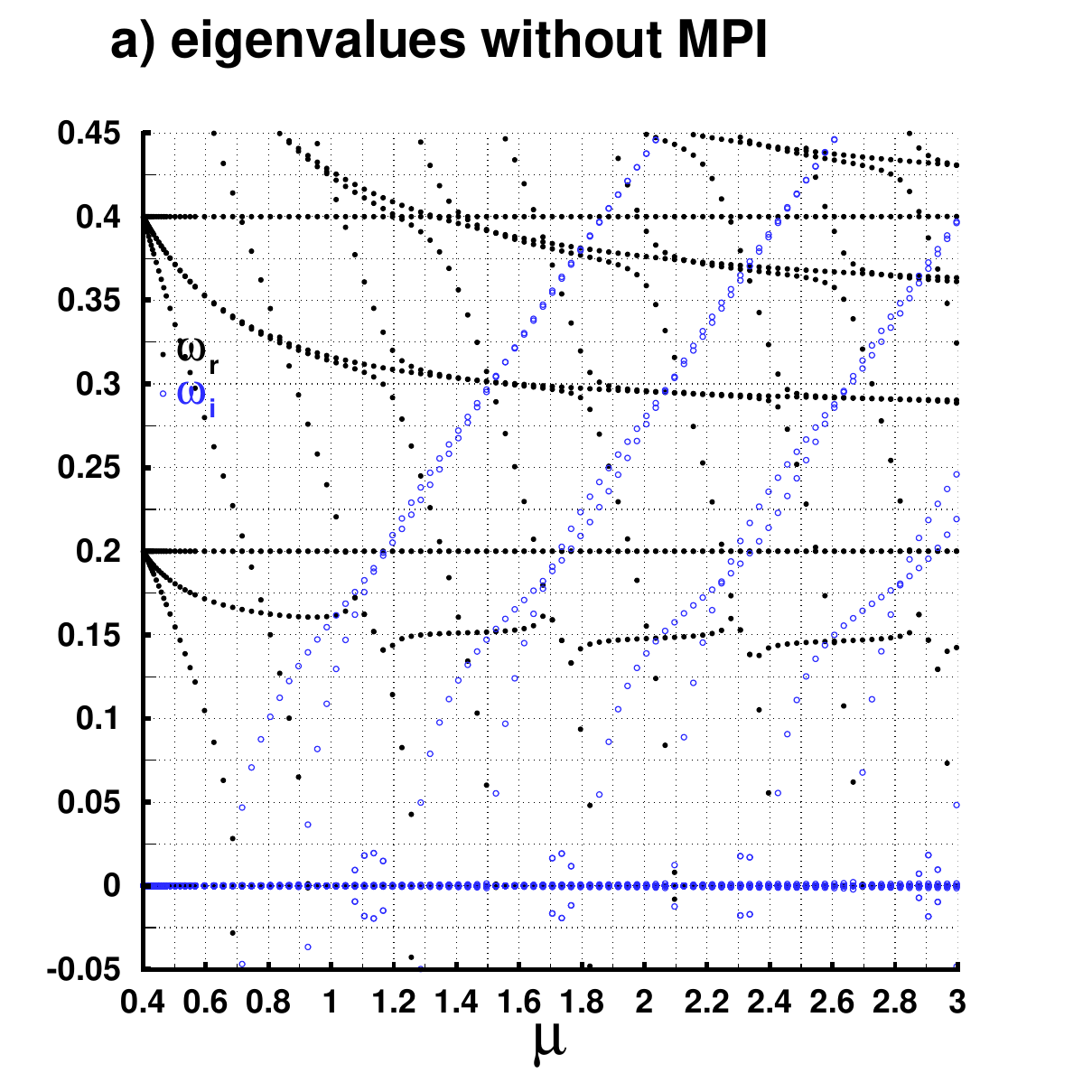}
	\label{fig:2D_DSH_e}
\end{subfigure}
\begin{subfigure}[t]{0.5\textwidth}
	\centering
	\includegraphics[width=\textwidth]{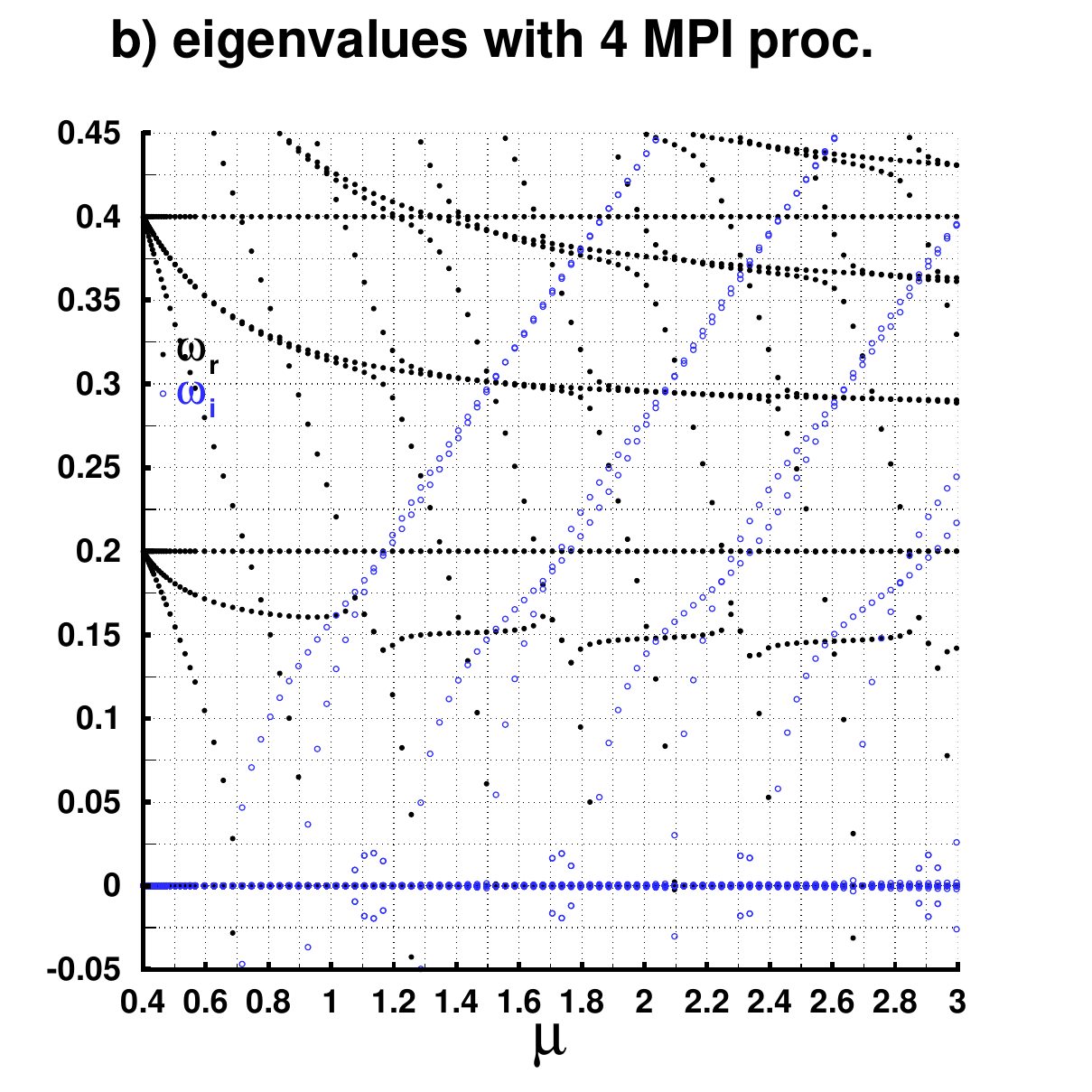}
	\label{fig:2D_DSH_f}
\end{subfigure}
	\caption{BdG results for the 2D dark soliton:
	a) real and imaginary parts of the eigenvalues, \ie $\omega_r$
	and $\omega_i$, respectively, as a function of $\mu$, and without
	using MPI; b) same as panel a) but using 4 MPI processors.~Blue open
	and dark filled circles in both panels depict $\omega_i$ and $\omega_r$,
	respectively.}
	\label{fig-2D-DSH-compeig}
\end{figure}

\begin{figure}[!h]
	\centering
	\begin{subfigure}[t]{0.4\textwidth}
		\centering
		\includegraphics[width=\textwidth]{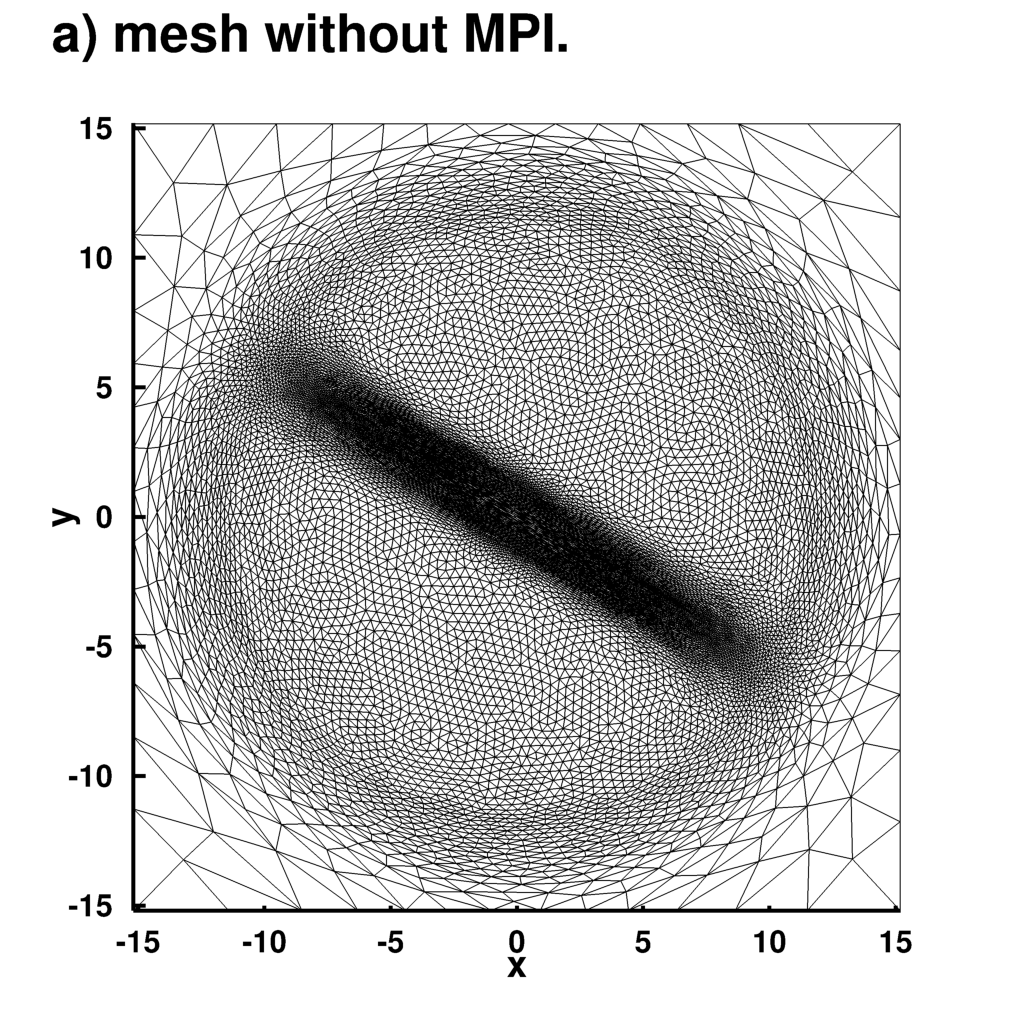}
		\label{fig:2D_DSH_a}
	\end{subfigure}
	\begin{subfigure}[t]{0.4\textwidth}
		\centering
		\includegraphics[width=\textwidth]{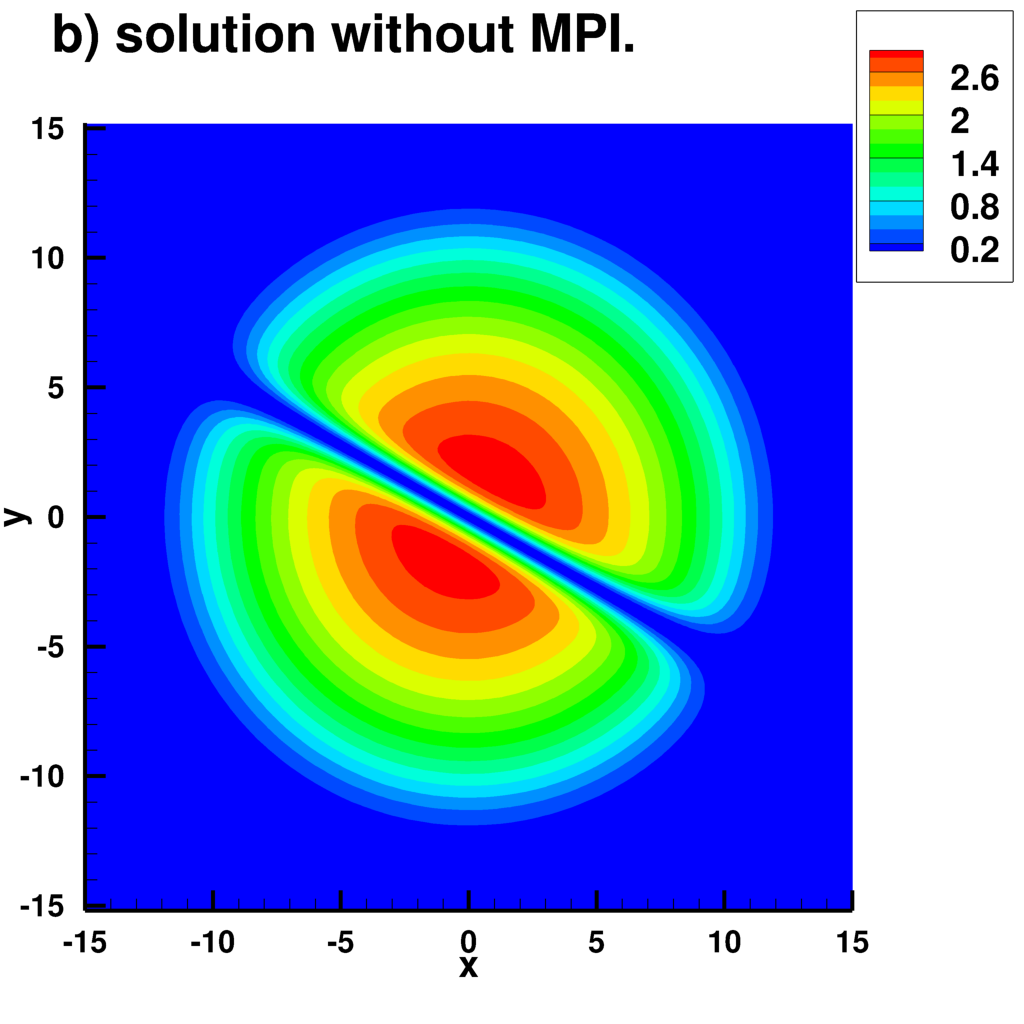}
		\label{fig:2D_DSH_b}
	\end{subfigure}
	\begin{subfigure}[t]{0.4\textwidth}
		\centering
		\includegraphics[width=\textwidth]{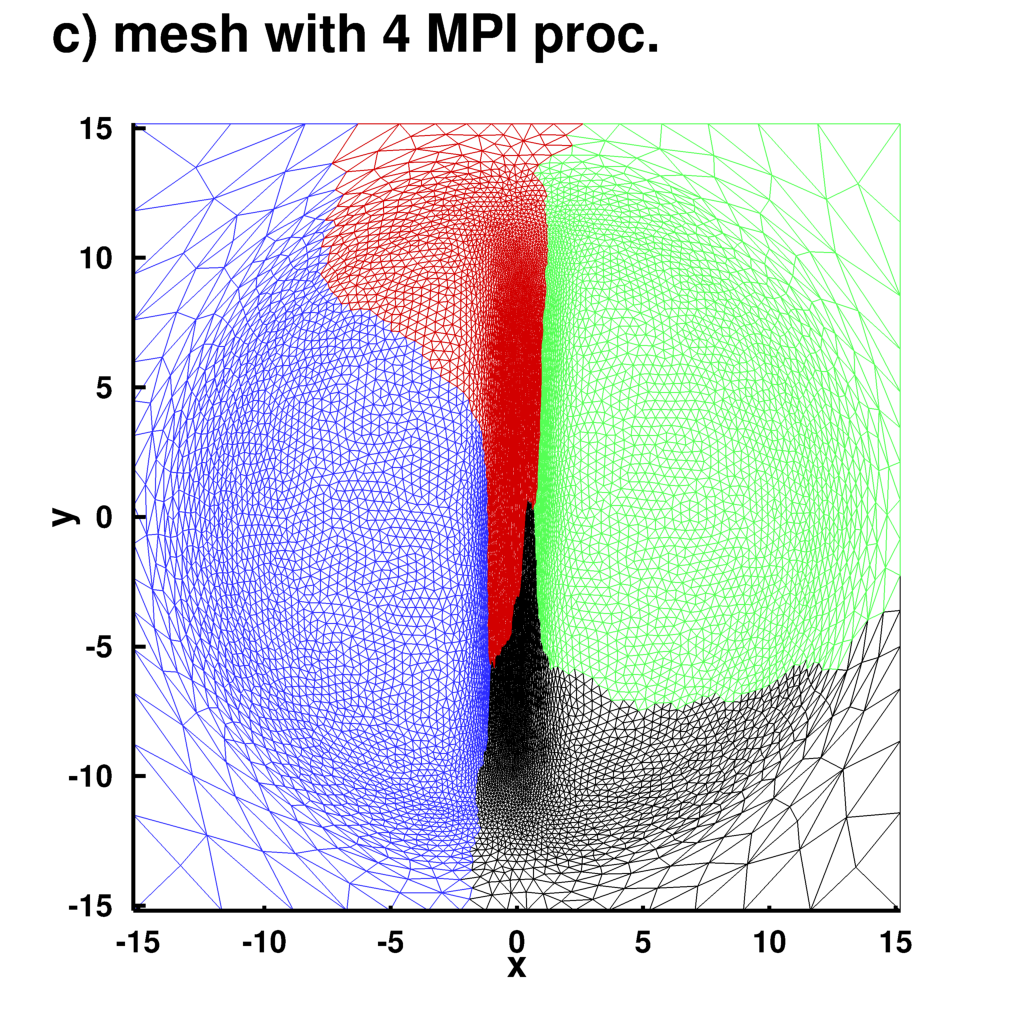}
		\label{fig:2D_DSH_c}
	\end{subfigure}
	\begin{subfigure}[t]{0.4\textwidth}
		\centering
		\includegraphics[width=\textwidth]{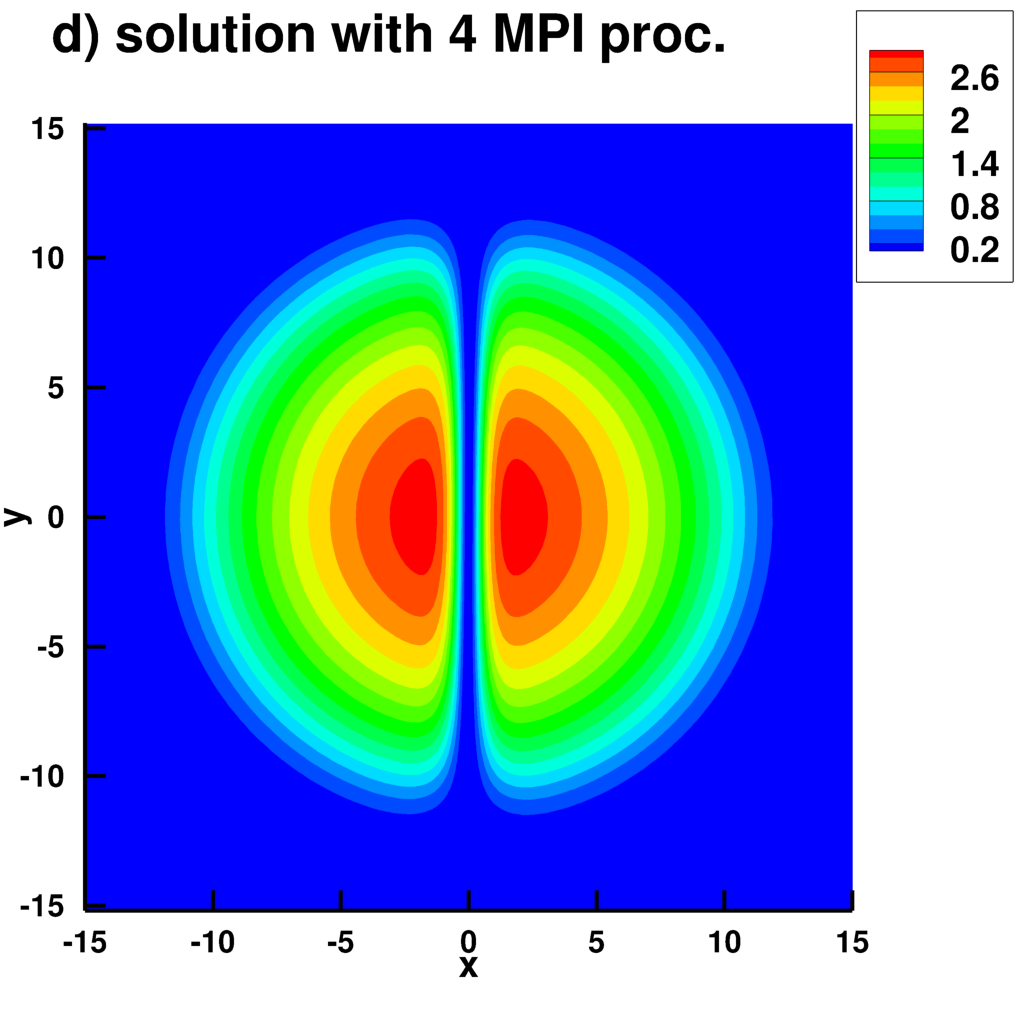}
		\label{fig:2D_DSH_d}
	\end{subfigure}
	\caption{2D dark soliton for $\mu=3$: without MPI a) adapted mesh, b) density $|\phi|^2$
	and with 4 MPI processors c) adapted mesh, d) density $|\phi|^2$.}
	\label{fig-2D-DSH-compsol}
\end{figure}

We present the numerical results we obtained from the toolbox in
Figs.  \ref{fig-2D-DSH-compeig}-b) and \ref{fig-2D-DSH-compsol}-c)-d), and we compare
them to the ones reported in \citep{sadaka_2023} in Figs.  \ref{fig-2D-DSH-compeig}-a)
and \ref{fig-2D-DSH-compsol}-a)-b). Note that we used the new $\mu$-continuation strategy, with adapted step
(see the last paragraphs in Sec. \ref{subsec-num-meth-GP2c}). In particular, the real
and imaginary parts of  $\omega$ as a function of $\mu$ are depicted in
panels a) and b) of Fig.  \ref{fig-2D-DSH-compeig} without MPI and with 4 MPI
processors, respectively.The results are identical to the results of
  \cite{middelkamp2010bifurcations} (note also the emergence of a cascade of
pitchfork, \ie symmetry-breaking bifurcations).   Panels a) and c) (resp. b) and d)) in
Fig.  \ref{fig-2D-DSH-compsol} show the final adapted
mesh and the respective density $|\phi|^2$ of the solution for $\mu=3$ without MPI
and with 4 MPI processors, respectively.  A few remarks are in order.
Per the discretized problem the toolbox solves, there exists a preferred
direction along which the configuration will tend to align itself. When we adapt
the mesh, this direction changes, and the wave function will then rotate. We
overcome this issue by allowing the toolbox to perform mesh adaptation at each
continuation step. Note that mesh adaptivity is performed at every
step in Newton's method, especially, when the norm of the correction is greater
than $0.1$.
%This permits to not only optimally adapt the the mesh while reducing
%the effects of the rotation, but also to reduce the computational time
%(see, Table  \ref{tab-cputime-1c}).
%Finally, Figs.  \ref{fig-2D-DSH}(e) and  \ref{fig-2D-DSH}(f)
%showcase the dependence of the number of atoms $N$ [cf.~Eq.  \eqref{eq-GP-N}]
%and the energy ${\mathcal E}$ [cf.~Eq.  \eqref{eq-NRJ}] of the system on $\mu$,
%respectively.

\clearpage

\subsection{3D case: Ground state}
\begin{figure}[!h]
	\centering
	\begin{subfigure}[t]{0.5\textwidth}
		\centering
		\includegraphics[width=\textwidth]{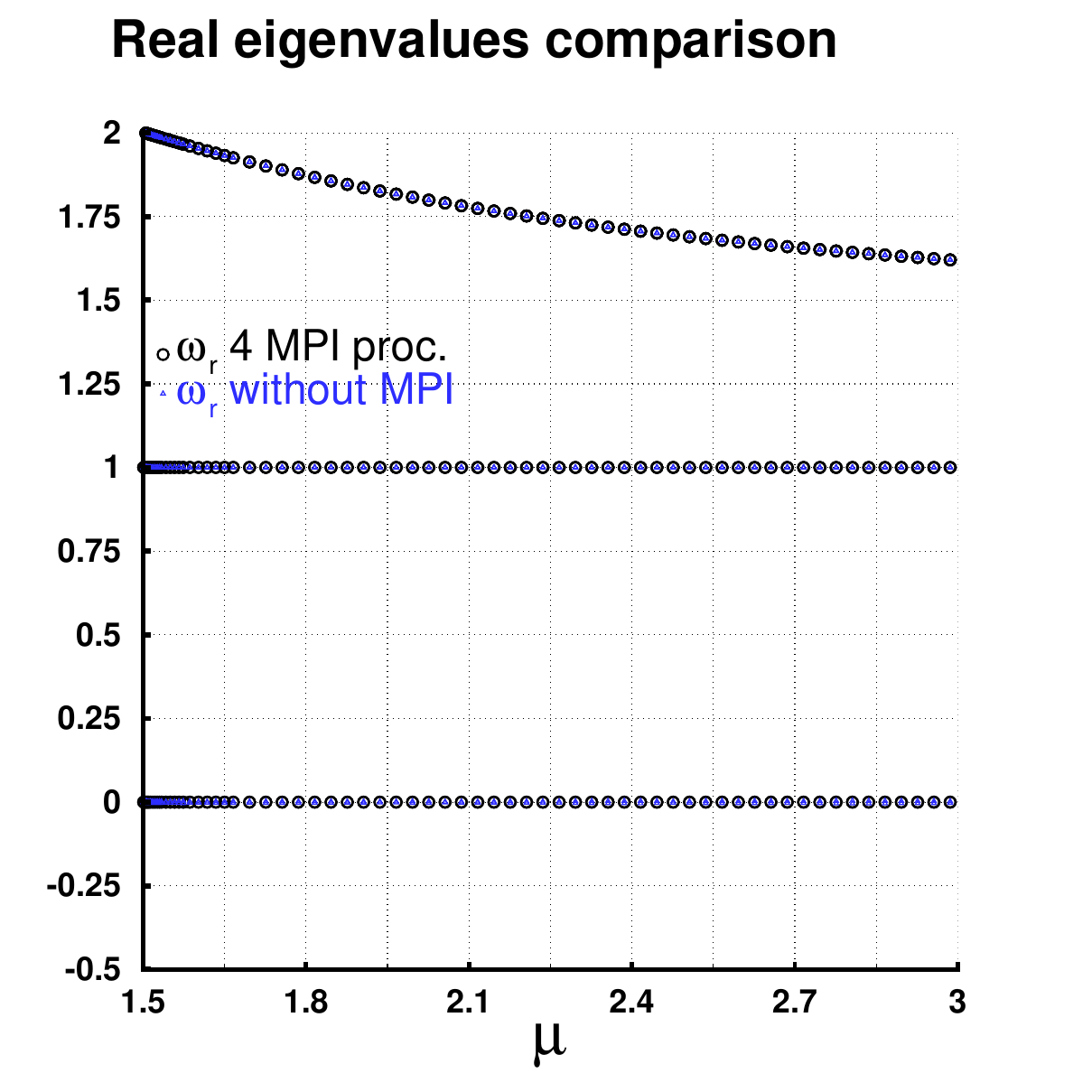}
		\label{fig:3D_TF_a}
	\end{subfigure}
%	\hfill
%		\begin{subfigure}[t]{0.45\textwidth}
%			\centering
%			\includegraphics[width=\textwidth]{\figpath/\figpath/fig03_3D_TF-new}
%			\label{fig:3D_TF_b}
%		\end{subfigure}
	\caption{3D ground state. Variation of the real part $\omega_r$ of the BdG spectrum
		as a function of $\mu$. Results obtained using 4 MPI
	processors (black triangles) and without MPI (blue open circles).%
	% is shown in the left panel,
	%whereas the right panel showcases the isosurfaces of the modulus
	%of four BdG modes.
	}
\label{fig-3D-TF}
\end{figure}

% ----------------------------------------------------------------------
Considering 3D configurations renders
the computation of the BdG spectrum a very challenging problem.
This is the
case because even with mesh adaptation, the number of degrees of freedom
is still high, and increases with the size of the BEC which is controlled
by the chemical potential $\mu$ (or, equivalently, by the number of atoms).
In our previous contribution  \citep{sadaka_2023}, as a prototypical case
to test the published toolbox therein, we considered the 3D ground state
whose BdG spectrum was computed for $\omega_\perp=1$. We argued that the
computation of more complex states requires the use of parallelization which
is capable of reducing the computational time and memory requirements. Indeed,
with the present parallel toolbox, we accomplish this goal. Illustratively,
we compute the BdG spectrum of the 3D ground state, and respective results are
shown presented in Fig.  \ref{fig-3D-TF}. The real part of the eigenvalues
($\omega_r$) as a function of $\mu$ computed without MPI (using the sequential
toolbox published in \citep{sadaka_2023}) and with 4 MPI processors (and the
$\mu$-adaptivity continuation strategy discussed previously) are shown in the
same panel. Both numerical
results are in full agreement with the numerical results reported in \citep{bisset2015robust}.
%whereas the right panel depicts a few associated BdG modes (see, the legends therein).
%These results are in full agreement with the sequential version of the present toolbox \citep{sadaka_2023},
%as well as the numerical results reported in \citep{bisset2015robust}.
%\clearpage

\subsection{Dark soliton in 3D}
\begin{figure}[!h]
	\centering
	\begin{subfigure}[t]{0.4\textwidth}
		\centering
		\includegraphics[width=\textwidth]{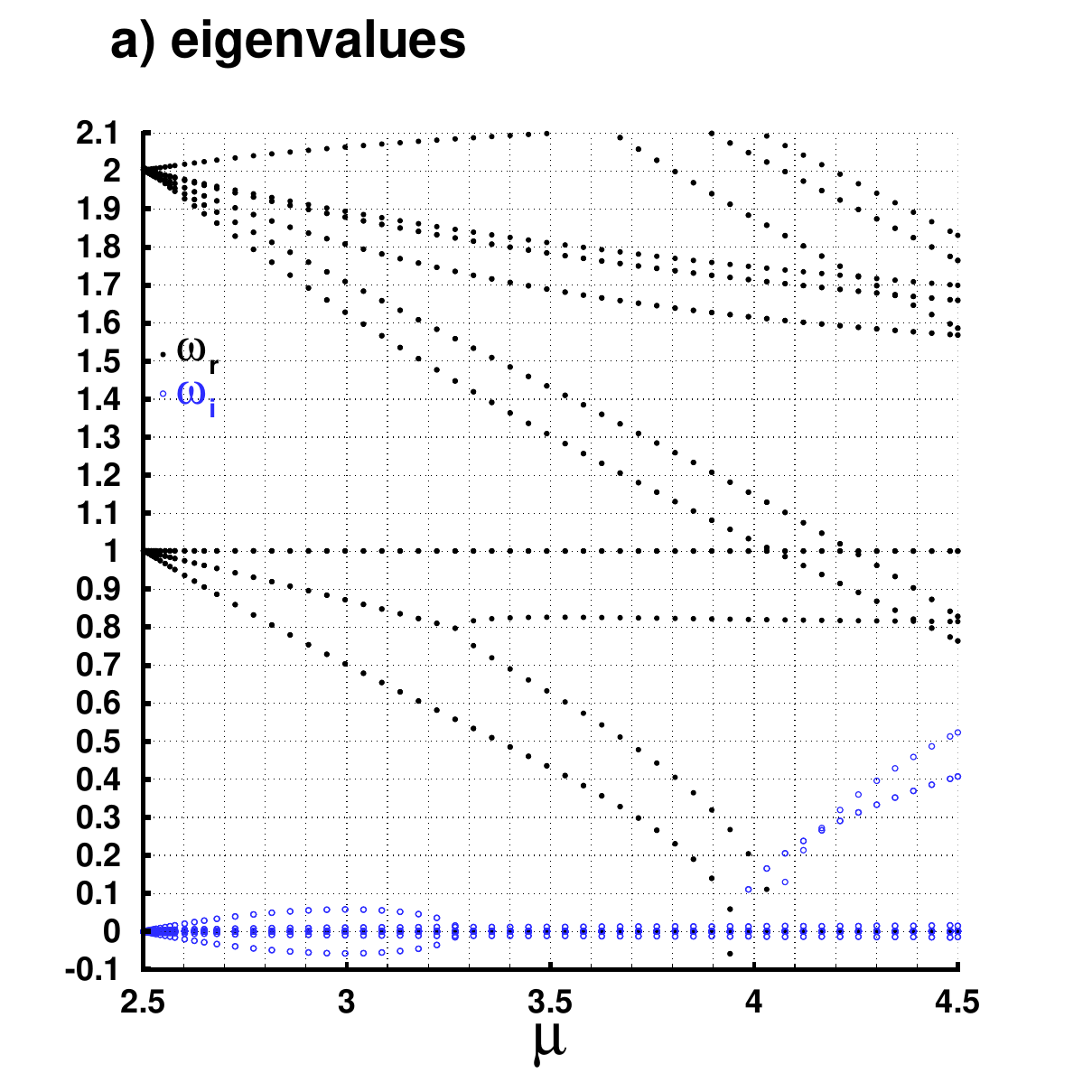}
		\label{fig:Herm_DS_a}
	\end{subfigure}
	\begin{subfigure}[t]{0.4\textwidth}
		\centering
		\includegraphics[width=\textwidth]{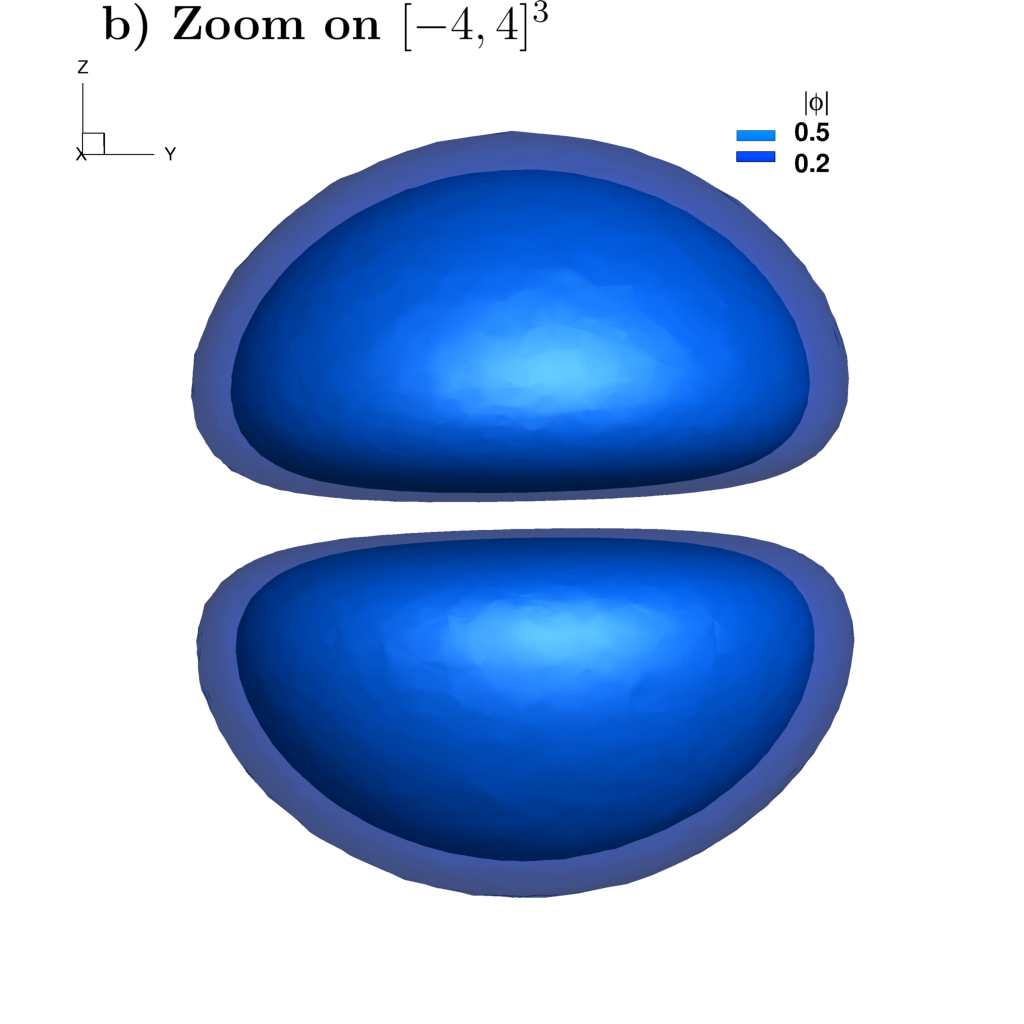}
		\label{fig:Herm_DS_b}
	\end{subfigure}
    \caption{3D dark soliton. a) The BdG spectrum and b) density $|\phi|^2$ for $\mu=4.5$. The
    computational domain is the cube $[-5.4,5.4]^3$.
    % The left panel depicts the eigenvalue
    %spectrum as a function of $\mu$ (real part $\omega_{r}$ is shown as
    %filled black circles and imaginary part $\omega_{i}$ as blue open circles)
    %whereas the right panel showcases the isosurface of the density $|\phi|^{2}$
    %for $\mu=4.5$.\\
    %a) eigenvalue spectrum (real parts as full lines and imaginary parts as dashed lines), b) density isosurface for the final state, c) nonzero element, d) number of unknown Wh.ndof,
    %e) cputime for each $\mu$.\\
    %GP : number of cores 28, job wall-clock 11:49:01, memory utilized estimated 21.05 GB.\\
    %BdG : number of cores 28, job wall-clock 02:01:40, memory utilized estimated 82.09 GB.
    }
	\label{fig-BdG-3D-DS}
\end{figure}

% ----------------------------------------------------------------------
We now proceed with another 3D case  to validate our
parallel toolbox. The existence and BdG analysis of the dark soliton in
3D was considered in \citep{bisset2015bifurcation}, where the azimuthal
symmetry of the state was taken into account in order to reduce the 3D
BdG problem to a 2D one. Here, we perform a full 3D BdG analysis by using
(similarly to \citep{bisset2015bifurcation}) an isotropic potential with
$\omega_\perp=1$. We note that the 3D dark soliton (or planar dark soliton)
can be constructed in the linear limit by the Cartesian eigenstate $|0,0,1\rangle$
(bearing a zero cut in the $z$ direction), and can be expressed in terms of Hermite
polynomials \citep{bisset2015bifurcation} (see also  \citep{boulle2020deflation}). This
state emanates from the linear limit at $\mu=5/2$, and it is degenerate; the eigenstates
$|1,0,0\rangle$ and $|0,1,0\rangle$ produce the same solution although they now have
zero cuts along the $x$ and $y$ directions, respectively. Upon using this eigenstate
as a seed to Newton's method, we performed a continuation over $\mu$. The results
are shown in Fig.  \ref{fig-BdG-3D-DS}: the
left panel depicts the BdG spectrum of the state (real and imaginary parts of
the eigenvalues are shown with black filled circles and blue open circles, respectively);
the right panel shows the isosurface of the density $|\phi|^2$ for
$\mu=4.5$. It can be discerned from the left panel that our results match perfectly the
ones published in \citep{bisset2015bifurcation} (see Fig.~1 therein).
%We conclude our discussion
%in this example by making some observations. The computation time for this case was 4,310\,s
%on 12 cores. The mesh adaptation that was employed herein was able to decrease the mesh
%size to 12510 tetrahedrons (or, equivalently, 36622 degrees of freedom).
% EGC: I may include this when I work on the paper again.
% It can be seen on Fig. \ref{fig-BdG-3D-DS} a) where the
% imaginary parts of these eigenvalues are around $4\times 10^{-3}$.

%\pagebreak

\pagebreak

\subsection{3D case: Vortex-lines and beyond}
\begin{figure}[!h]
	\centering
	\begin{subfigure}[t]{0.4\textwidth}
		\centering
		\includegraphics[width=\textwidth]{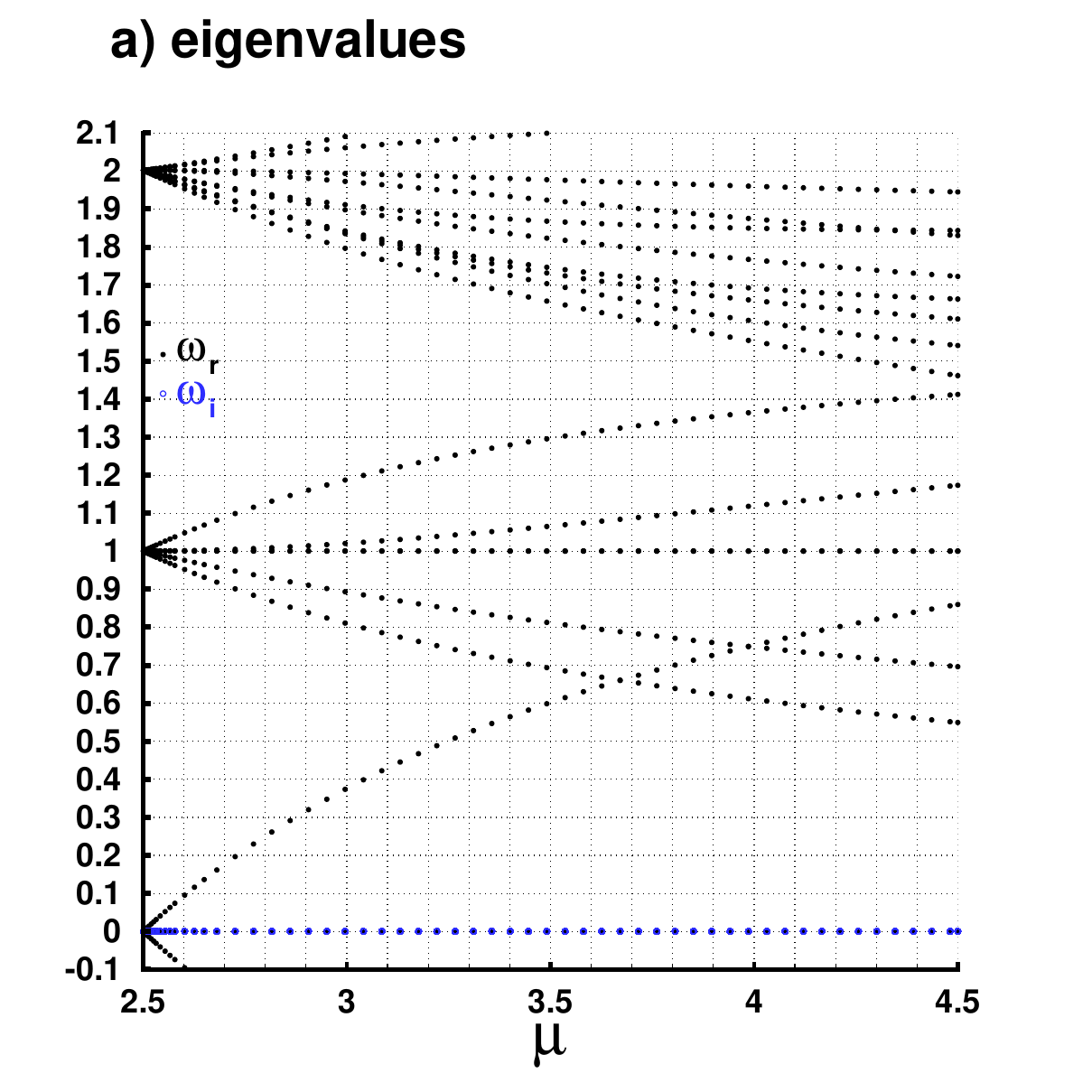}
		\label{fig:Lag_0_1_0_a}
	\end{subfigure}
	\begin{subfigure}[t]{0.4\textwidth}
		\centering
		\includegraphics[width=\textwidth]{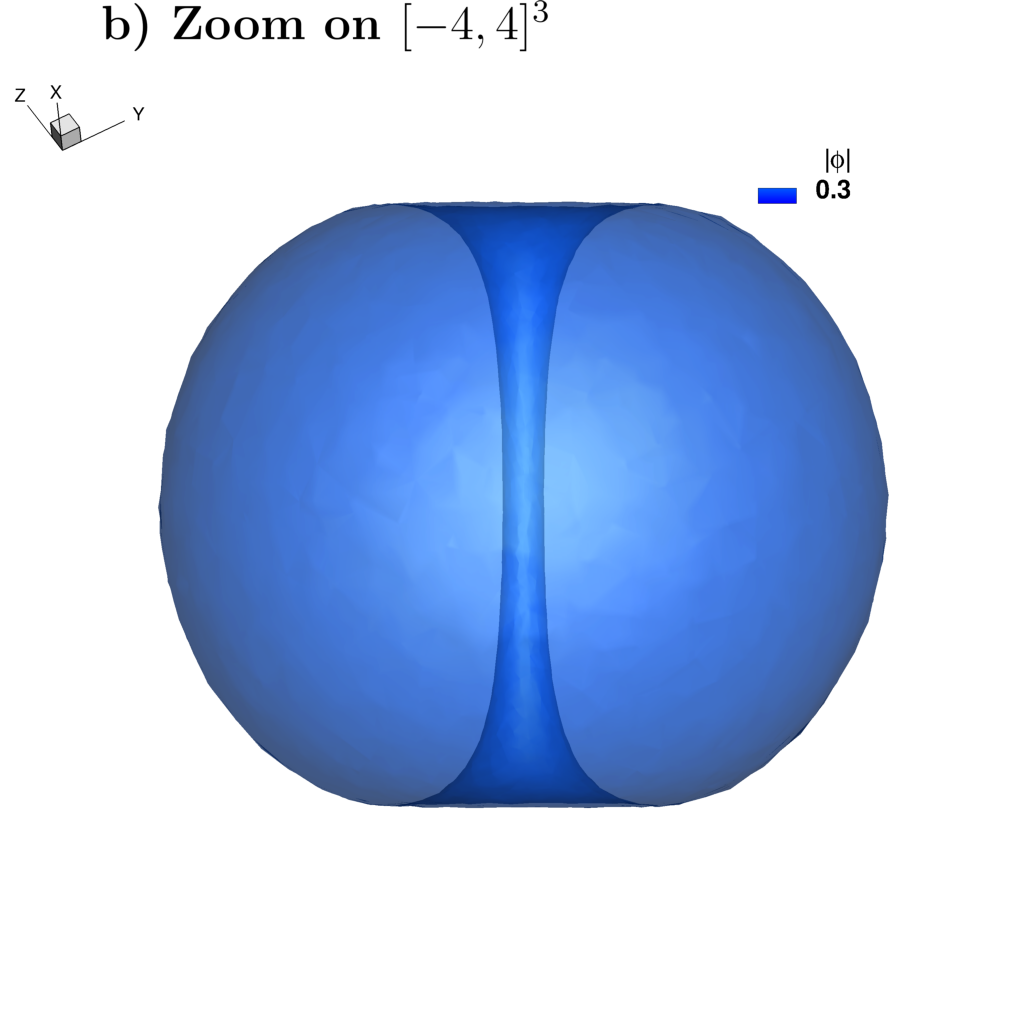}
		\label{fig:Lag_0_1_0_b}
	\end{subfigure}
    \caption{3D single-charged vortex-line (VL) configuration inside a cube $[-5.4,5.4]^3$.%
    ~The format of the figure is the same as in Fig.  \ref{fig-BdG-3D-DS} with the density $|\phi|^2$
    shown in b) for $\mu=4.5$.
    % The left panel depicts the real part $\omega_r$ of the eigenvalues as a function of $\mu$, whereas
    % the right panel showcases the isosurface of the density $|\phi|^{2}$ for $\mu=X$.
    }
	\label{fig-BdG-3D-Lag_0_1_0}
\end{figure}

\begin{figure}[!h]
	\centering
	\begin{subfigure}[t]{0.4\textwidth}
		\centering
		\includegraphics[width=\textwidth]{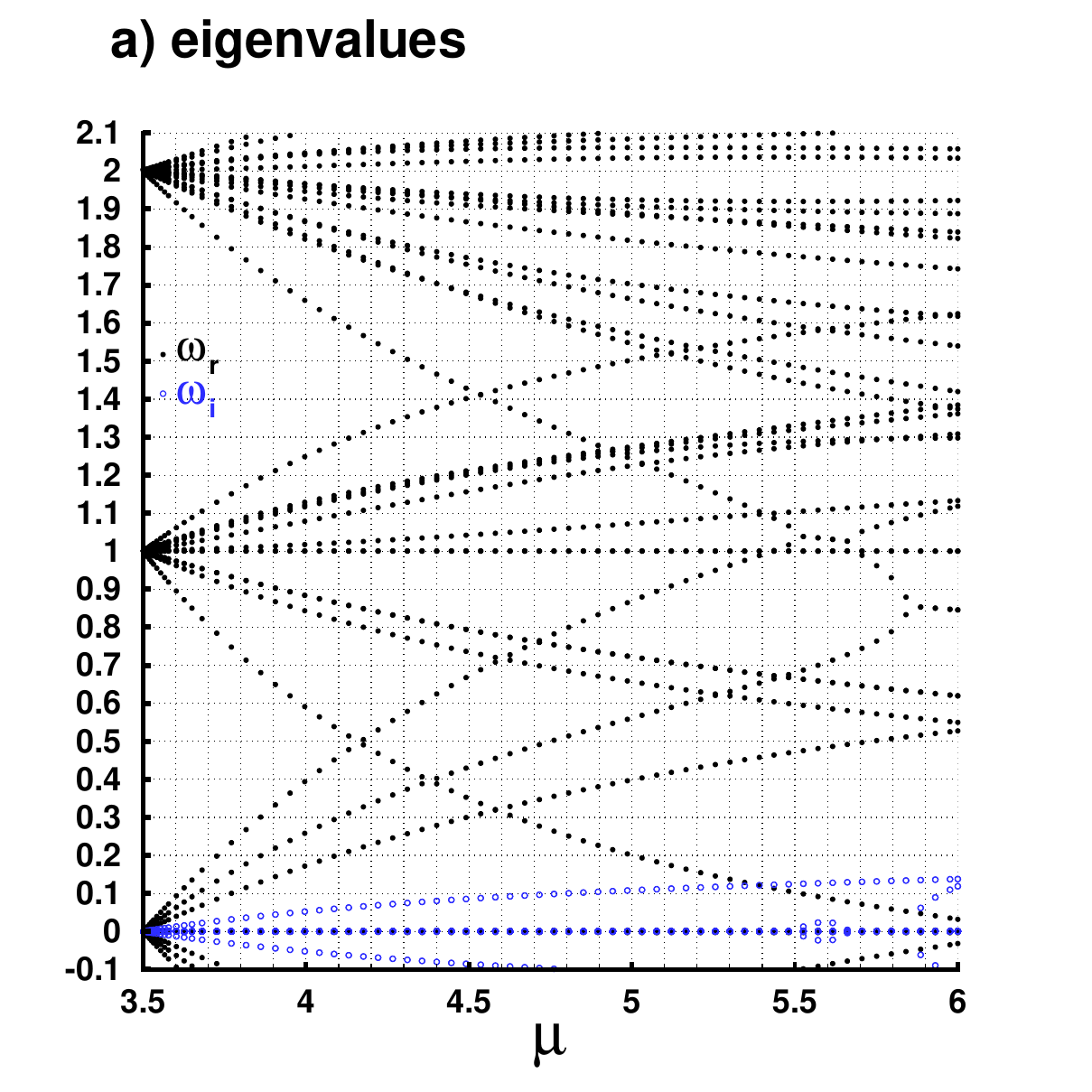}
		\label{fig:Herm_e_a}
	\end{subfigure}
	\begin{subfigure}[t]{0.4\textwidth}
		\centering
		\includegraphics[width=\textwidth]{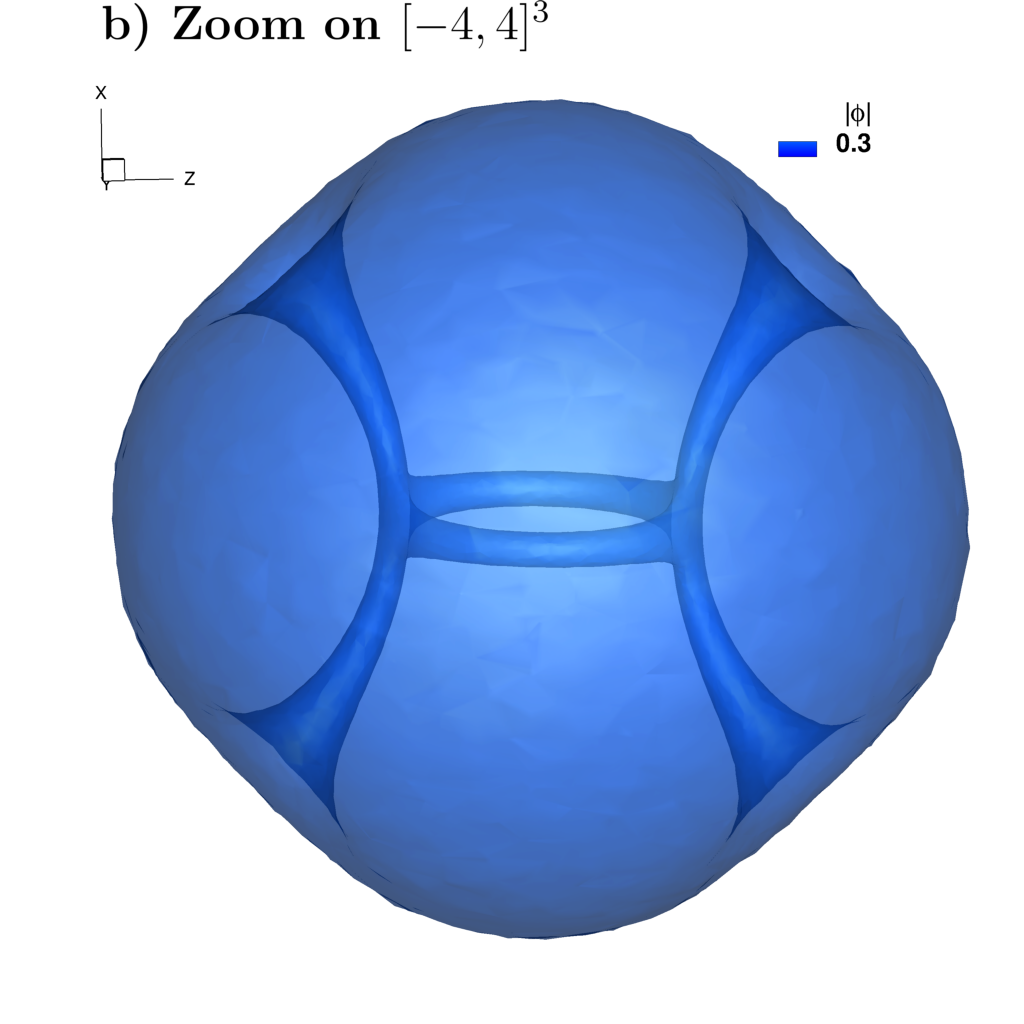}
		\label{fig:Herm_e_b}
	\end{subfigure}
    \caption{3D vortex-ring (VR) with two (oppositely charged) VL
     handles configuration inside a cube $[-6.23,6.23]^3$. The format of
     the figure is the same as in Fig.  \ref{fig-BdG-3D-DS} with the density
     $|\phi|^2$ shown in b) for $\mu=6$.
    %The left panel depicts the real part $\omega_r$ of the eigenvalues as a function of $\mu$,
    %whereas the right panel showcases the isosurface of the density $|\phi|^{2}$ for $\mu=X$.
    }
	\label{fig-BdG-3D-Herm_e}
\end{figure}

% ----------------------------------------------------------------------
We conclude our series of validation cases for the single-component GP
equation in 3D by considering two extra cases (we set $\omega_\perp = 1$ as previously):
a single-charged vortex-line (VL) state \citep{bisset2015robust}, and a
vortex-ring (VR) configuration bearing two (oppositely charged) VL
handles  \cite{boulle2020deflation}. The former state bifurcates from the linear
limit at $\mu=2.5$ (\ie 1st-excited state), and can be classified in
terms of cylindrical coordinates as $|0,1,0\rangle_{\mathrm{cyl}}=r^2 L_{0}^{1}(x^2+y^2)e^{\ii\theta}%
e^{-(x^{2}+y^{2}+z^2)/2}$  \cite{boulle2020deflation} (where $L_{0}^{1}$
stands for the associated Laguerre polynomial). Similarly, the vortex-ring
(VR) with two handles bifurcates at $\mu=3.5$ from the linear limit,
and is constructed by the combination of Hermite polynomials (in Cartesian
coordinates) $|2,0,0\rangle+|0,2,0\rangle+\ii|1,0,1\rangle$. Our toolbox was
capable of tracing branches of solutions for both cases where the respective
results are shown in Figs.  \ref{fig-BdG-3D-Lag_0_1_0} and  \ref{fig-BdG-3D-Herm_e},
respectively. In particular, the panel a) in the figure depicts the BdG
spectra of the pertinent states that match with the findings in
 \citep{bisset2015robust} and  \cite{boulle2020deflation}. Panel b) shows two snapshots
of isosurfaces of the densities $|\phi|^{2}$ of the solutions for $\mu=4.5$ and $\mu=6$, respectively.


\clearpage

\section{Validation test cases for the two-component BEC}\label{sec-valid2c}
% ----------------------------------------------------------------------
We now move on to the study of
localized configurations in two-component GP equations  \eqref{eq-scal-GP2c-stat}.
Admittedly, the study of their existence, and more crucially, their
BdG spectrum (upon solving Eqs.  \eqref{eq-BdG2c-M}-\eqref{eq-BdG2c-M1})
places them in an one level harder category. Indeed, the size
of the BdG problem for 2D and (even more) 3D configurations becomes quite
large, especially when one wants to provide a detailed and accurate description
of the spectral properties of such configurations. However, the combination
of mesh adaptivity implemented in \ff with parallelization tools provided
by \texttt{PETSc} makes the present toolbox a great candidate to compute such challenging 3D cases.
Table  \ref{tab-cputime-2c-BdG} summarizes the test cases we considered,
and has the same format as Table  \ref{tab-cputime-1c-3D}.

%\clearpage
\begin{table}[!h]
\resizebox{\textwidth}{!}{%
\begin{tabular}{l|c|c|c|c|c|c|}
\cline{2-7}
\qquad GP test cases & Processes  &  CPU time  & niter & ndof & nt & maxRSS\\ \hline
\multicolumn{1}{|l|}{2D vortex-antidark state}          & 4 & 00:00:17 & 46 & 30,506 & 3,744 & 1,11 Gb \\
\multicolumn{1}{|l|}{2D ring-antidark state}          & 4 & 00:00:09 & 105 & 37,847 & 4,634 & 4,53 Gb \\
\multicolumn{1}{|l|}{2D Hermite LL 1}          & 4 & 00:00:12 & 81 & 27,775 & 3,399 & 0,34 Gb \\
\multicolumn{1}{|l|}{2D Hermite LL 7}          & 4 & 00:00:29 & 81 & 58,081 & 7,175 & 2,22 Gb \\
\multicolumn{1}{|l|}{2D Hermite LL 8}          & 4 & 00:00:28 & 81 & 41,717 & 5,131 & 1,42 Gb \\
\multicolumn{1}{|l|}{3D Hermite LL 1}          & 28 & 00:02:18 & 81 & 24,698 & 3,936 & 1,22 Gb \\
\multicolumn{1}{|l|}{3D Hermite LL 7}          & 28 & 00:03:52 & 81 & 24,578 & 3,935 & 1,16 Gb \\
\hline
\end{tabular}%
}
\vspace{.5cm}
%\end{table}

%\begin{table}[h]
\resizebox{\textwidth}{!}{%
\begin{tabular}{l|c|c|c|c|c|c|}
\cline{2-7}
\qquad BdG test cases & Processes  &  CPU time  & niter & ndof & nnz & maxRSS\\ \hline
\multicolumn{1}{|l|}{2D vortex-antidark state}          & 4 & 00:00:31 & 46 & 27,787 & 5,125,301 & 0,78 Gb \\
\multicolumn{1}{|l|}{2D ring-antidark state}          & 4 & 00:00:41 & 91 & 37,791 & 6,946,743 & 1,02 Gb \\
\multicolumn{1}{|l|}{2D Hermite LL 1}          & 4 & 00:00:33 & 81 & 25,303 & 4,644,479 & 0,75 Gb \\
\multicolumn{1}{|l|}{2D Hermite LL 7}          & 4 & 00:01:03 & 81 & 55,155 & 10,136,839 & 1,43 Gb \\
\multicolumn{1}{|l|}{2D Hermite LL 8}          & 4 & 00:00:37 & 81 & 38,882 & 7,142,816 & 1,11 Gb \\
\multicolumn{1}{|l|}{3D Hermite LL 1}          & 28 & 00:03:56 & 81 & 72,202 & 32,963,342 & 6,70 Gb \\
\multicolumn{1}{|l|}{3D Hermite LL 7}          & 28 & 00:03:21 & 81 & 71,627 & 32,681,221 & 7,03 Gb \\
\hline
\end{tabular}%
}
\caption{
Same as Table  \ref{tab-cputime-1c-3D}, but for the two-component GP and BdG
problems (again, with mesh adaptivity). Note that 100 and 60 eigenvalues
were computed for all 2D and 3D test cases, respectively.
The continuations (and thus BdG computations) were performed over $\beta_{12}$
for the 2D vortex-antidark and 2D ring-antidark states whereas for the rest
of the cases, over $\mu_{2}$.
%These computations were performed
%on the CRIANN Computing Center and MATRICS platform utilizing an Intel
%Broadwell E5-2680 v4 @ 2.40GHz (14 cores per socket) architecture with two
%sockets per node and 128 GB of DDR4 2400 MHz RAM.~An Intel Omnipath 100Gb/s low
%latency network was used for communications.
}
\label{tab-cputime-2c-BdG}
\end{table}

\subsection{2D two-component case: vortex dark-antidark configurations}

For illustration purposes, we consider two test cases taken from \citep{dan-2016-PRA}
corresponding to 2D vortex-antidark and dark-antidark ring solutions
(see also Tab.  \ref{tab-cputime-2c-BdG} for a summary of our results).
Such bound modes emerge in two-component GP equations
% (see, Eqs.  \eqref{eq-scal-GP2c-stat})
due to the inter-component interaction. Indeed, a dark soliton or a
vortex (or a ring) in $\phi_{1}$ will induce an effective potential
through the inter-component nonlinearity which itself ``traps'' a
localized mode in $\phi_{2}$. As a consequence, atoms in $\phi_{2}$ ``fill-in''
the density dip of $\phi_{1}$ through this (effective) trapping
process. We consider the GP system  \eqref{eq-scal-GP2c-stat}
in the case of repulsive inter-component interactions with miscibility
condition $0 \leq \beta_{12} < \sqrt{\beta_{11} \beta_{22}}$
which ensures that the two components co-exist outside the dark-antidark
state. To simplify the case study (and following  \cite{sadaka_2023}),
we set $\beta_{11}=\beta_{22}=\beta=1$, $\beta_{12}=\beta_{21}$, and
$0<\beta_{12} < \beta$ since the ratio between non-linear interaction
coefficients matters only.

\begin{figure}[!h]
	\centering
	\includegraphics[width=.45\textwidth]{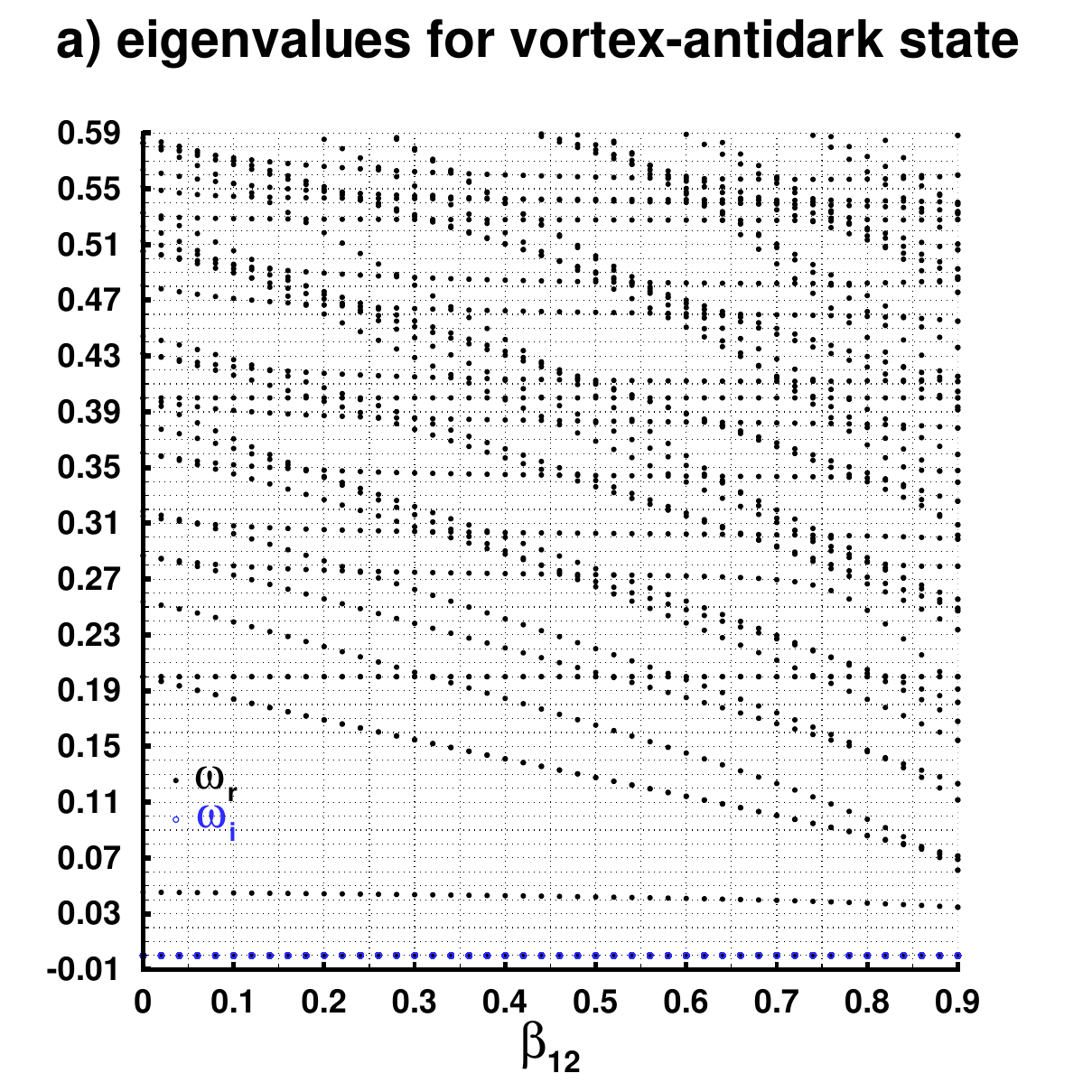}
	\includegraphics[width=.45\textwidth]{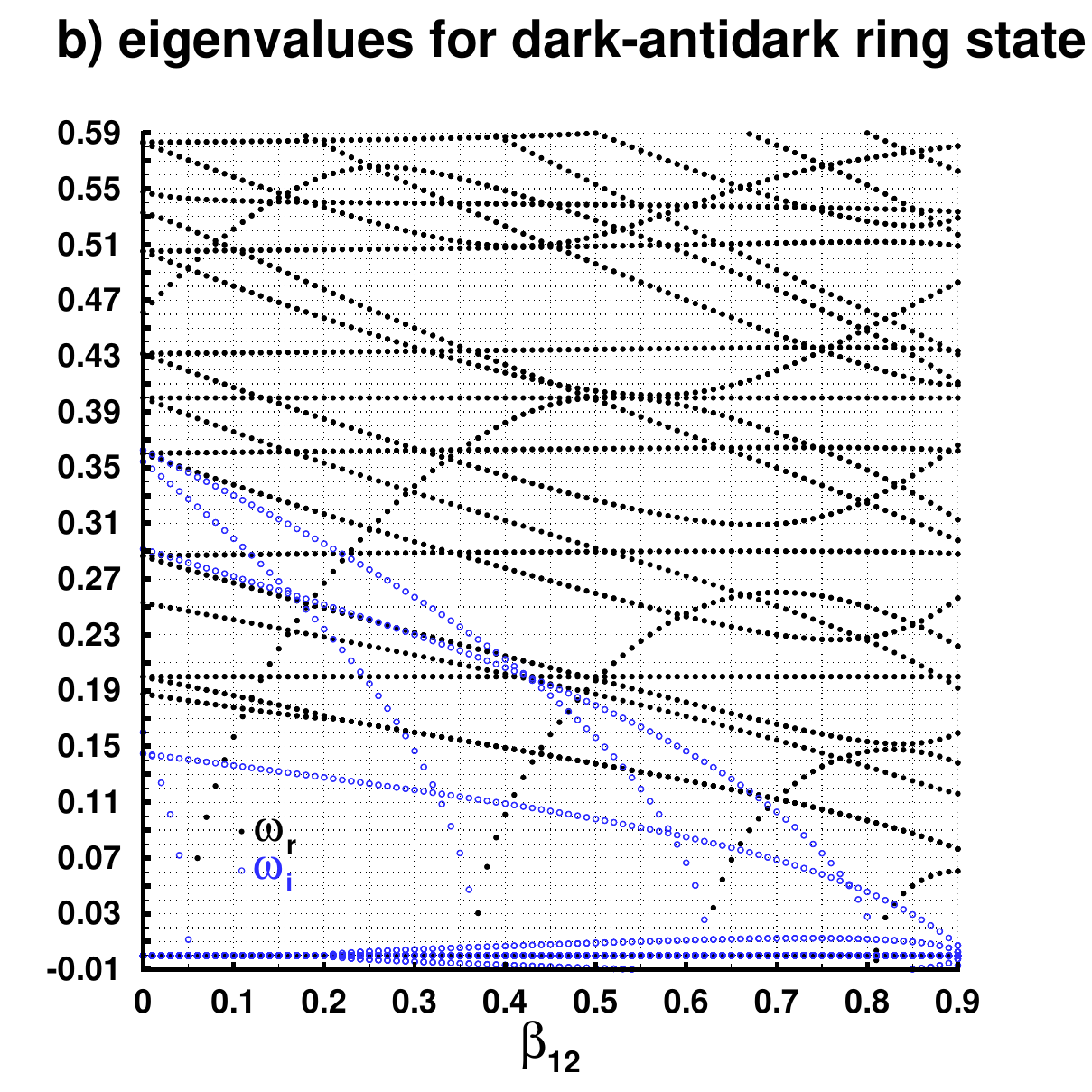}
	\caption{2D two-component vortex dark-antidark configurations. The BdG spectrum of a) vortex-antidark
		and b) dark-antidark ring states as a function of $\beta_{12}$. Black
		filled and blue open circles correspond to the real ($\omega_{r}$) and
		imaginary parts ($\omega_{i}$) of the eigenvalues, respectively.}
	\label{fig-2D-VS-TF-and-RS-TF-Eig}
\end{figure}

\begin{figure}[!h]
	\centering
	\begin{subfigure}[t]{0.3\textwidth}
		\centering
		\includegraphics[width=\textwidth]{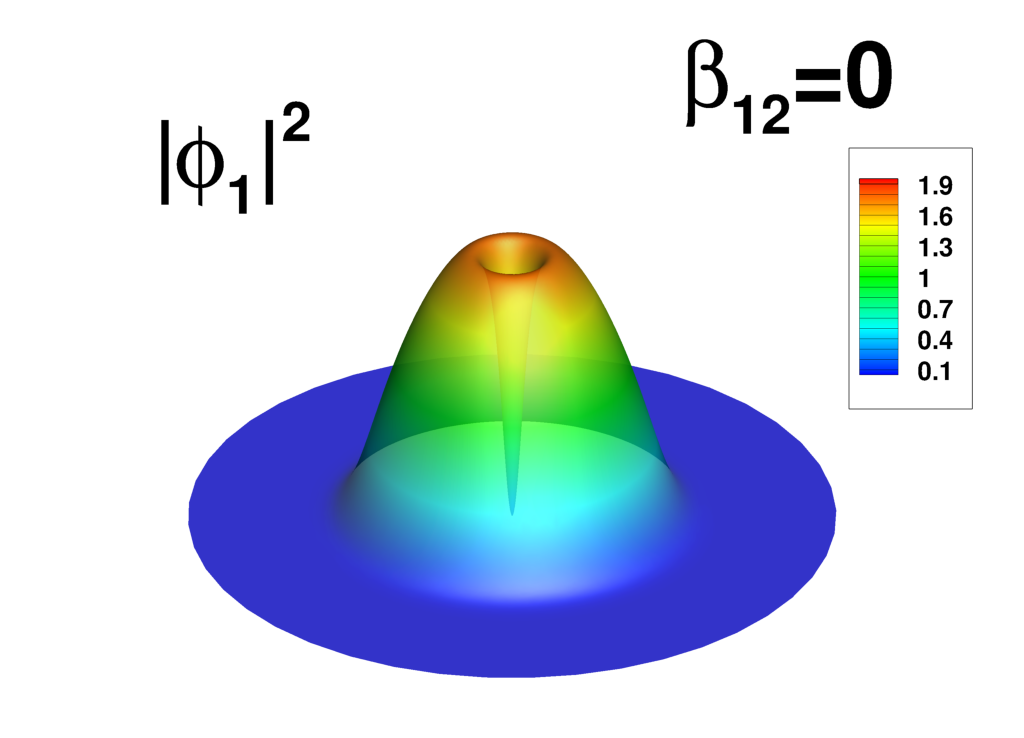}
		\label{fig:2D_VS_TF_a}
	\end{subfigure}
	\begin{subfigure}[t]{0.3\textwidth}
		\centering
		\includegraphics[width=\textwidth]{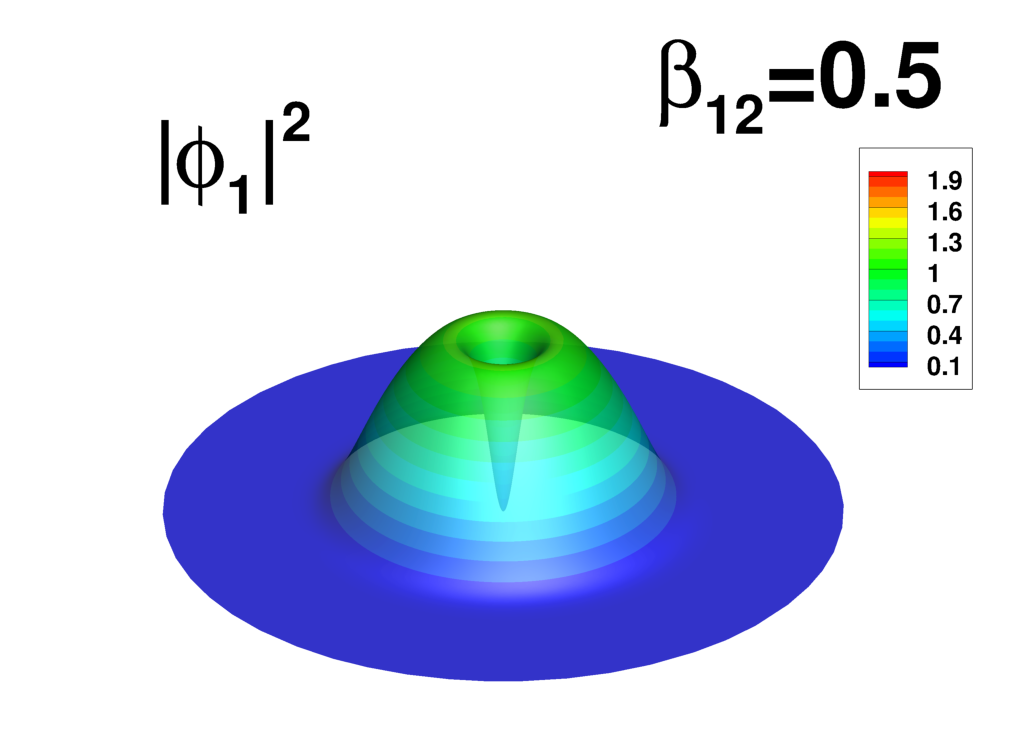}
		\label{fig:2D_VS_TF_b}
	\end{subfigure}
	\begin{subfigure}[t]{0.3\textwidth}
		\centering
		\includegraphics[width=\textwidth]{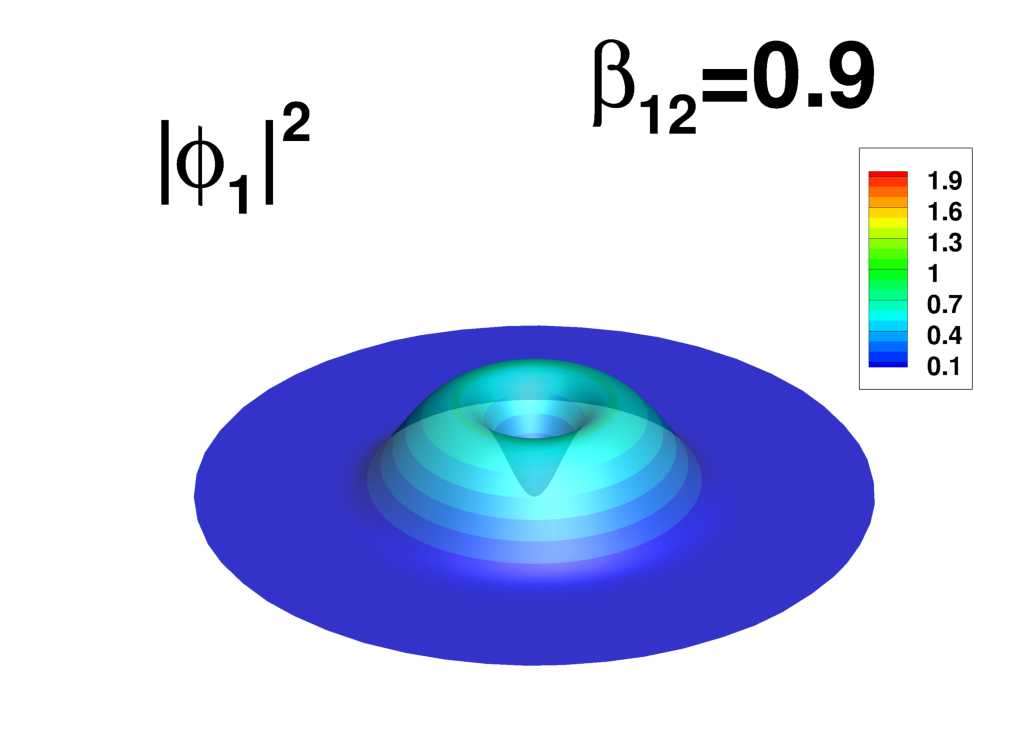}
		\label{fig:2D_VS_TF_c}
	\end{subfigure}
	\begin{subfigure}[t]{0.3\textwidth}
		\centering
		\includegraphics[width=\textwidth]{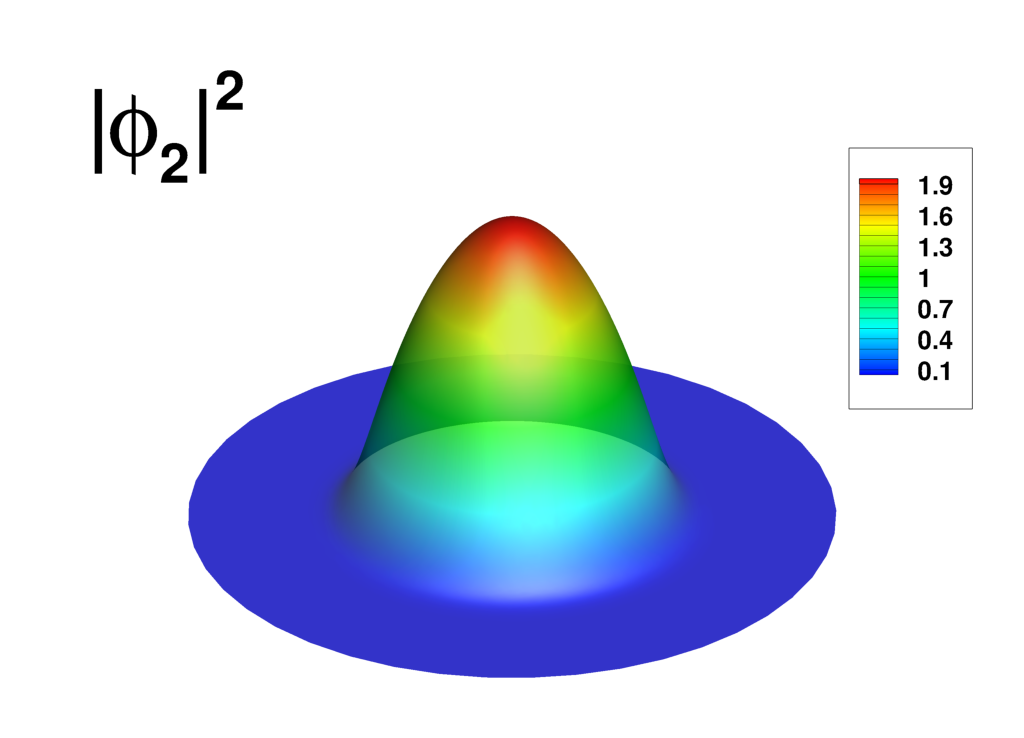}
		\label{fig:2D_VS_TF_d}
	\end{subfigure}
	\begin{subfigure}[t]{0.3\textwidth}
		\centering
		\includegraphics[width=\textwidth]{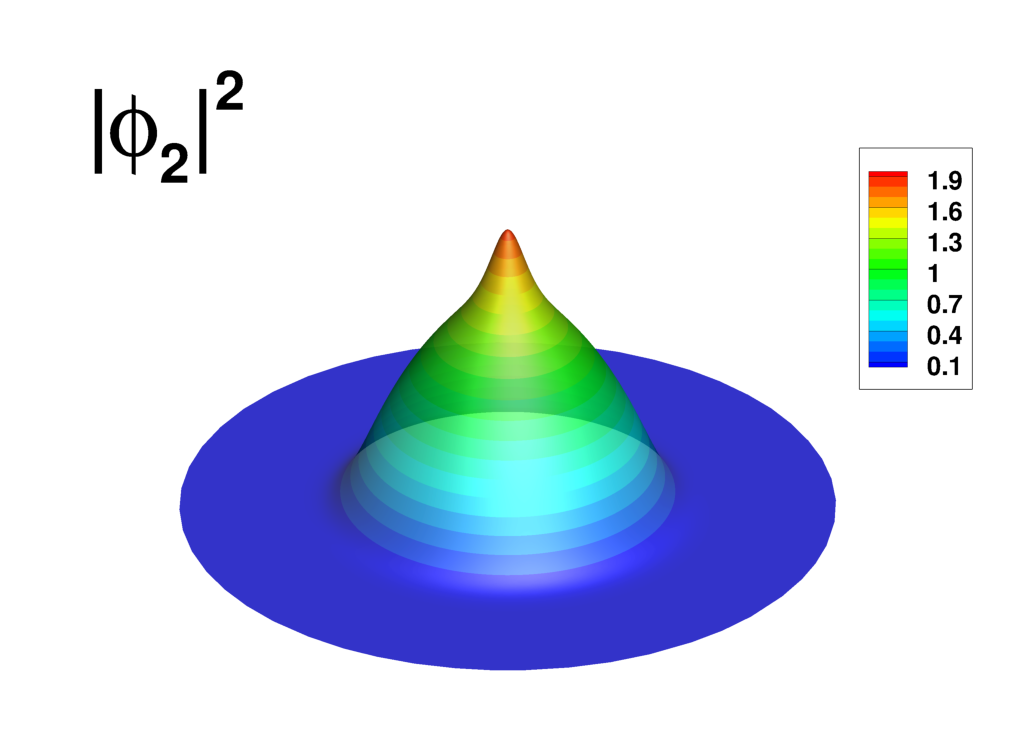}
		\label{fig:2D_VS_TF_e}
	\end{subfigure}
	\begin{subfigure}[t]{0.3\textwidth}
		\centering
		\includegraphics[width=\textwidth]{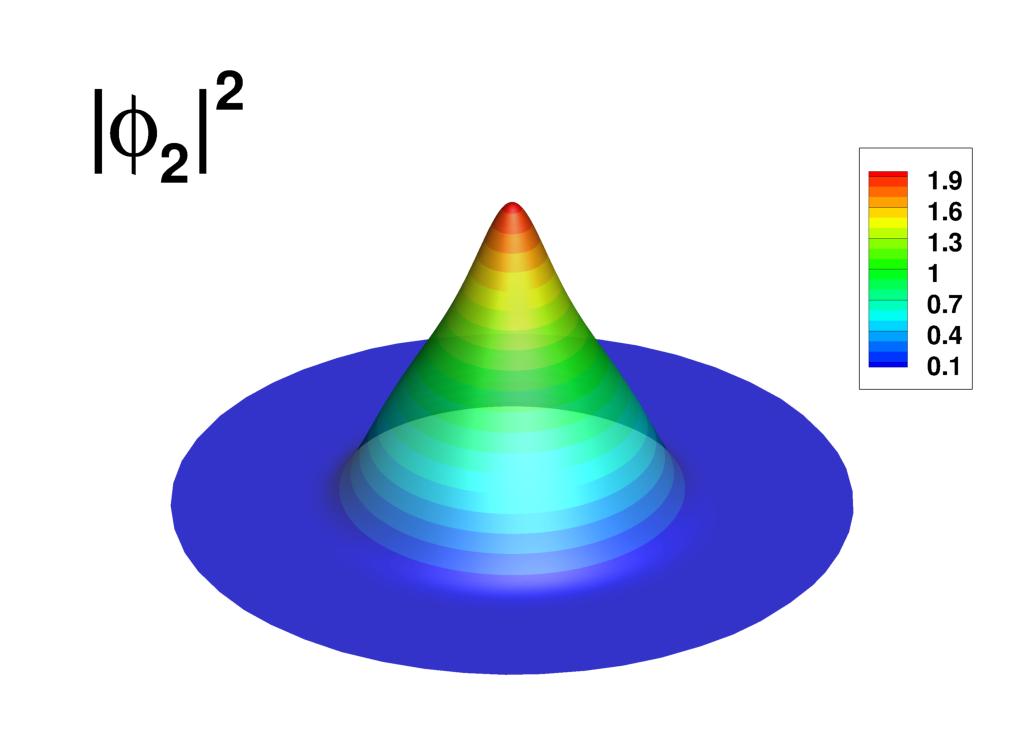}
		\label{fig:2D_VS_TF_f}
	\end{subfigure}
	\caption{2D two-component vortex-antidark case. Densities of components $|\phi_{1}|^{2}$
		(top) and $|\phi_{2}|^{2}$ (bottom) corresponding to the vortex-antidark bound mode
		for different values of $\beta_{12}$. The computational
		domain is a disk centered at the origin and with radius 19.
	}
	\label{fig-2D-VS-TF}
\end{figure}

\begin{figure}[h!]
	\centering
	\begin{subfigure}[t]{0.3\textwidth}
		\centering
		\includegraphics[width=\textwidth]{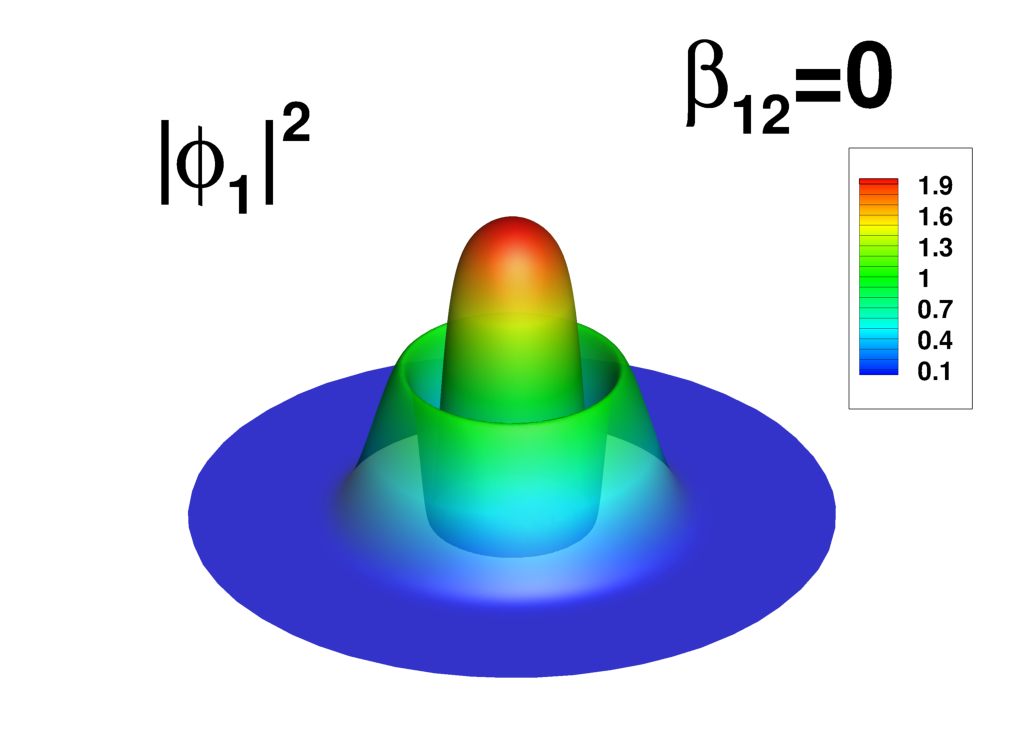}
		\label{fig:2D_RS_TF_a}
	\end{subfigure}
	\begin{subfigure}[t]{0.3\textwidth}
		\centering
		\includegraphics[width=\textwidth]{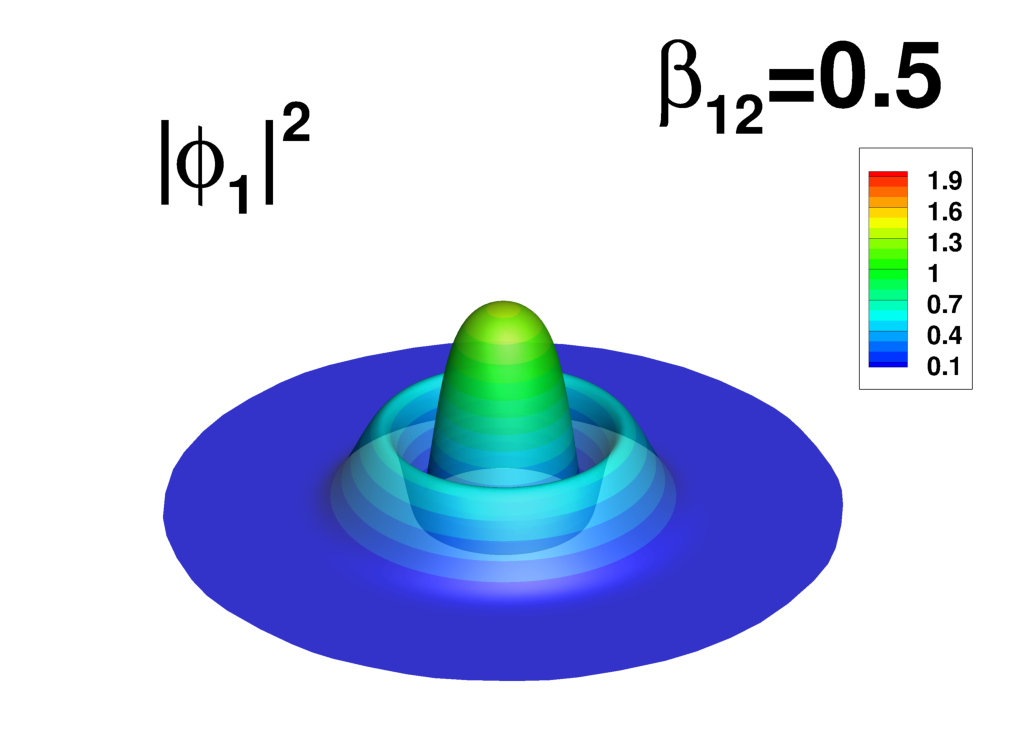}
		\label{fig:2D_RS_TF_b}
	\end{subfigure}
	\begin{subfigure}[t]{0.3\textwidth}
		\centering
		\includegraphics[width=\textwidth]{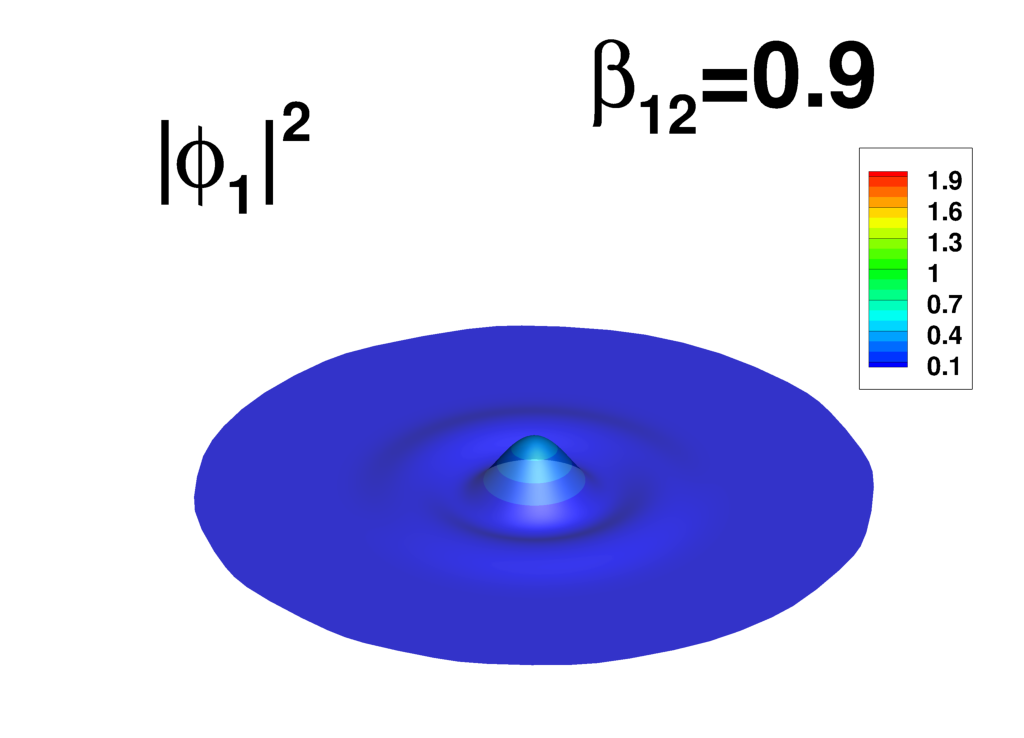}
		\label{fig:2D_RS_TF_c}
	\end{subfigure}
	\begin{subfigure}[t]{0.3\textwidth}
		\centering
		\includegraphics[width=\textwidth]{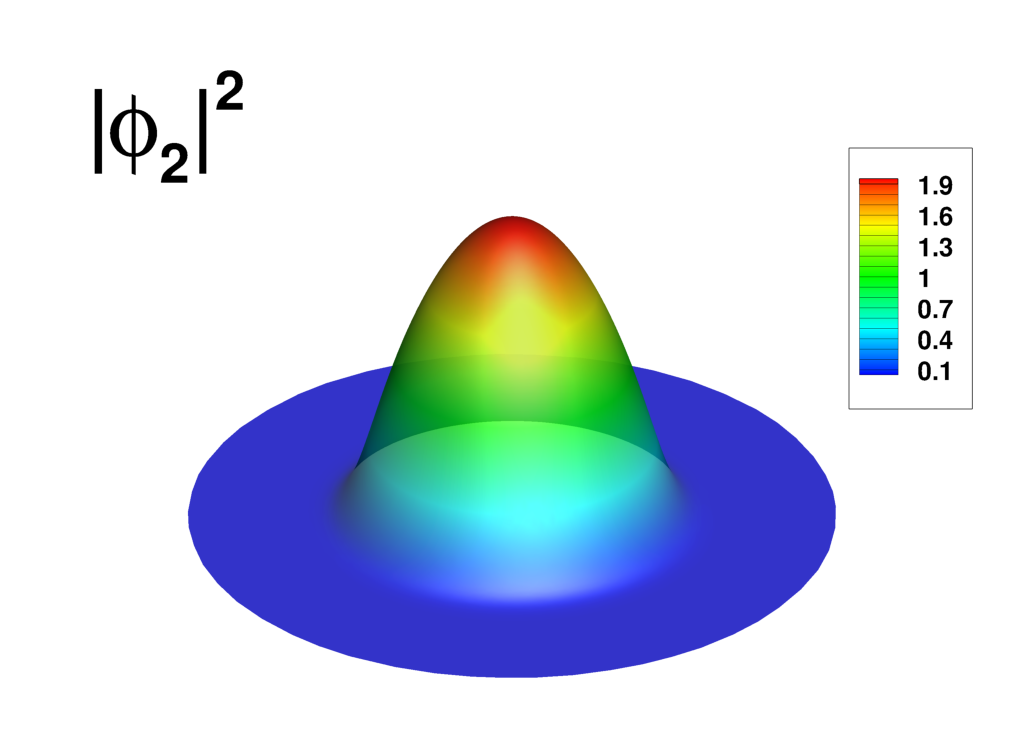}
		\label{fig:2D_RS_TF_d}
	\end{subfigure}
	\begin{subfigure}[t]{0.3\textwidth}
		\centering
		\includegraphics[width=\textwidth]{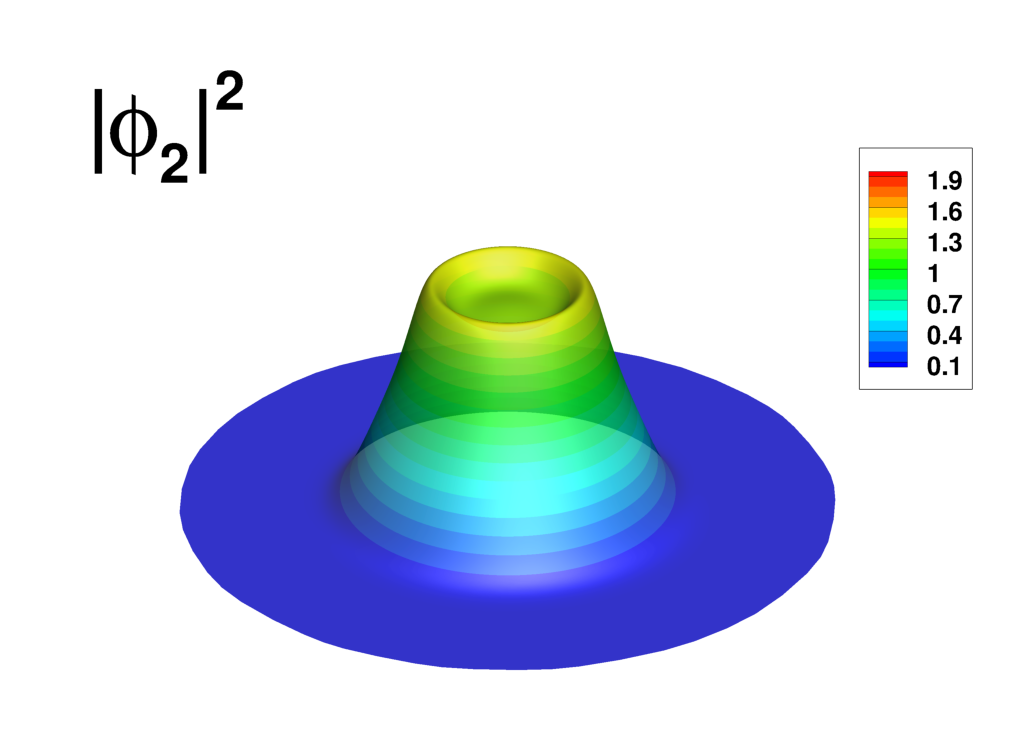}
		\label{fig:2D_RS_TF_e}
	\end{subfigure}
	\begin{subfigure}[t]{0.3\textwidth}
		\centering
		\includegraphics[width=\textwidth]{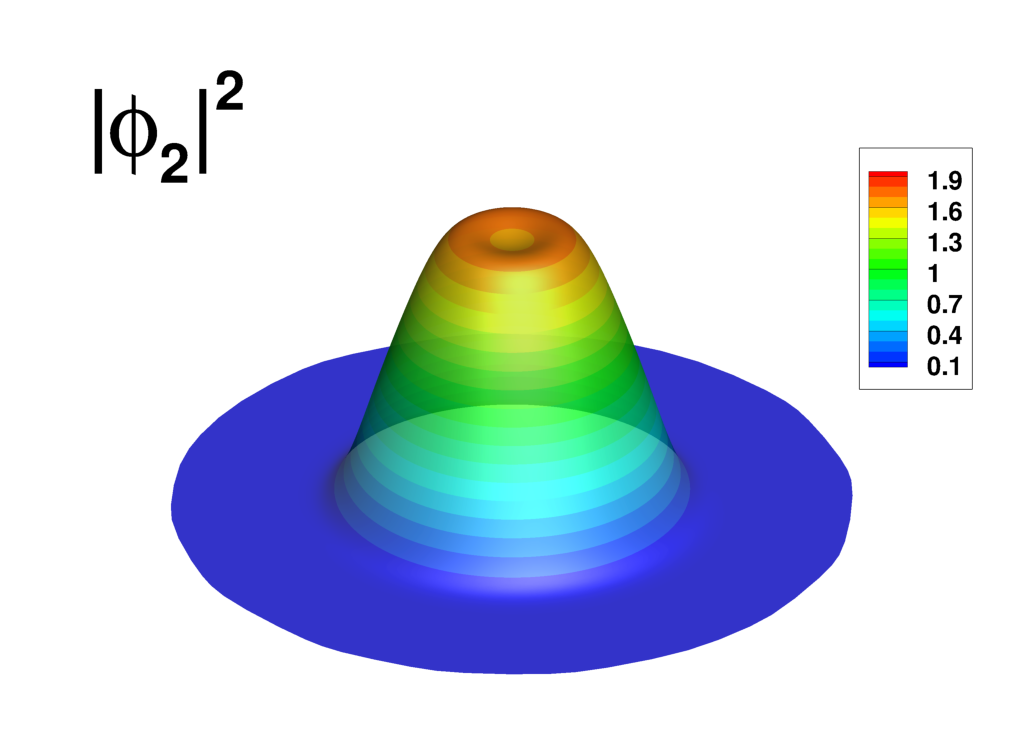}
		\label{fig:2D_RS_TF_f}
	\end{subfigure}
	\caption{2D two-component dark-antidark ring  case. Same legend and computational domain as in Fig.  \ref{fig-2D-VS-TF}.}
	\label{fig-2D-RS-TF}
\end{figure}

For this two-component case in 2D, we compute the BdG spectra of the
vortex-antidark and dark-antidark ring solutions studied in \citep{dan-2016-PRA}.
The respective BdG results we obtained from our toolbox are summarized
in panels a) and b) of Fig.  \ref{fig-2D-VS-TF-and-RS-TF-Eig}, and densities
of bound modes are shown in Figs.  \ref{fig-2D-VS-TF} (vortex-antidark states)
and  \ref{fig-2D-RS-TF} (dark-antidark states).~Note that $\phi_{1}$ contains
a ring soliton whereas $\phi_{2}$ corresponds to the ground state in Fig.  \ref{fig-2D-RS-TF}%
.~Finally, we highlight the additional feature of our toolbox in considering
distinct values of the interaction coefficients $\beta_{ij}, 1 \leq i,j \leq 2$.%
~This way, the user has the flexibility to adjust these values according to
other configurations of interest (and potentially, experimental setups) for
performing distinct numerical studies other than those reported in \citep{dan-2016-PRA}.

\subsection{2D two-component case: from soliton necklaces to multipoles}

The last 2D cases that we considered for testing our parallel toolbox 
stem from   \cite{charalampidis2020bifurcation}. At first, we construct
the ground state of Eq.  \eqref{eq-scal-GP-stat} by
\begin{align}
\phi = \sqrt{\frac{\omega_\perp}{2\pi}} %
H_0(\sqrt{\omega_\perp}x)H_0(\sqrt{\omega_\perp}y)e^{-\frac{1}{2}\omega_\perp(x^2+y^2)},
\end{align}
and use it to seed  Newton's method with $\beta=1.03$ and $\omegap=0.2$. The branch of
the ground state is traced from $\mu\approx 0.202$, \ie close to the linear
limit where this state bifurcates from, until $\mu=1$.  The terminal profile
$\phi$, now called $\phi_{1}$, is extracted while setting
$\mu_{1}=\mu=1$.~Then, we focus on Eq.  \eqref{eq-scal-GP2c-stat} with
$\beta_{11}=1.03$, $\beta_{22}=0.97$, and $\beta_{12}=1$
($\omegap=0.2$ and $\mu_{1}=1$ are as before). Following
the approach discussed in  \cite{PhysRevE.91.012924}, we plug the (terminal)
profile $\phi_{1}$ into the equation for $\phi_{2}$ [cf.~Eq.  \eqref{eq-scal-GP2c-stat}],
and linearize it with respect to $\phi_{2}$. This process results in the
following eigenvalue problem for $(\mu_{2},\phi_{2})$
\begin{align}
-\frac{1}{2}\nabla^{2}\phi_{2}+%
C_\eff\phi_{2}=\mu_{2}\phi_{2}
\label{eq-bdg-ll}
\end{align}
with $C_\eff=C_\trap+\beta_{21}|\phi_{1}|^{2}$ being the effective
potential  \cite{PhysRevE.91.012924} which is responsible for ``trapping''
bound modes in the $\phi_{2}$ component of Eq.  \eqref{eq-scal-GP2c-stat}.
Upon solving Eq.  \eqref{eq-bdg-ll} numerically, we obtain eigenvalue-eigenvector
pairs $(\mu_{2},\phi_{2})$, that together with $(\mu_{1},\phi_{1})$ form
the initial guess that we seed to Newton's method. We then trace
branches of bound modes of the coupled system of Eq.  \eqref{eq-scal-GP2c-stat},
over the principal continuation parameter $\mu_{2}$. Note
that upon selecting a pair $(\mu_{2},\phi_{2})$, we perform continuation
from $\mu_2$ until $\mu_{2}+0.4$ in all the cases that we present next.

\begin{figure}[!h]
	\centering
	\centering
 	\includegraphics[width=0.32\textwidth]{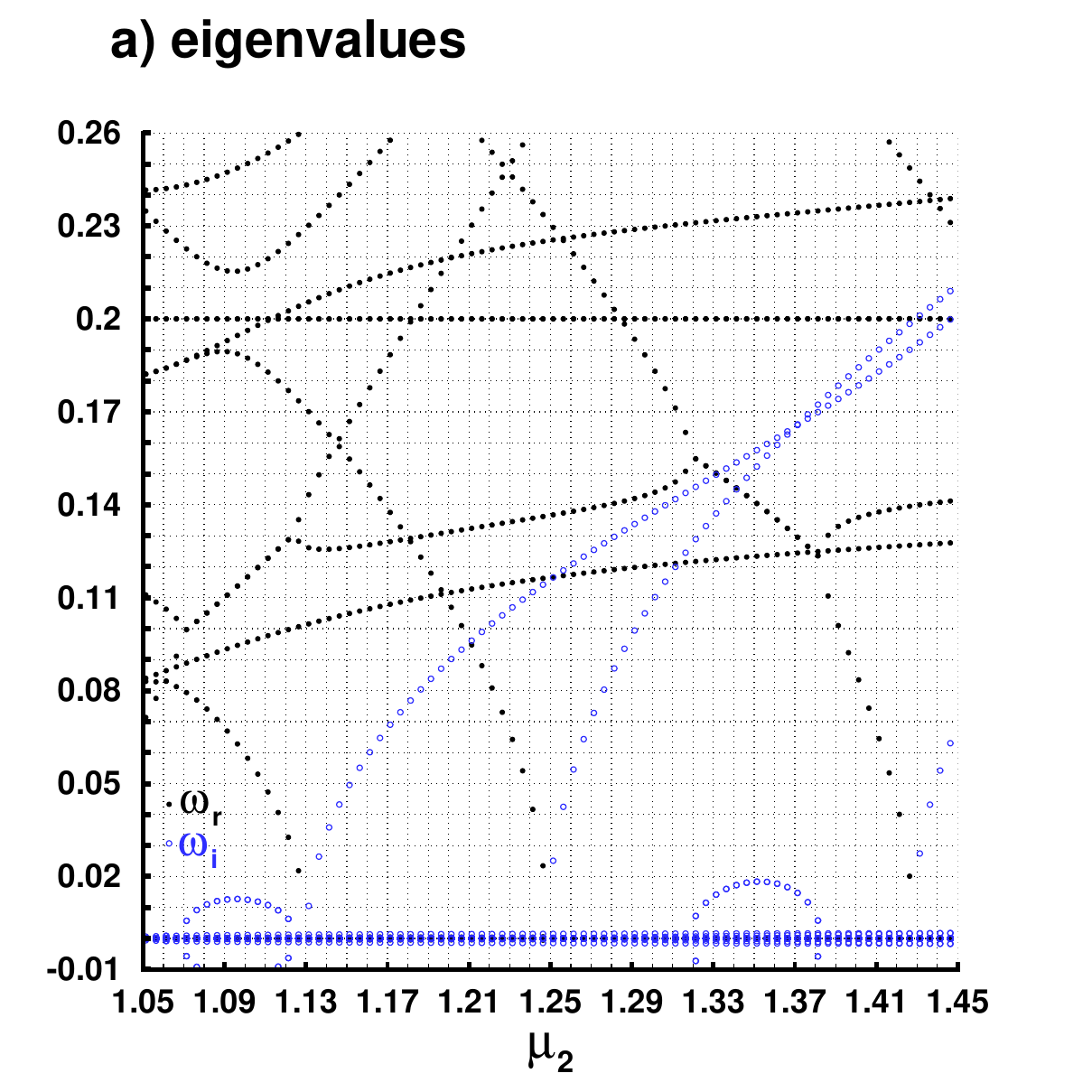}
 	\includegraphics[width=0.32\textwidth]{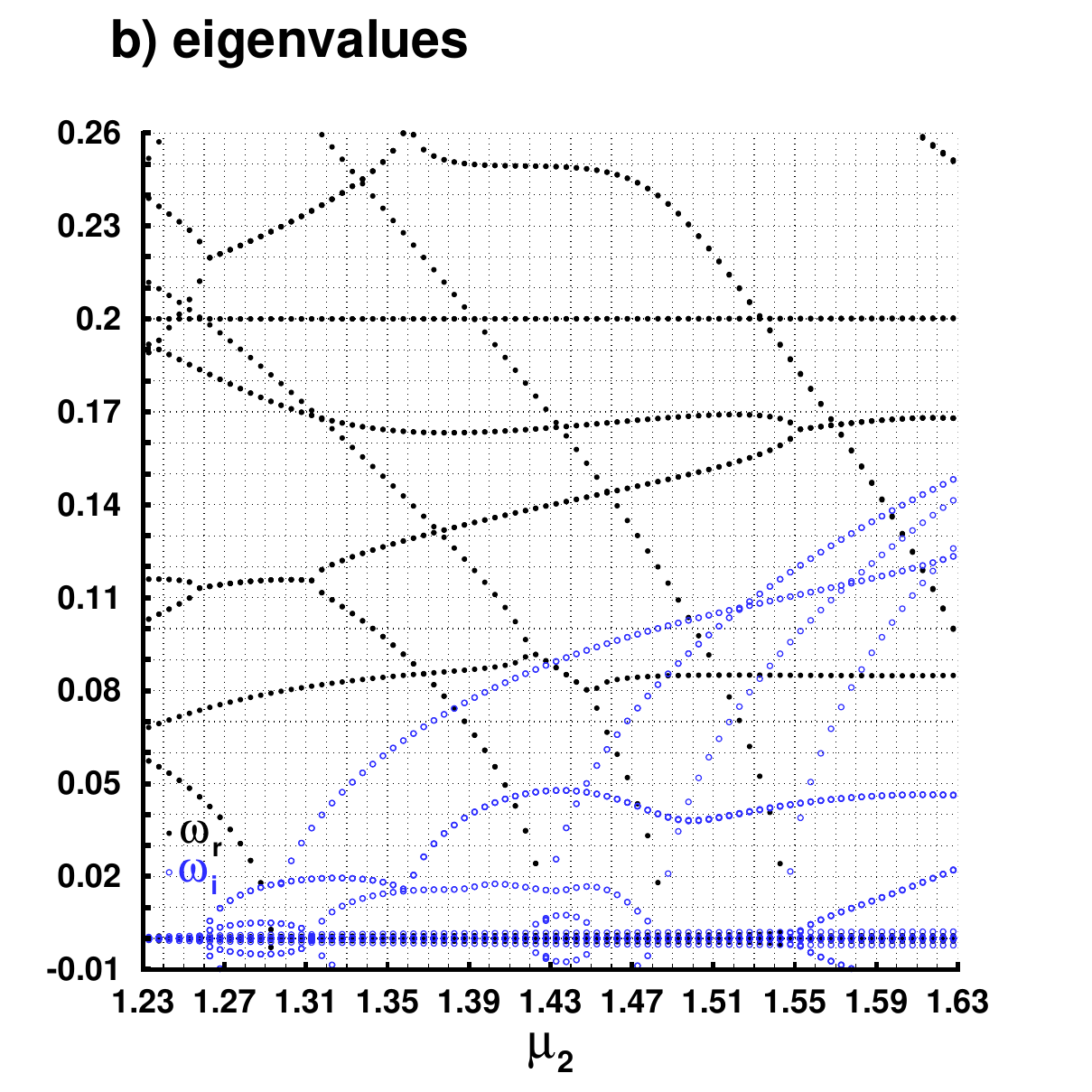}
 	\includegraphics[width=0.32\textwidth]{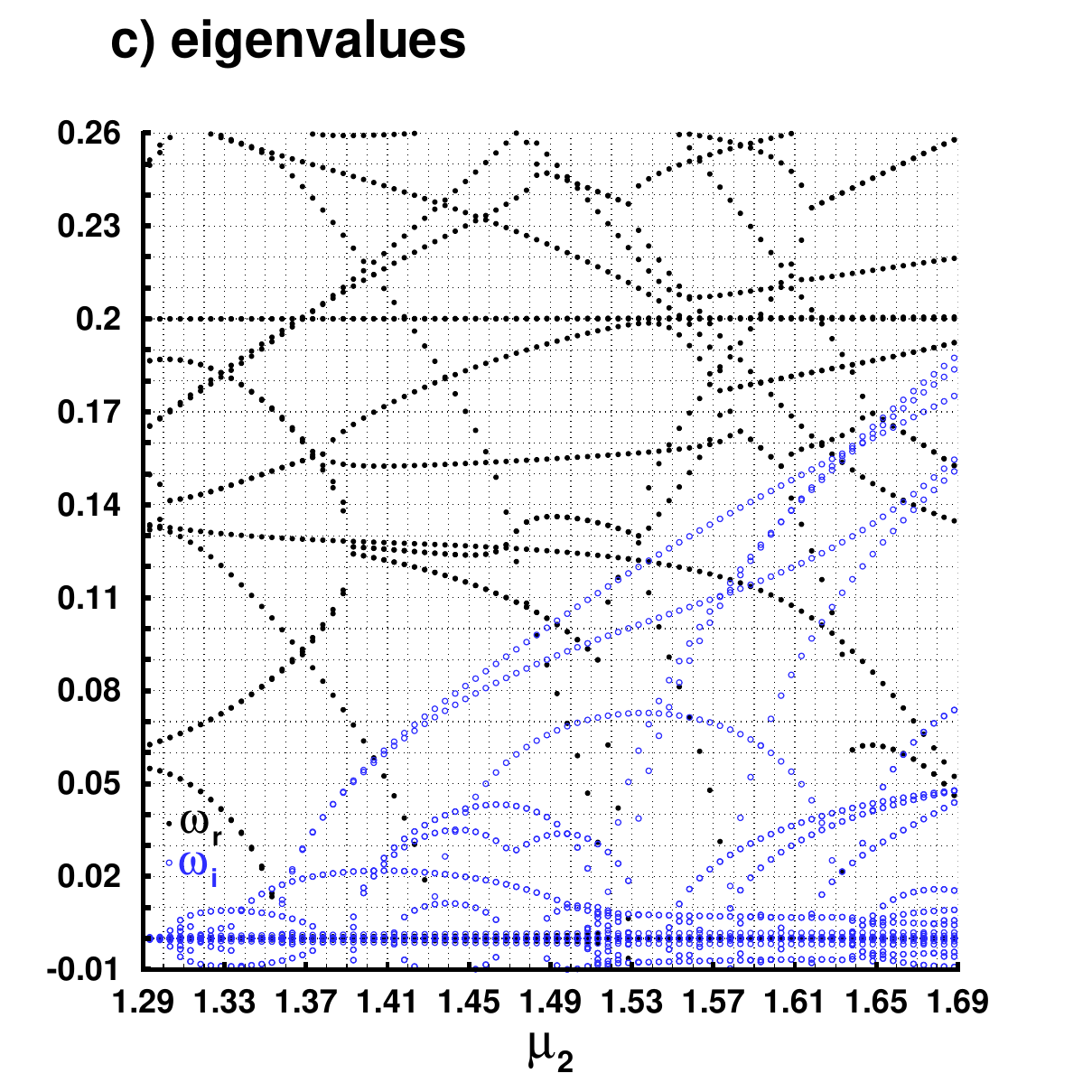}
% 	\begin{subfigure}[t]{0.45\textwidth}
% 		\centering
% 		\includegraphics[width=\textwidth]{\figpath/2D_Hermite_LL_1}
% 		\label{fig:2D-Herm-LL-1}
% 	\end{subfigure}
% 	\begin{subfigure}[t]{0.45\textwidth}
% 		\centering
% 		\includegraphics[width=\textwidth]{\figpath/2D_Hermite_LL_7}
% 		\label{fig:2D-Herm-LL-7}
% 	\end{subfigure}
% 	\begin{subfigure}[t]{0.45\textwidth}
% 		\centering
% 		\includegraphics[width=\textwidth]{\figpath/2D_Hermite_LL_8}
% 		\label{fig:2D-Herm-LL-8}
% 	\end{subfigure}
	\caption{
	2D two-component soliton and necklace configurations. The BdG spectra as functions of $\mu_{2}$
	corresponding to a) the dark-bright soliton branch, b) ground
	state ($\phi_{1}$) and soliton necklace ($\phi_{2}$) branch,
	and c) the multipole branch. Each of these branches bifurcate
	from $(\mu_{1},\mu_{2})\approx (1,1.05133)$, $(1,1.23276)$, and $(1,1.29325)$.
	Their respective density profiles (for different values of $\mu_{2}$)
	are shown in Figs.  \ref{fig-2D-HermLL_1}, \ref{fig-2D-HermLL_7},
	and  \ref{fig-2D-HermLL_8}.
% 	2D two-component case: eigenvalue of dark-bright soliton stripe
% 	(Fig. (2a), Fig. (15a), Fig. (16c) \\
% 	a) : $\mu_1=1, \mu_2=1.05133$, b) : $\mu_1=1, \mu_2=1.23276$, c) : $\mu_1=1, \mu_2=1.29325$.
	}
	\label{fig-2D-Herm-LL-1-7}
\end{figure}

\begin{figure}[!h]
	\centering
	\begin{subfigure}[t]{0.3\textwidth}
		\centering
		\includegraphics[width=\textwidth]{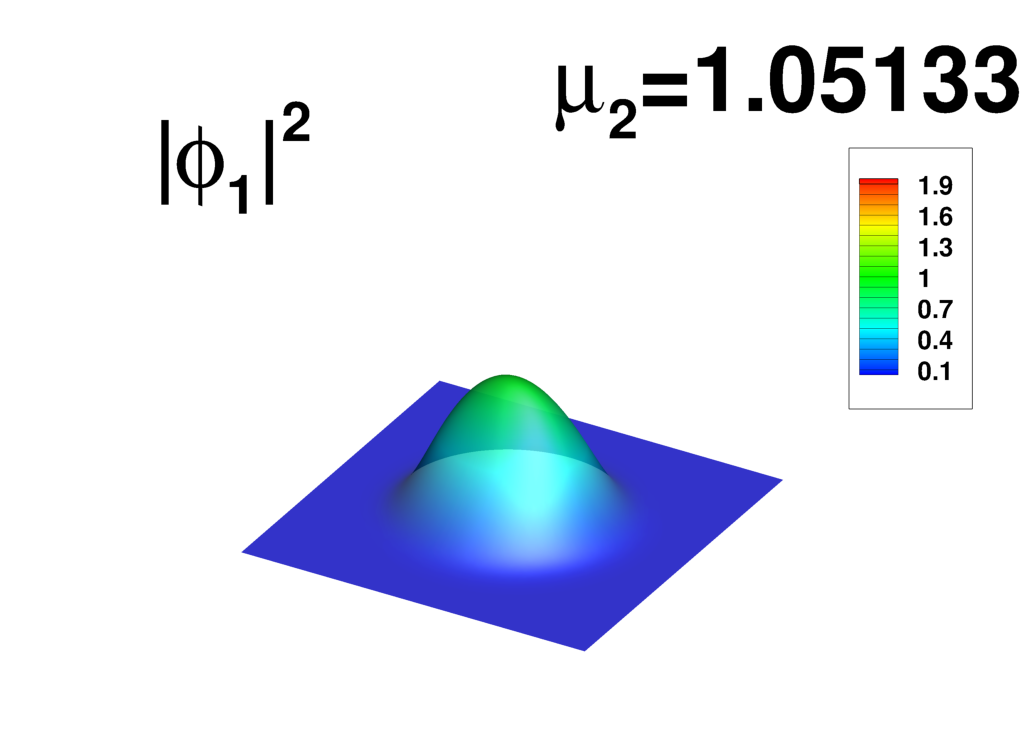}
		\label{fig:2D_HermLL_SOL_1_a}
	\end{subfigure}
	\begin{subfigure}[t]{0.3\textwidth}
		\centering
		\includegraphics[width=\textwidth]{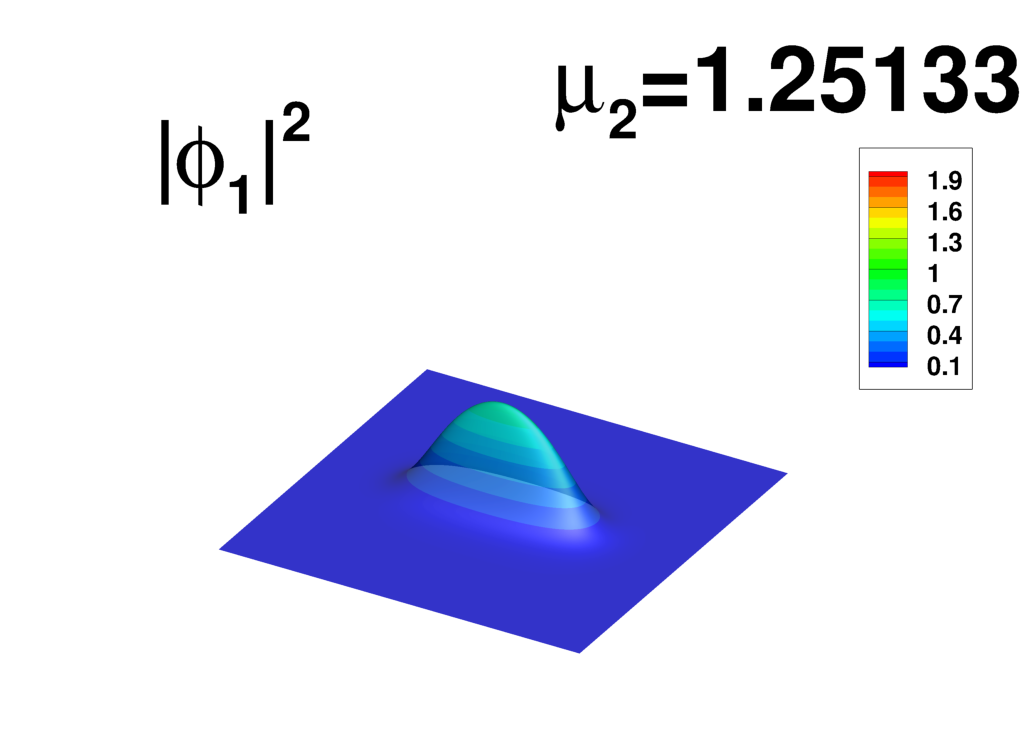}
		\label{fig:2D_HermLL_SOL_1_b}
	\end{subfigure}
	\begin{subfigure}[t]{0.3\textwidth}
		\centering
		\includegraphics[width=\textwidth]{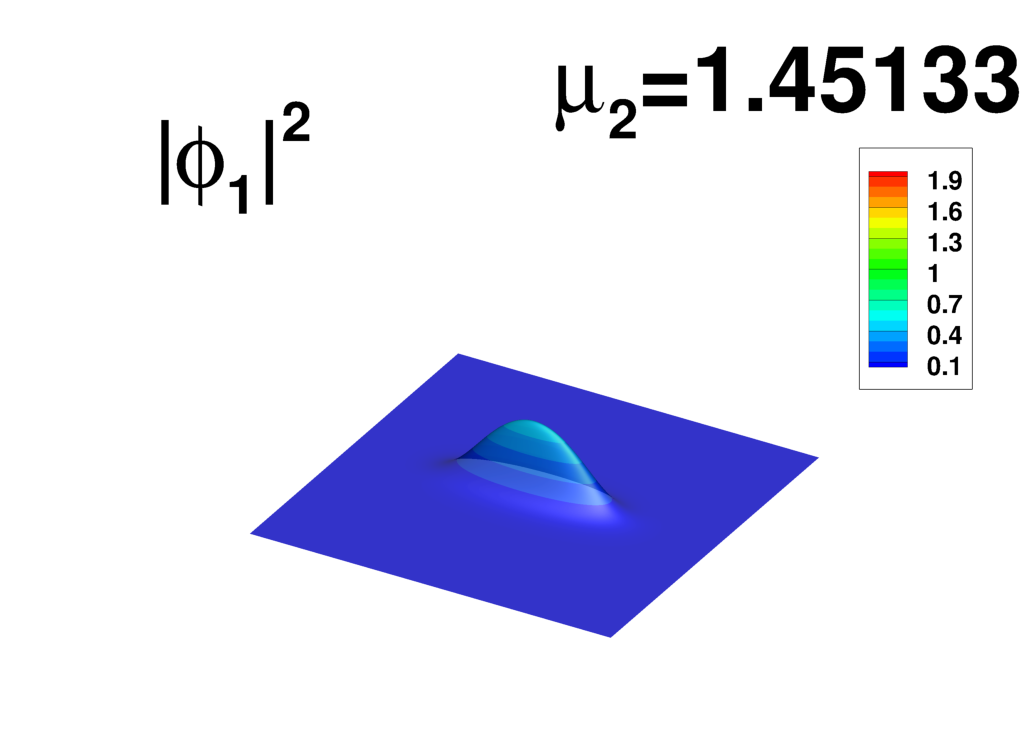}
		\label{fig:2D_HermLL_SOL_1_c}
	\end{subfigure}
	\begin{subfigure}[t]{0.3\textwidth}
		\centering
		\includegraphics[width=\textwidth]{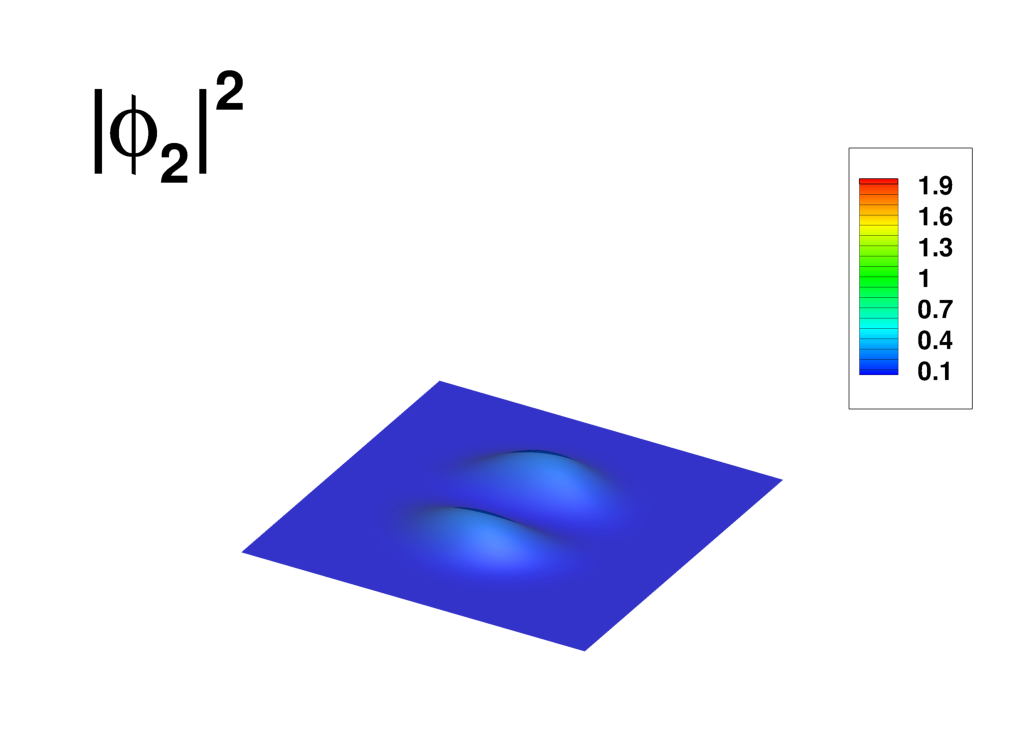}
		\label{fig:2D_HermLL_SOL_1_d}
	\end{subfigure}
	\begin{subfigure}[t]{0.3\textwidth}
		\centering
		\includegraphics[width=\textwidth]{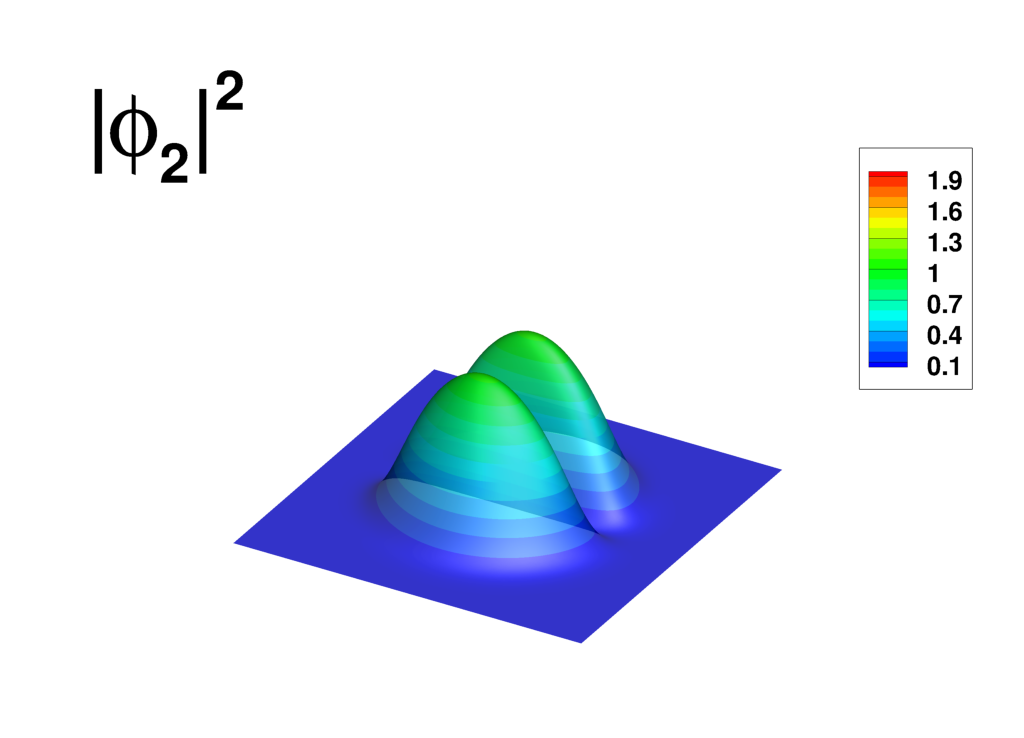}
		\label{fig:2D_HermLL_SOL_1_e}
	\end{subfigure}
	\begin{subfigure}[t]{0.3\textwidth}
		\centering
		\includegraphics[width=\textwidth]{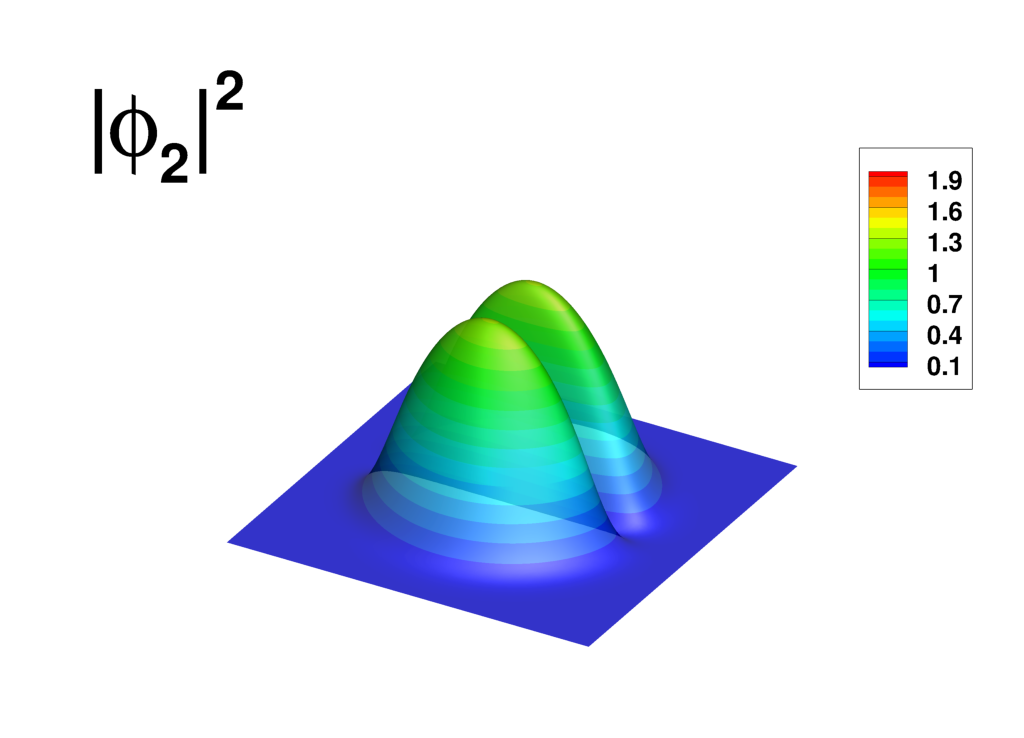}
		\label{fig:2D_HermLL_SOL_1_f}
	\end{subfigure}
	\caption{2D two-component dark-bright soliton. Density profiles of each of the components,
	\ie $|\phi_{1}|^{2}$ (top row) and $|\phi_{2}|^{2}$ (bottom row), for different values of $\mu_{2}$. 
	For this case $\mu_{1}=1$ (see text), and the computational
	domain 
 is $[-11.62,11.62]^2$.}
	\label{fig-2D-HermLL_1}
\end{figure}

\begin{figure}[!h]
	\centering
	\begin{subfigure}[t]{0.3\textwidth}
		\centering
		\includegraphics[width=\textwidth]{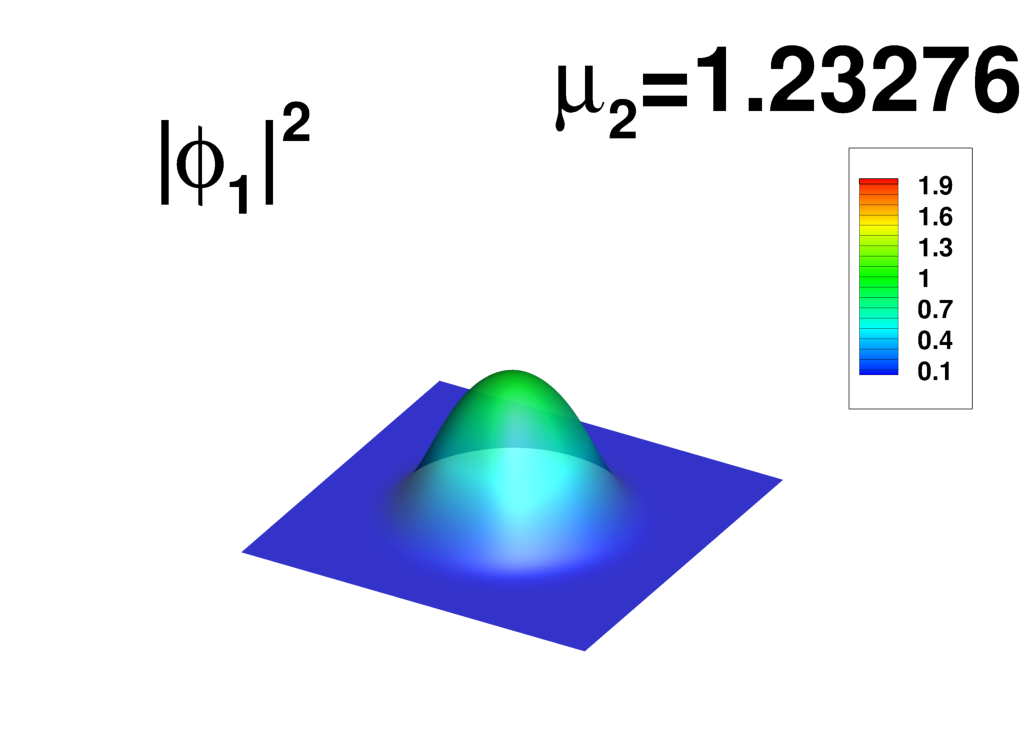}
		\label{fig:2D_HermLL_SOL_7_a}
	\end{subfigure}
	\begin{subfigure}[t]{0.3\textwidth}
		\centering
		\includegraphics[width=\textwidth]{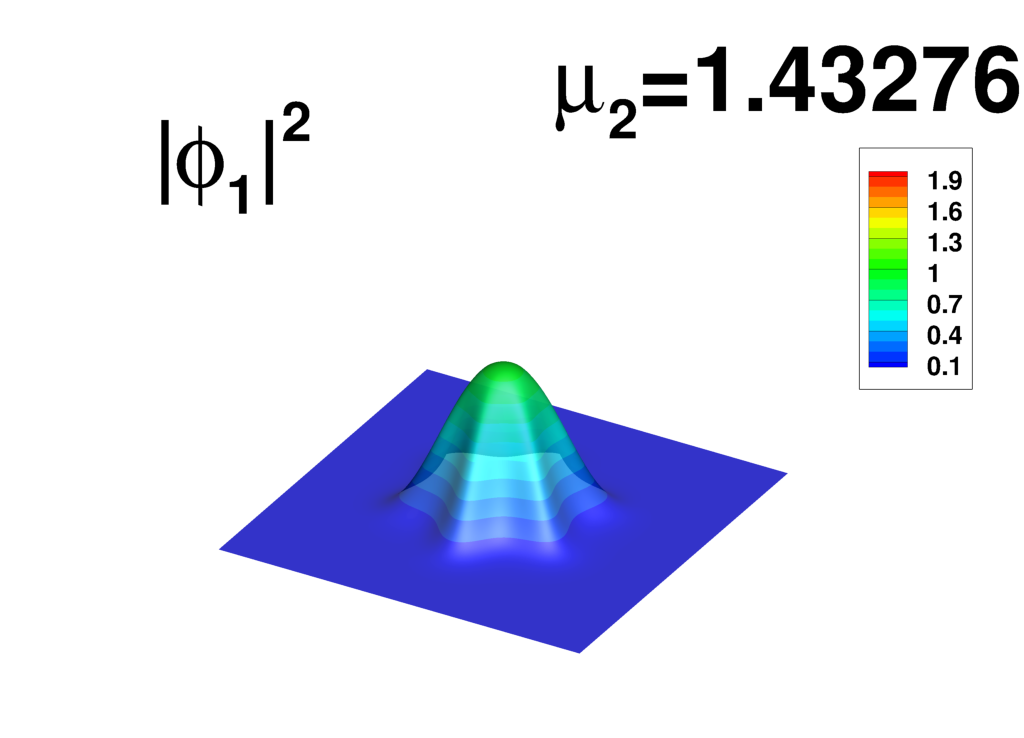}
		\label{fig:2D_HermLL_SOL_7_b}
	\end{subfigure}
	\begin{subfigure}[t]{0.3\textwidth}
		\centering
		\includegraphics[width=\textwidth]{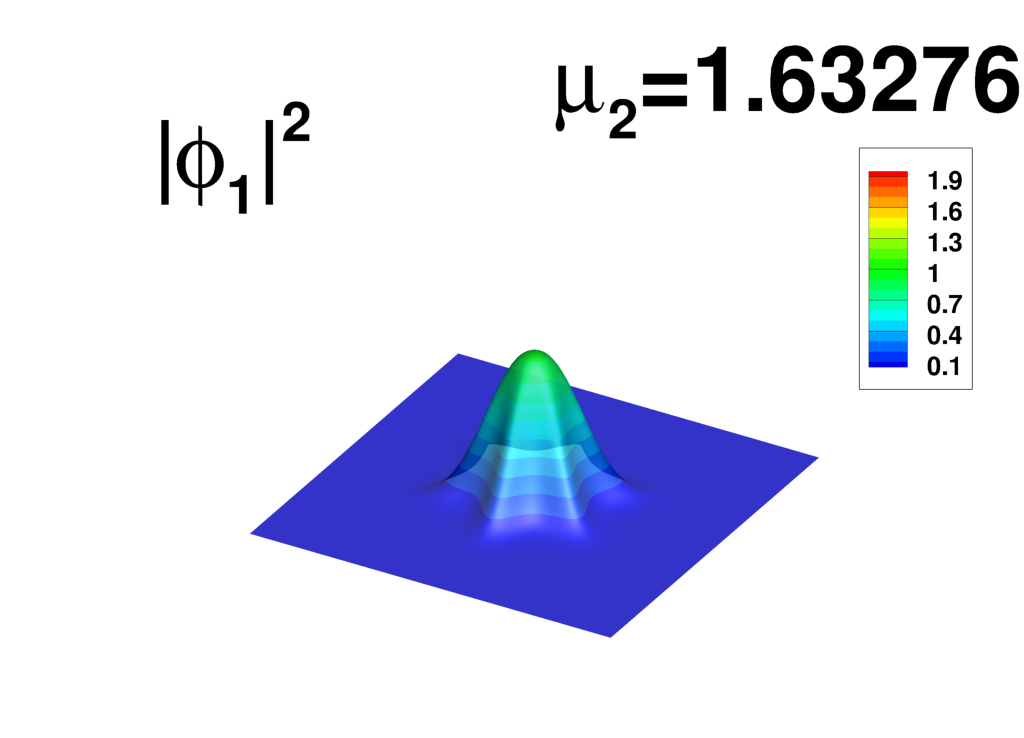}
		\label{fig:2D_HermLL_SOL_7_c}
	\end{subfigure}
	\begin{subfigure}[t]{0.3\textwidth}
		\centering
		\includegraphics[width=\textwidth]{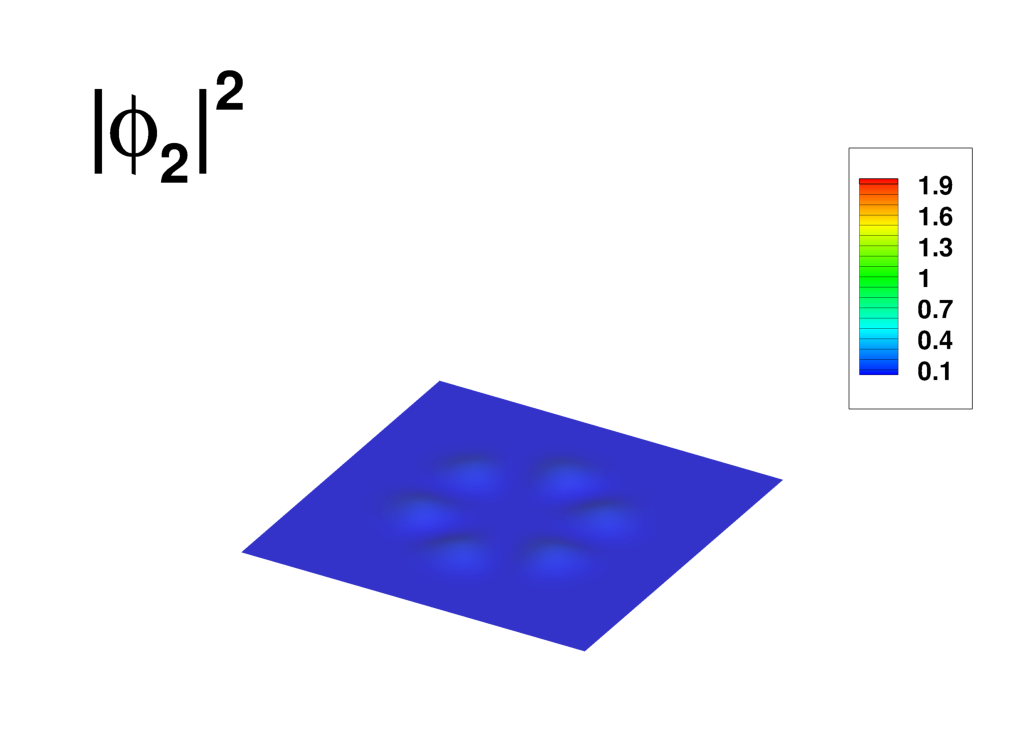}
		\label{fig:2D_HermLL_SOL_7_d}
	\end{subfigure}
	\begin{subfigure}[t]{0.3\textwidth}
		\centering
		\includegraphics[width=\textwidth]{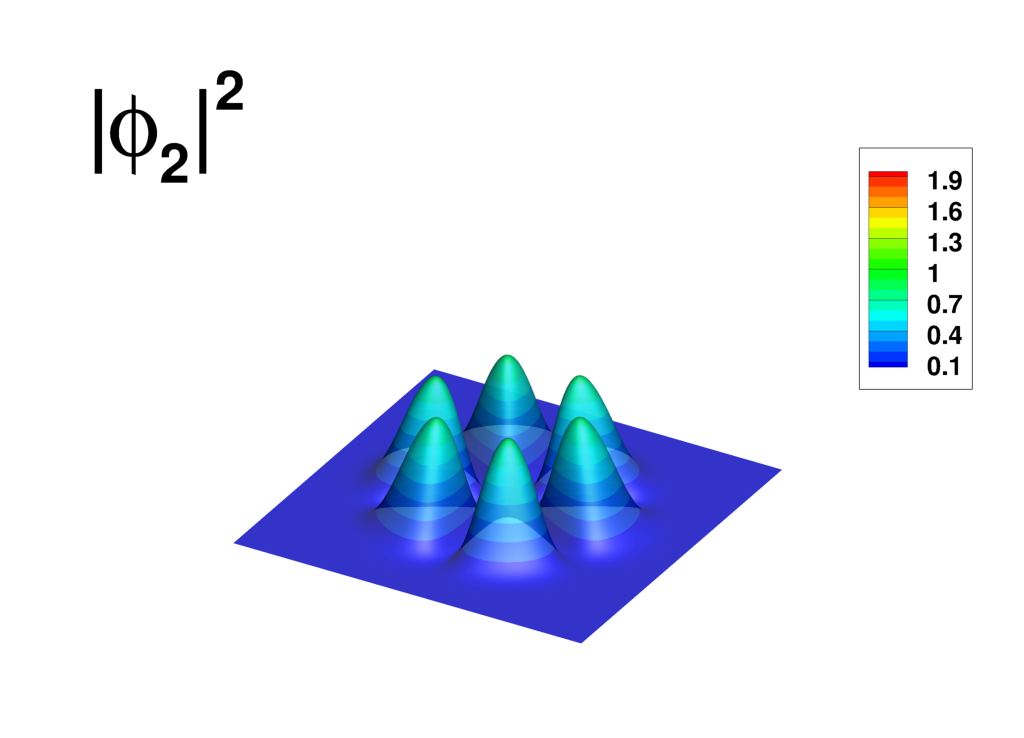}
		\label{fig:2D_HermLL_SOL_7_e}
	\end{subfigure}
	\begin{subfigure}[t]{0.3\textwidth}
		\centering
		\includegraphics[width=\textwidth]{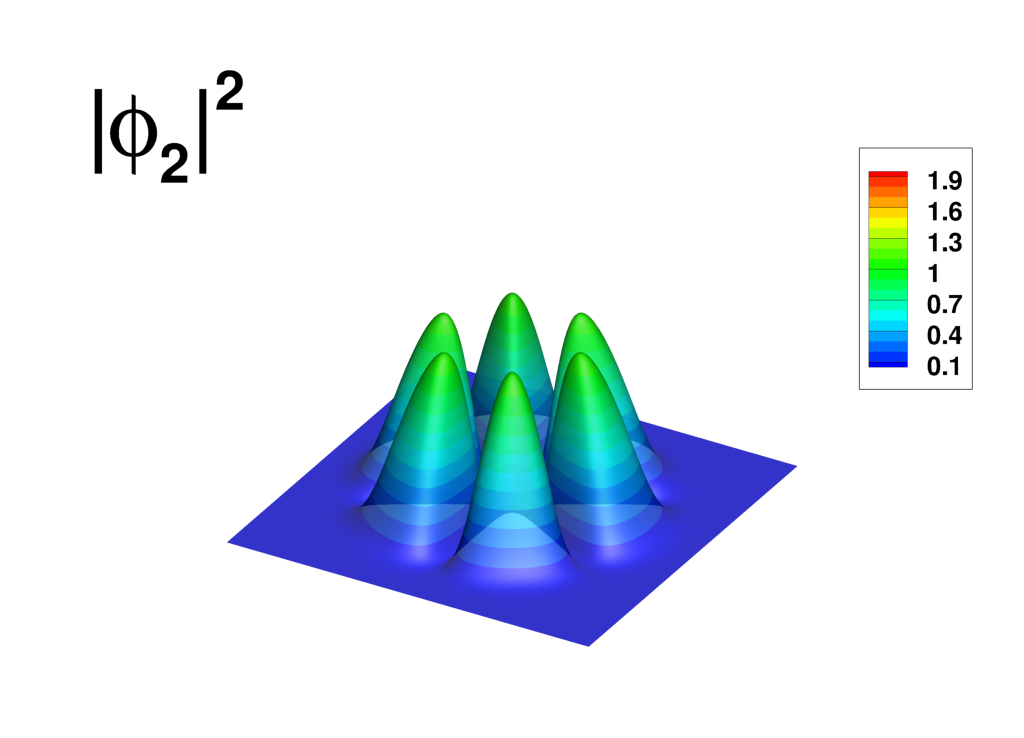}
		\label{fig:2D_HermLL_SOL_7_f}
	\end{subfigure}
	\caption{2D two-component ground-state-soliton-necklace case. Same caption and computational domain as in Fig.  \ref{fig-2D-HermLL_1}.
		As the chemical
	potential $\mu_{2}$ increases, the imprint of
	$\phi_{2}$ on $\phi_{1}$ becomes more apparent (see the third column).}
	\label{fig-2D-HermLL_7}
\end{figure}

\begin{figure}[!h]
	\centering
	\begin{subfigure}[t]{0.3\textwidth}
		\centering
		\includegraphics[width=\textwidth]{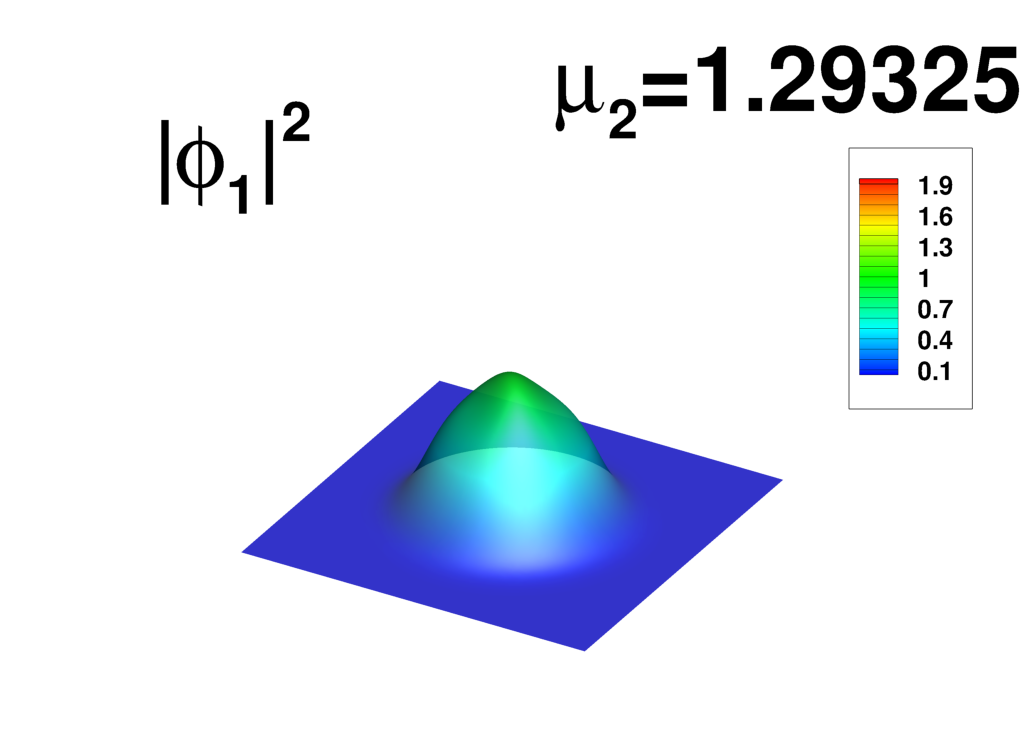}
		\label{fig:2D_HermLL_SOL_8_a}
	\end{subfigure}
	\begin{subfigure}[t]{0.3\textwidth}
		\centering
		\includegraphics[width=\textwidth]{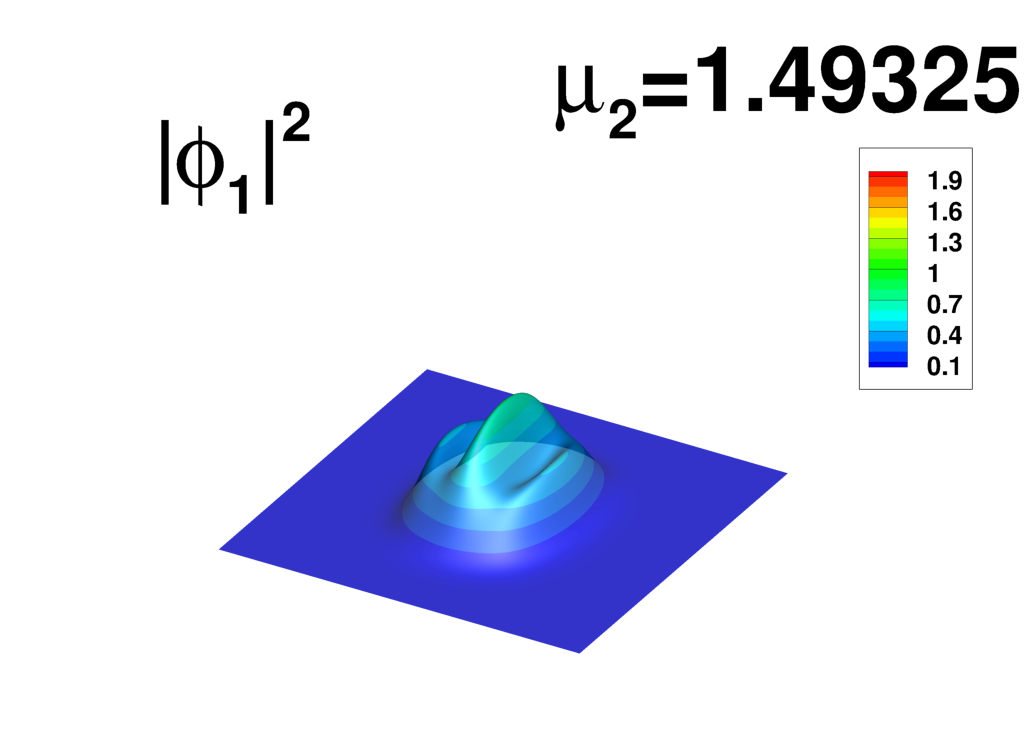}
		\label{fig:2D_HermLL_SOL_8_b}
	\end{subfigure}
	\begin{subfigure}[t]{0.3\textwidth}
		\centering
		\includegraphics[width=\textwidth]{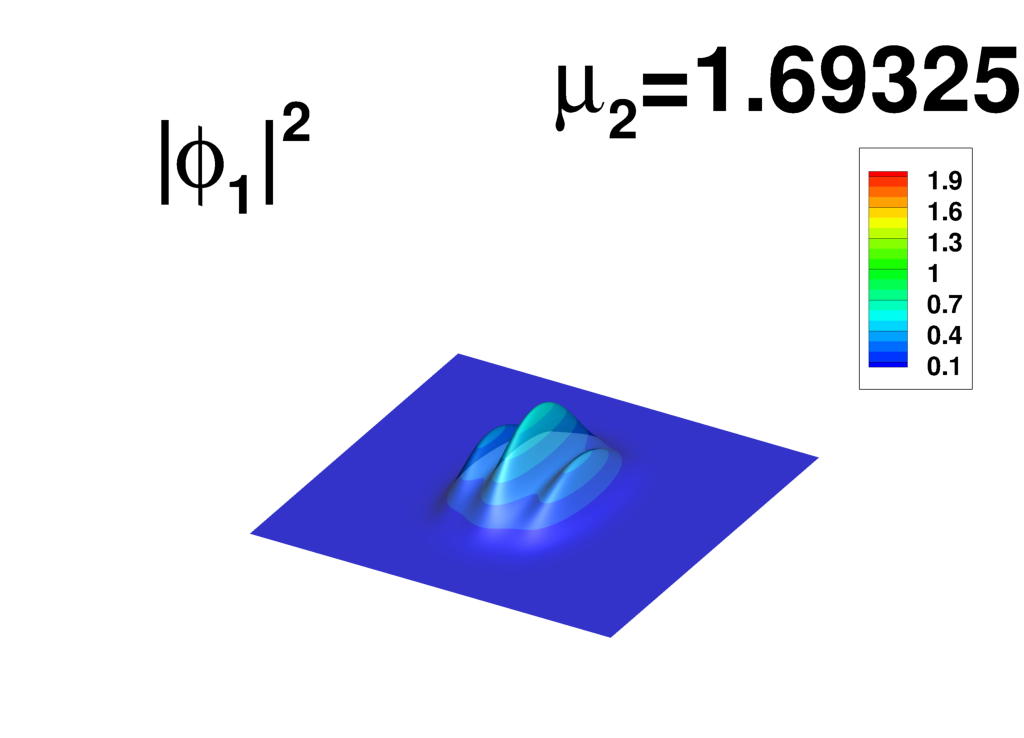}
		\label{fig:2D_HermLL_SOL_8_c}
	\end{subfigure}
	\begin{subfigure}[t]{0.3\textwidth}
		\centering
		\includegraphics[width=\textwidth]{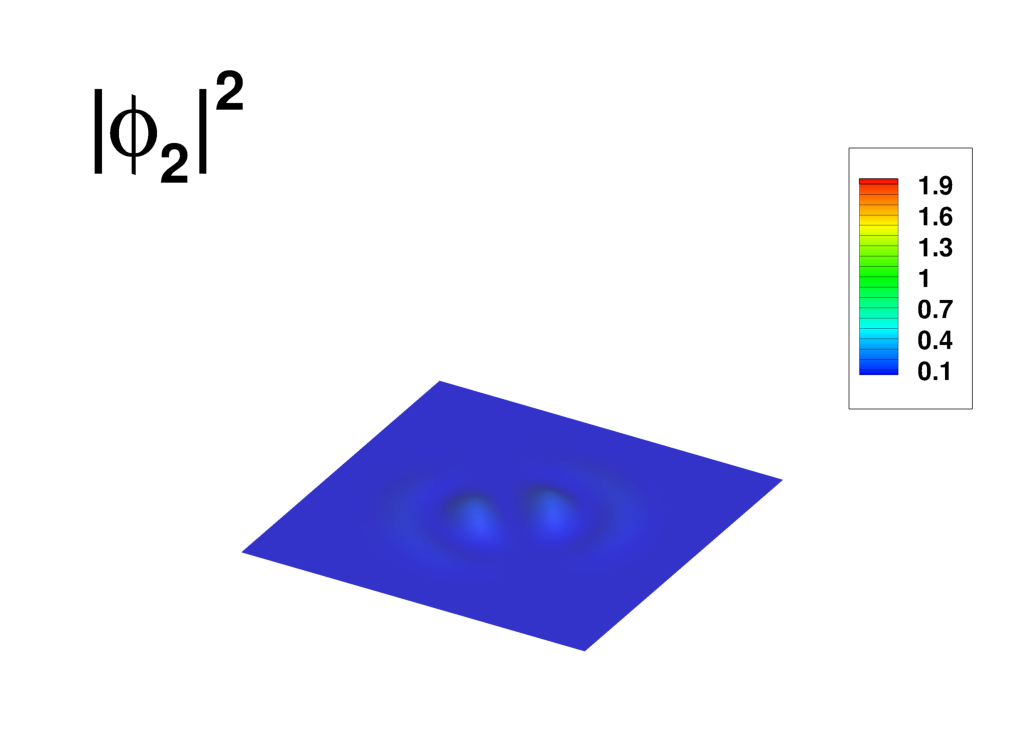}
		\label{fig:2D_HermLL_SOL_8_d}
	\end{subfigure}
	\begin{subfigure}[t]{0.3\textwidth}
		\centering
		\includegraphics[width=\textwidth]{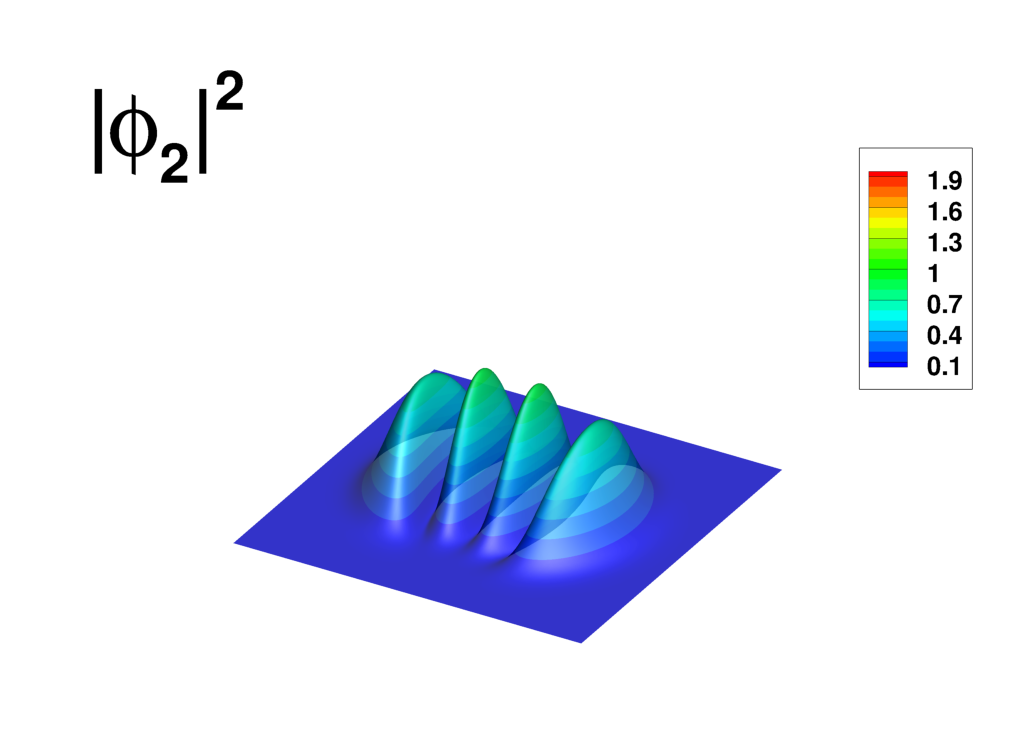}
		\label{fig:2D_HermLL_SOL_8_e}
	\end{subfigure}
	\begin{subfigure}[t]{0.3\textwidth}
		\centering
		\includegraphics[width=\textwidth]{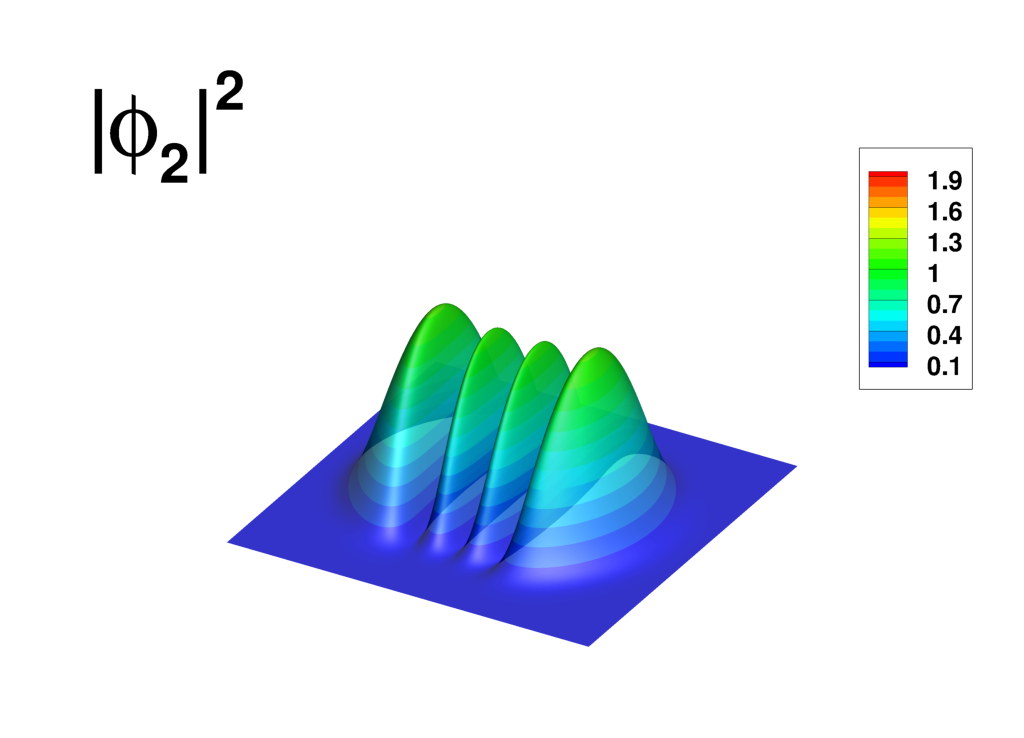}
		\label{fig:2D_HermLL_SOL_8_f}
	\end{subfigure}
% 	\begin{subfigure}[t]{0.3\textwidth}
% 		\centering
% 		\includegraphics[width=\textwidth]{\figpath/2D_HermLL_SOL_8_i}
% 		\label{fig:2D_HermLL_SOL_8_i}
% 	\end{subfigure}
% 	\begin{subfigure}[t]{0.3\textwidth}
% 		\centering
% 		\includegraphics[width=\textwidth]{\figpath/2D_HermLL_SOL_8_j}
% 		\label{fig:2D_HermLL_SOL_8_j}
% 	\end{subfigure}
% 	\begin{subfigure}[t]{0.3\textwidth}
% 		\centering
% 		\includegraphics[width=\textwidth]{\figpath/2D_HermLL_SOL_8_k}
% 		\label{fig:2D_HermLL_SOL_8_k}
% 	\end{subfigure}
	\caption{2D two-component groud-state-multipole case. Same caption and computational domain as in Fig.  \ref{fig-2D-HermLL_1}.
		The ground
	state in depicted by $\phi_{1}$ and the multipole state by $\phi_{2}$.
	}
	\label{fig-2D-HermLL_8}
\end{figure}

The results for these test cases are depicted in Fig.  \ref{fig-2D-Herm-LL-1-7}
and Figs.  \ref{fig-2D-HermLL_1}-\ref{fig-2D-HermLL_8} showcasing the
BdG spectra and density profiles of the components, respectively.
Specifically, the panels a), b), and c) of Fig.  \ref{fig-2D-Herm-LL-1-7}
are associated with density profiles of Figs.  \ref{fig-2D-HermLL_1},
\ref{fig-2D-HermLL_7}, and  \ref{fig-2D-HermLL_8}, each of the latter
shown for different values of $\mu_{2}$. The bound mode depicted in
Fig.  \ref{fig-2D-HermLL_1} corresponds
to the dark-bright branch that bifurcates from the linear limit at
$\mu_2\approx 1.05133$.~This state is unstable over $\mu_{2}$ (see,
panel a) of Fig.  \ref{fig-2D-Herm-LL-1-7}) except from a very narrow
window of stability close to the linear limit  (see also Fig.~2(a) in
  \cite{charalampidis2020bifurcation}, and references therein). Our toolbox
traced the branch of Fig.  \ref{fig-2D-HermLL_7} that bifurcates at
$\mu_2\approx 1.23276$, and involves a soliton necklace in $\phi_{2}$
(note its imprint on $\phi_{1}$). Similar to the case of
Fig.  \ref{fig-2D-Herm-LL-1-7}, this state is unstable
(see also Fig.~1(a) in  \cite{charalampidis2020bifurcation})
but features a  very narrow window of stability as is shown in
panel b) of Fig.  \ref{fig-2D-Herm-LL-1-7}. Finally, the so-called
multipole branch (see Fig.~16(c) in  \cite{charalampidis2020bifurcation})
was traced by our toolbox, with BdG spectrum and density profiles
shown in the panel c) of Fig.  \ref{fig-2D-Herm-LL-1-7}, and
Fig.  \ref{fig-2D-HermLL_8}, respectively. This state bifurcates
from the linear limit at $\mu_2=1.29325$, and can be described
by the combination of Cartesian (\ie Hermitian) eigenstates
$|2,1\rangle+|0,3\rangle$.~All the cases that we discussed here
match perfectly with the numerical results of  \cite{charalampidis2020bifurcation}.

%\clearpage
%\pagebreak

\subsection{3D two-component case: ground state, planar dark soliton, and vortex-ring state}\label{sec-3D-twoc}

We conclude our discussion on test cases for validating our
toolbox by presenting two 3D two-component configurations.
Note that we employ  the same ``trapping''
technique that was discussed for the 2D, two-component cases,
where the $\phi_{1}$ carries the ground state, but now
in 3D. The latter is continued over $\mu$ from its linear limit,
\ie $\mu\approx 1.501$, and in the single-component case first
[cf.~Eq.  \eqref{eq-scal-GP-stat}] with $\beta = 1.03$ and $\omegap = 1$,
until $\mu=2$. Then, the eigenvalue problem \eqref{eq-bdg-ll}
is solved to obtain the eigenvalue-eigenvector pairs $(\mu_{2},\phi_{2})$.%
 This way, and upon selecting an eigenvalue-eigenvector pair of our
choice, we trace branches of 3D bound modes of Eq.  \eqref{eq-scal-GP2c-stat}
by performing continuation over $\mu_{2}$ (while setting $\beta_{11}=1.03$,
$\beta_{22}=0.97$, $\beta_{12}=1$) for fixed $\omegap=1$ and $\mu_{1}=2$.
We stop the continuation process when the continuation parameter reaches
$\mu_{2}+0.4$, \ie being 0.4 units far away from the respective linear
limit of $\phi_{2}$.

\begin{figure}[!h]
	\centering
	\begin{subfigure}[t]{0.45\textwidth}
		\centering
		\includegraphics[width=\textwidth]{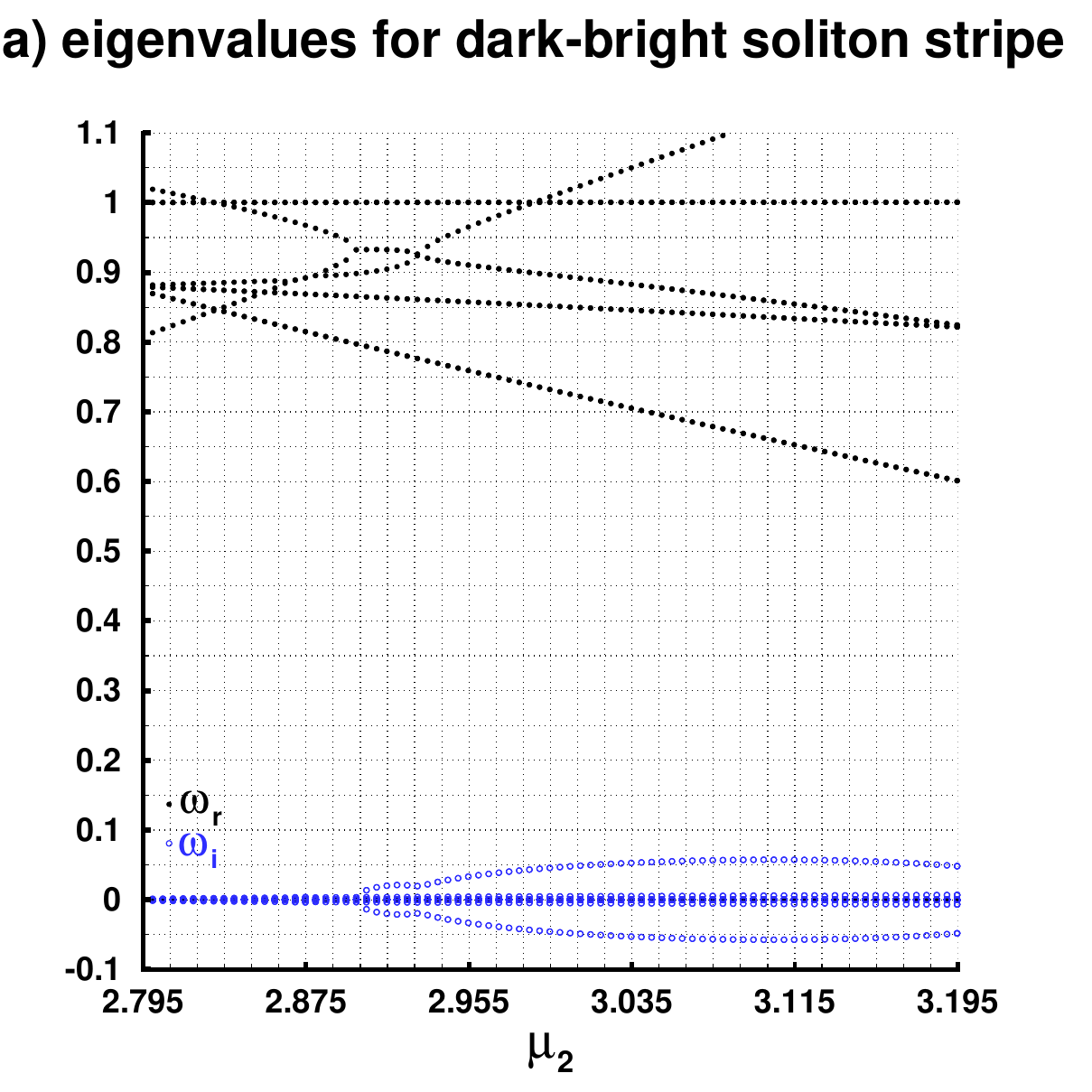}
		\label{fig:3D-Herm-LL-1}
	\end{subfigure}
	\begin{subfigure}[t]{0.45\textwidth}
		\centering
		\includegraphics[width=\textwidth]{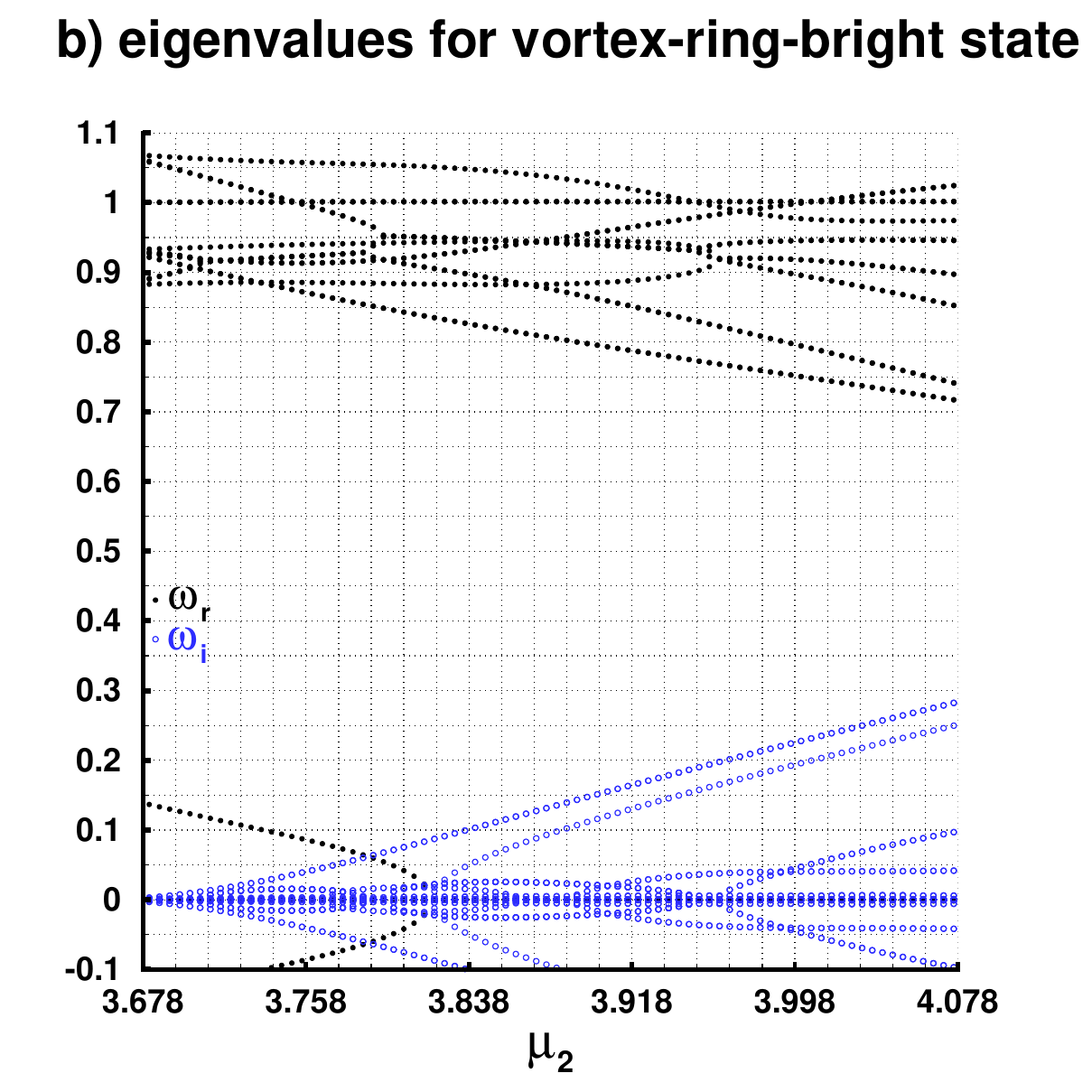}
		\label{fig:3D-Herm-LL-7}
	\end{subfigure}
	\caption{3D two-component cases. The BdG spectra over $\mu_{2}$
	for a) the dark-bright soliton stripe (with the ground state
	in $\phi_{1}$ and the planar dark soliton in $\phi_{2}$),
	and b) vortex-ring-bright state (with the vortex-ring state in
	$\phi_{2}$).~These states bifurcate from $(\mu_{1},\mu_{2})=(2,2.79467)$
	and $(2,3.67602)$, respectively. Their density
	profiles (for distinct values of $\mu_{2}$) are shown in Figs.  \ref{fig-3D-HermLL_1}
	and  \ref{fig-3D-HermLL_7}.}
	\label{fig-3D-Herm-LL-1-7}
\end{figure}

\begin{figure}[!h]
	\centering
	\begin{subfigure}[t]{0.3\textwidth}
		\centering
		\includegraphics[width=\textwidth]{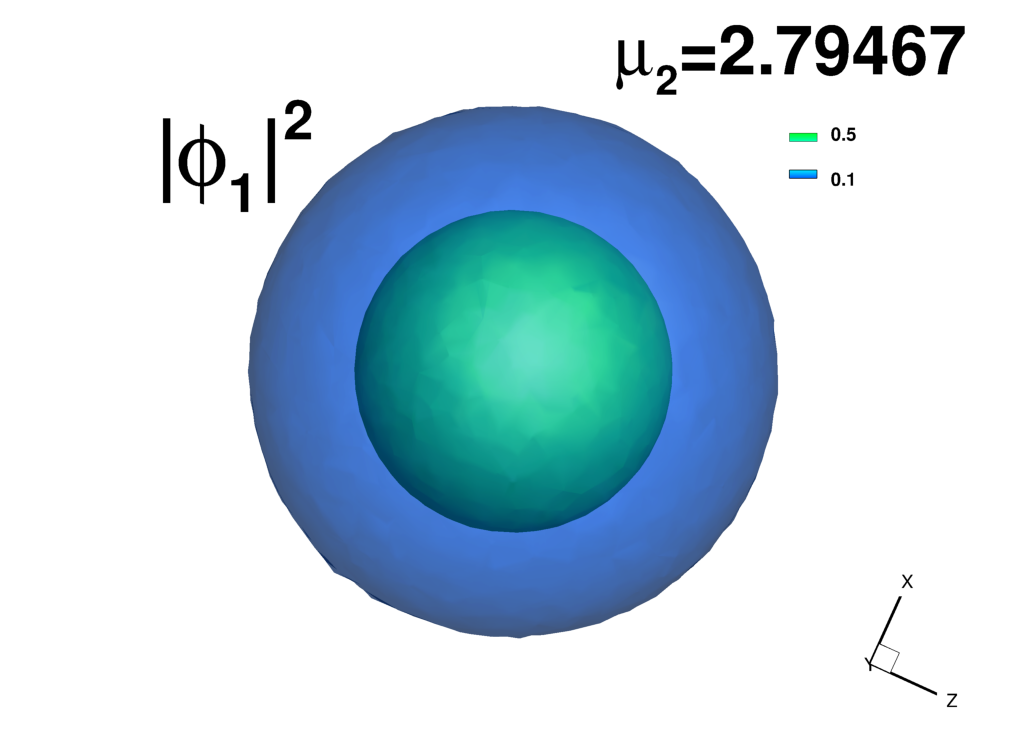}
		\label{fig:3D_HermLL_SOL_1_a}
	\end{subfigure}
	\begin{subfigure}[t]{0.3\textwidth}
		\centering
		\includegraphics[width=\textwidth]{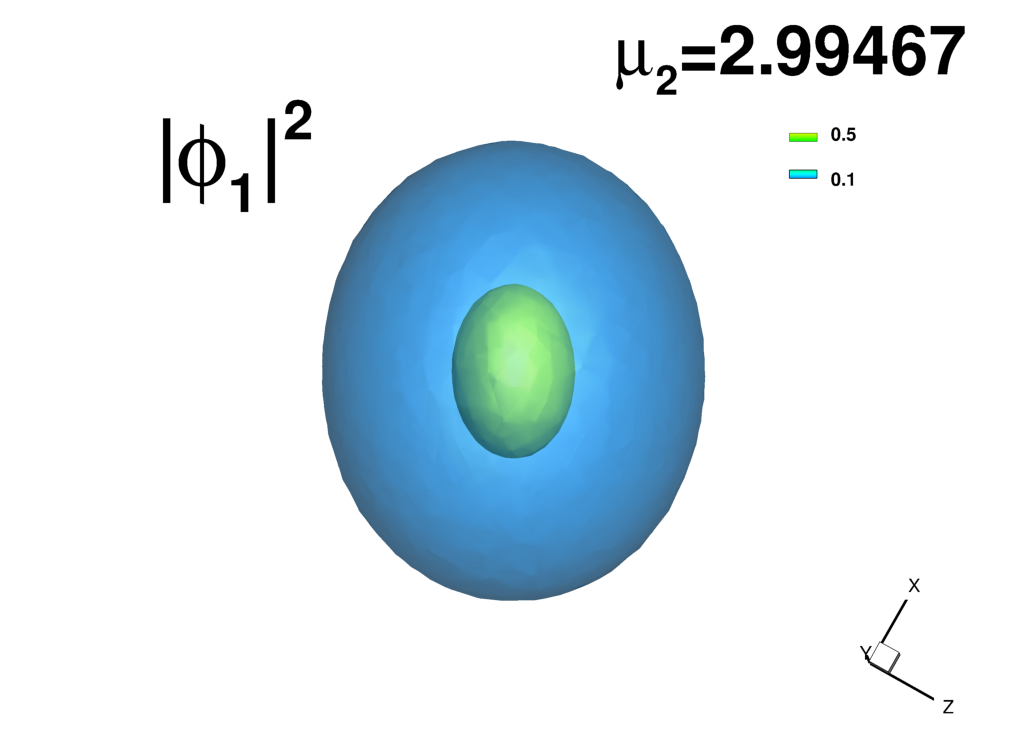}
		\label{fig:3D_HermLL_SOL_1_b}
	\end{subfigure}
	\begin{subfigure}[t]{0.3\textwidth}
		\centering
		\includegraphics[width=\textwidth]{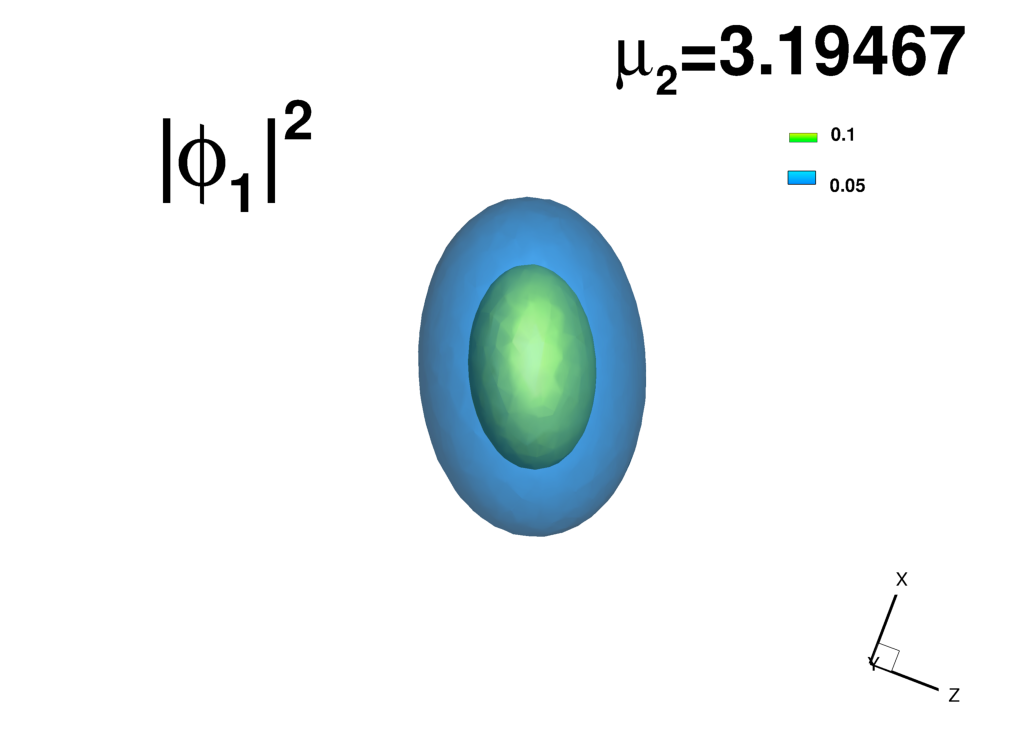}
		\label{fig:3D_HermLL_SOL_1_c}
	\end{subfigure}
	\begin{subfigure}[t]{0.3\textwidth}
		\centering
		\includegraphics[width=\textwidth]{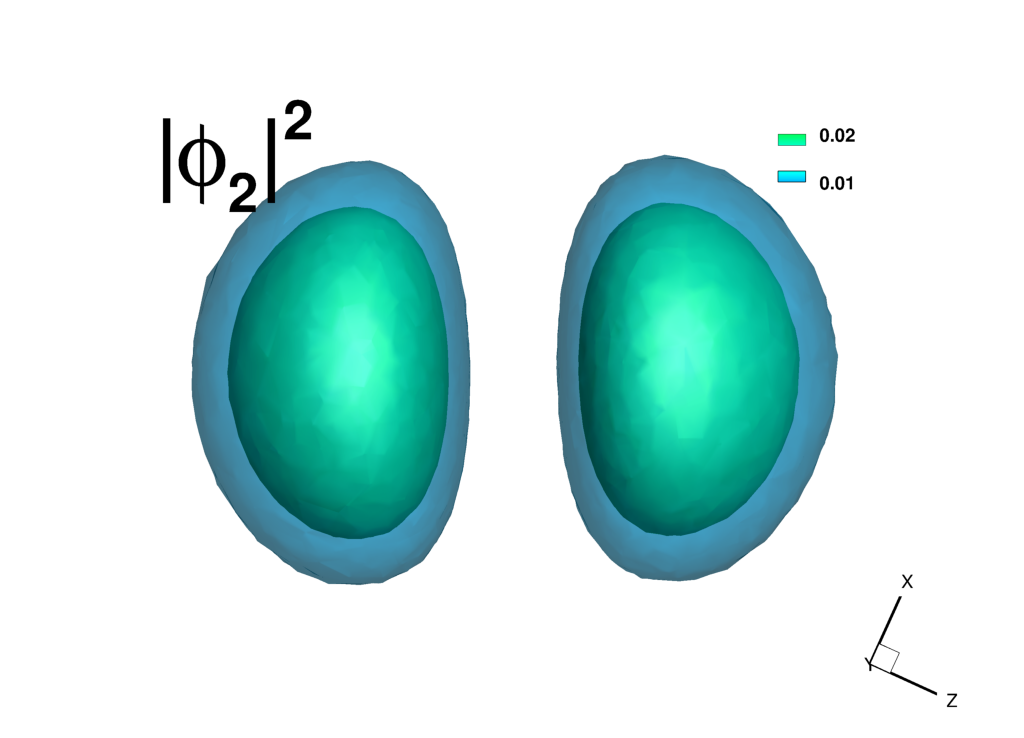}
		\label{fig:3D_HermLL_SOL_1_d}
	\end{subfigure}
	\begin{subfigure}[t]{0.3\textwidth}
		\centering
		\includegraphics[width=\textwidth]{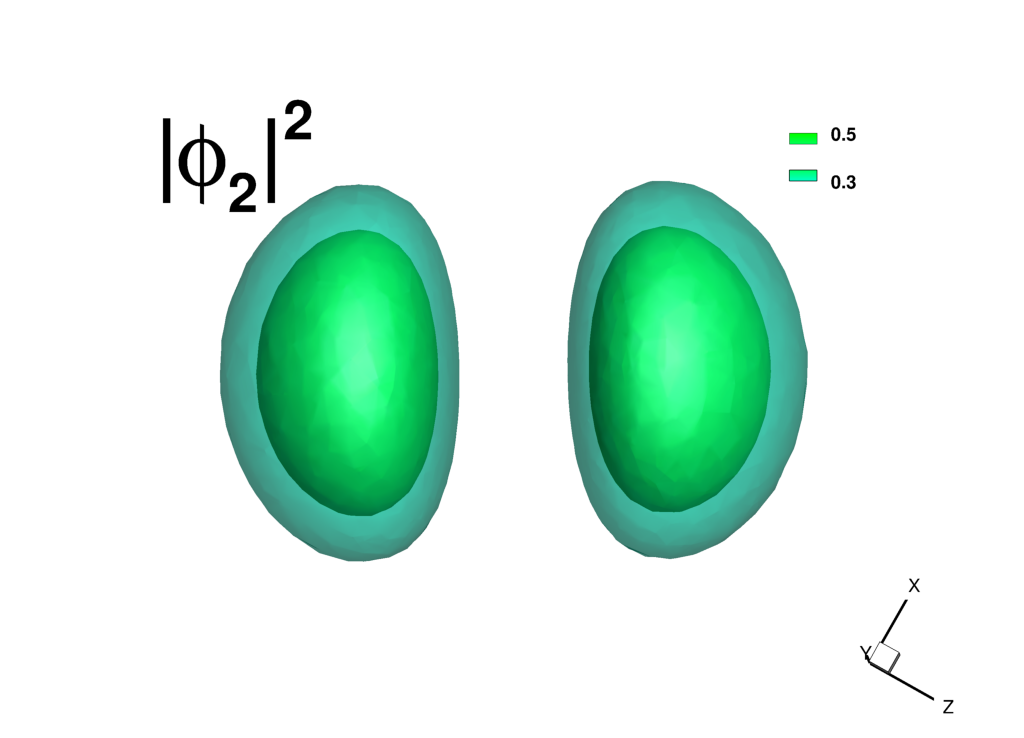}
		\label{fig:3D_HermLL_SOL_1_e}
	\end{subfigure}
	\begin{subfigure}[t]{0.3\textwidth}
		\centering
		\includegraphics[width=\textwidth]{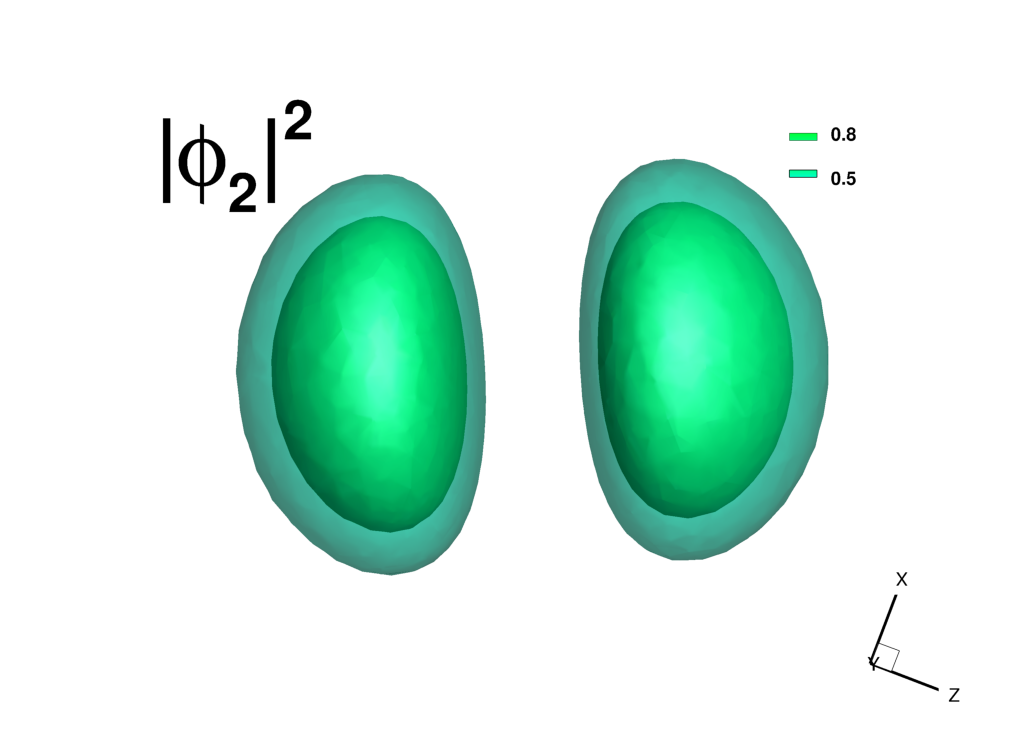}
		\label{fig:3D_HermLL_SOL_1_f}
	\end{subfigure}
	\caption{3D two-component dark-bright soliton stripe state. Density profiles  $|\phi_{1}|^{2}$ and $|\phi_{2}|^{2}$ are shown
	in the top and bottom rows, respectively, and for different values
	of $\mu_{2}$.
	Note that the $\phi_{1}$ component carries the ground state whereas
	the $\phi_{2}$ one the planar dark soliton (with a zero cut in the
	$y$ direction). The computational domain was the cube
	$[-4.4,4.4]^3$, and the above profiles are zoom-ins in $[-2,2]^3$.
	}
	\label{fig-3D-HermLL_1}
\end{figure}

\begin{figure}[!h]
	\centering
	\begin{subfigure}[t]{0.3\textwidth}
		\centering
		\includegraphics[width=\textwidth]{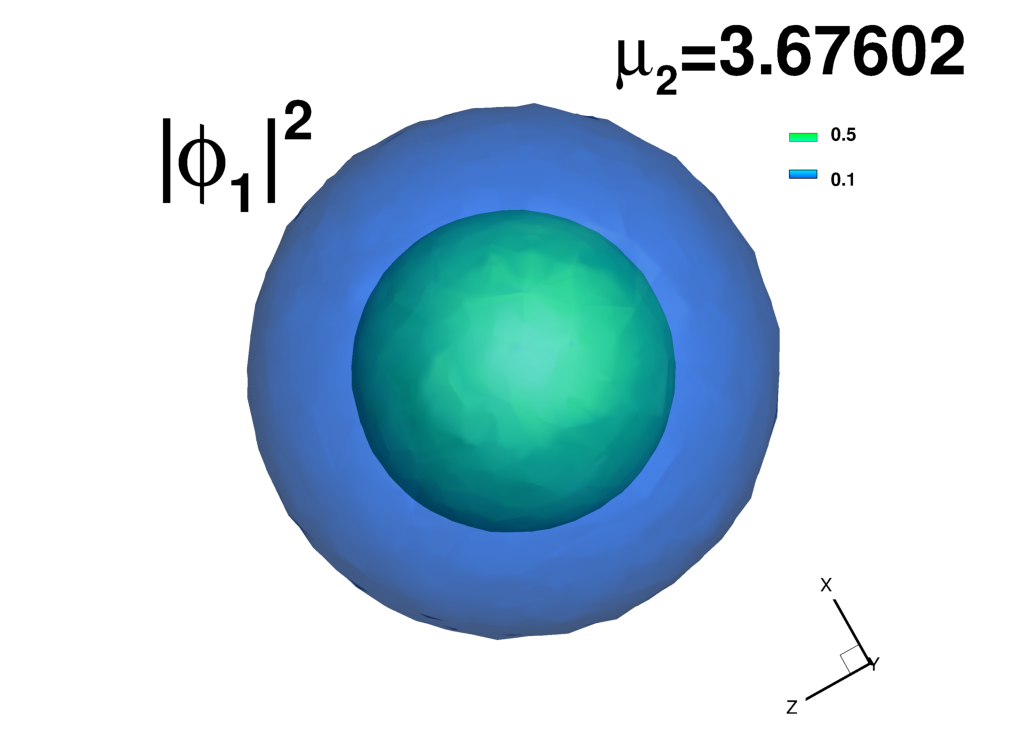}
		\label{fig:3D_HermLL_SOL_7_a}
	\end{subfigure}
	\begin{subfigure}[t]{0.3\textwidth}
		\centering
		\includegraphics[width=\textwidth]{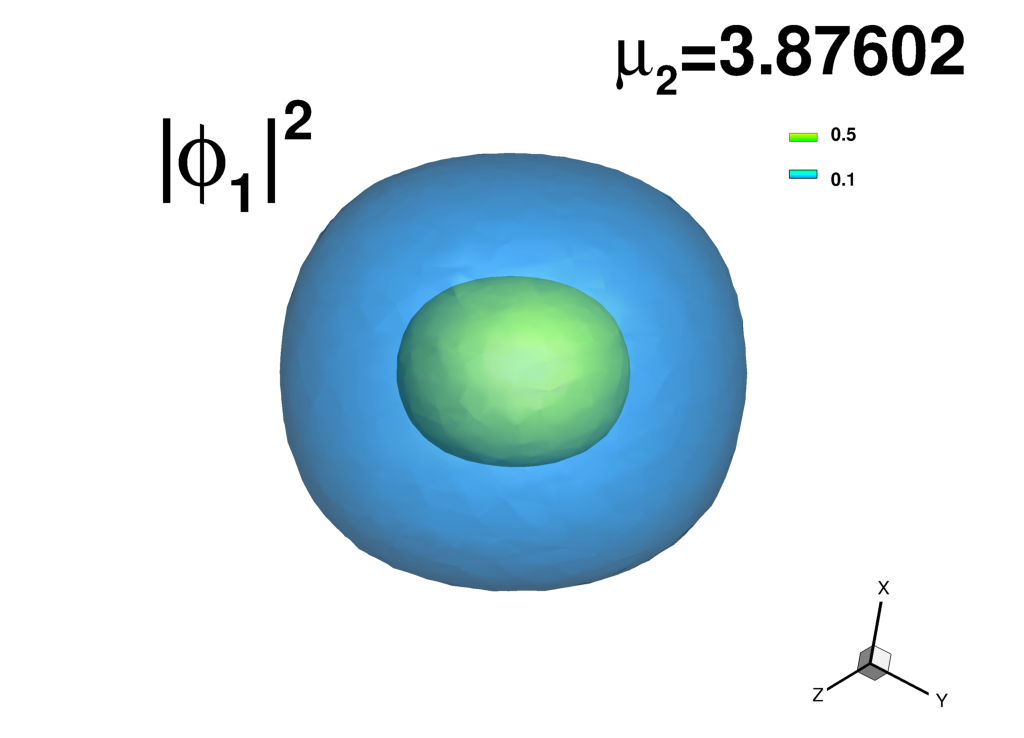}
		\label{fig:3D_HermLL_SOL_7_b}
	\end{subfigure}
	\begin{subfigure}[t]{0.3\textwidth}
		\centering
		\includegraphics[width=\textwidth]{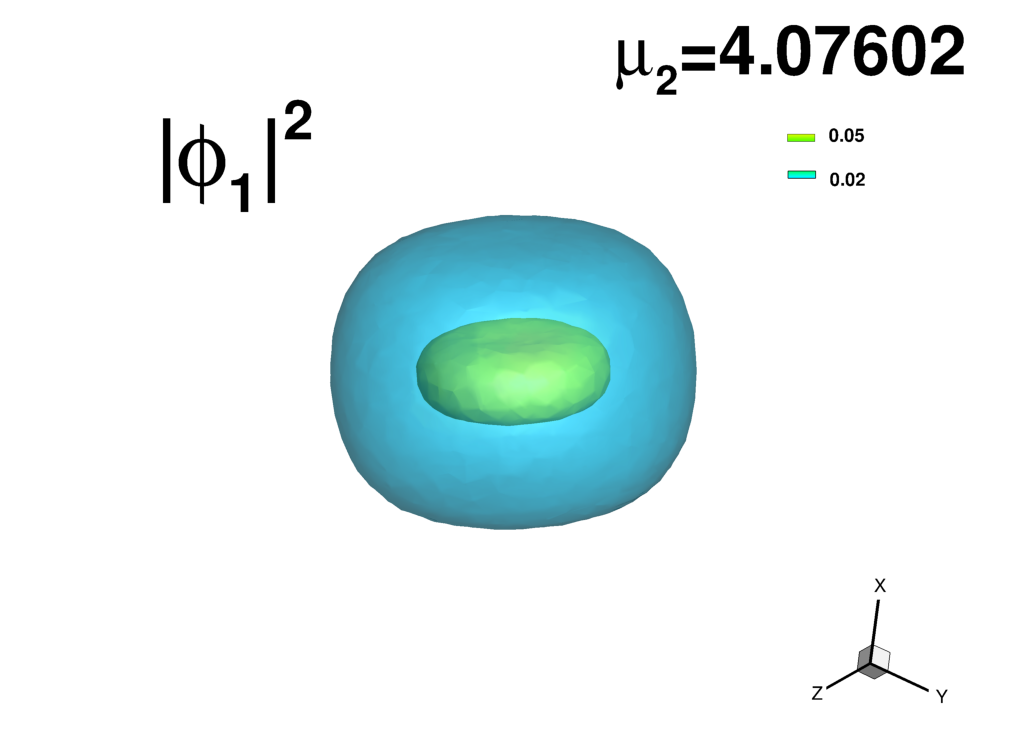}
		\label{fig:3D_HermLL_SOL_7_c}
	\end{subfigure}
	\begin{subfigure}[t]{0.3\textwidth}
		\centering
		\includegraphics[width=\textwidth]{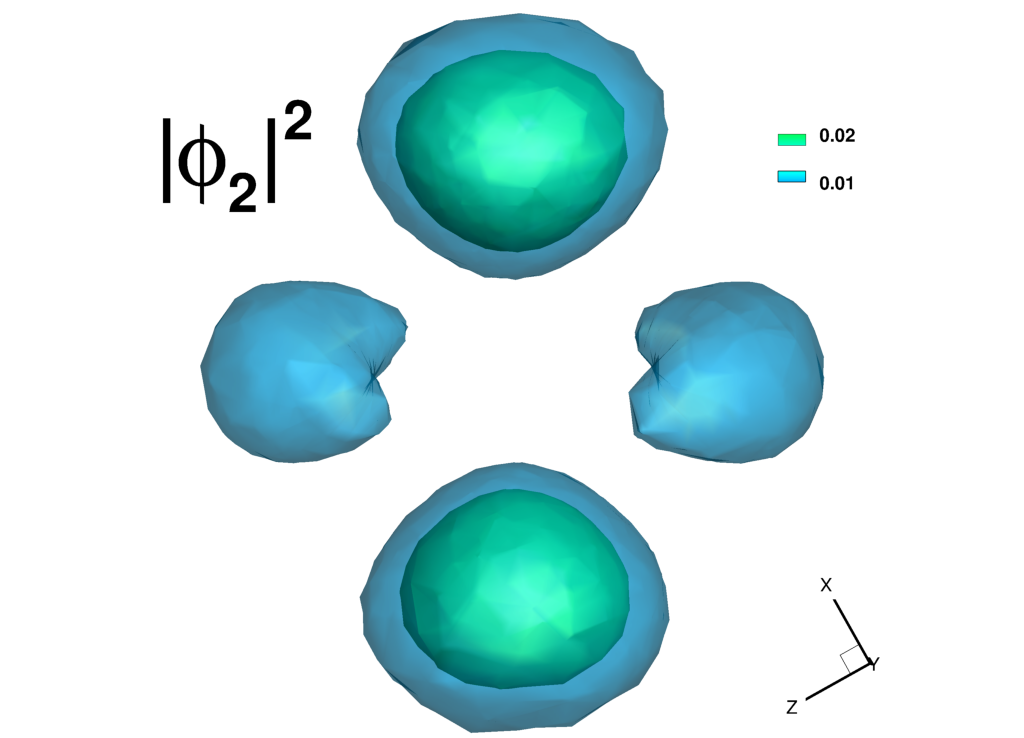}
		\label{fig:3D_HermLL_SOL_7_d}
	\end{subfigure}
	\begin{subfigure}[t]{0.3\textwidth}
		\centering
		\includegraphics[width=\textwidth]{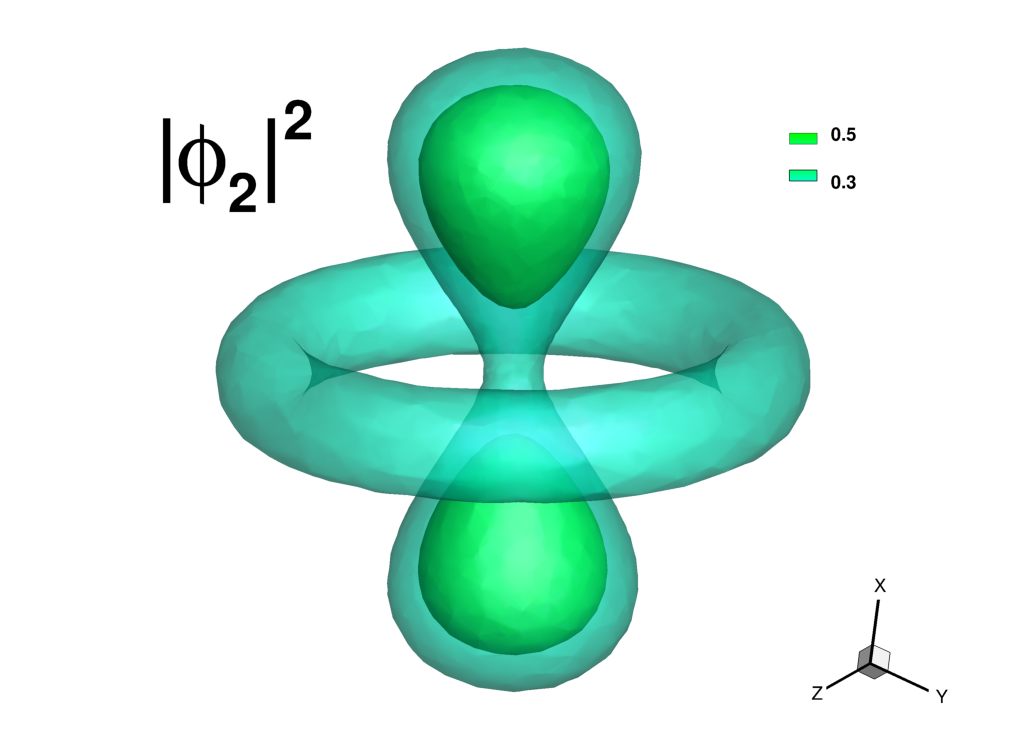}
		\label{fig:3D_HermLL_SOL_7_e}
	\end{subfigure}
	\begin{subfigure}[t]{0.3\textwidth}
		\centering
		\includegraphics[width=\textwidth]{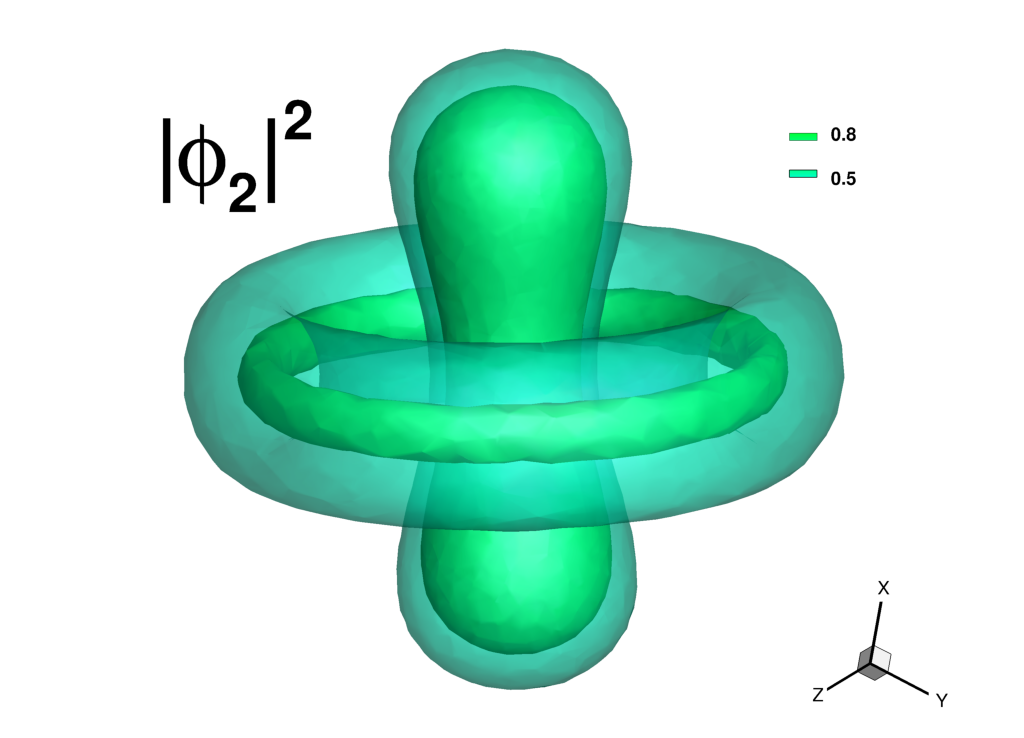}
		\label{fig:3D_HermLL_SOL_7_f}
	\end{subfigure}
	\caption{3D two-component vortex-ring-bright state. Same caption and computational domain as in Fig.  \ref{fig-3D-HermLL_1}.
		Note that the $\phi_{1}$
	component carries (again) the ground state whereas the $\phi_{2}$
	one the vortex-ring state.}
	\label{fig-3D-HermLL_7}
\end{figure}

The first case we considered involves the ground state in $\phi_{1}$,
and the planar dark soliton  \cite{pgk_siam_book} in $\phi_{2}$, \ie
a dark-bright soliton stripe in 3D. The latter bifurcates from its linear
limit at $\mu_2\approx 2.79467$, and can be classified in terms of Cartesian
eigenfunctions as $|0,1,0\rangle$. Our numerical results on its BdG spectrum
depicted in panel a) of Fig.  \ref{fig-3D-Herm-LL-1-7} show that the state
is stable from its inception until $\mu_{2}\approx 2.902$ when
it becomes unstable.
Snapshots of densities
$|\phi_{1}|^{2}$ and $|\phi_{2}|^{2}$
are shown in the top and bottom rows of Fig.  \ref{fig-3D-HermLL_1},
respectively, for different values of $\mu_{2}$.
Finally, the panel b) of Fig.  \ref{fig-3D-Herm-LL-1-7}
and Fig.  \ref{fig-3D-HermLL_7} present the BdG spectrum and density
profiles corresponding to the bound mode involving the ground state
in $\phi_{1}$, and the vortex-ring state in $\phi_{2}$, \ie a vortex-ring-bright
state. This configuration (obtained for $\mu_{1}=2$), bifurcates from the
linear limit of $\phi_{2}$ at $\mu_2=3.67602$, and is generically unstable
as shown in the panel b) of Fig.  \ref{fig-3D-Herm-LL-1-7}.
Note
that in the one-component setting, the vortex-ring state can be classified
in terms of a combination of Cartesian eigenstates as
$\frac{1}{\sqrt{2}}\left(|2,0,0\rangle>+|0,2,0\rangle\right)+i|0,0,1\rangle$%
  \cite{boulle2020deflation}.  Density profiles $|\phi_{1}|^{2}$
and $|\phi_{2}|^{2}$ of this bound mode are shown for different values of
$\mu_{2}$ in the top and bottom rows,
of Fig.  \ref{fig-3D-HermLL_7}.


\clearpage

%%%%%%%%%%%%%%%%%%%%%%%%%%%%%%%%%%%%%%%%%%%%%%%%%%%
\section{Description of the programs }\label{sec-desc-prog}

In this section, we first describe the architecture of the
programs and the organization of the provided files. We then
present the input parameters and the structure of the output files.

\subsection{Program architecture}

Codes and data files forming the BdG problem with the domain
decomposition method (DDM) are stored in the
\texttt{FFEM$\_$BdG$\_$ddm$\_$toolbox} directory. The latter
is organized in two main subdirectories:
\texttt{BdG\_1comp\_ddm} and \texttt{BdG\_2comp\_ddm}, corresponding
to the one- and two-component codes.~Each subdirectory contains two
main files: {\em FFEM\_GP\_\$case\_ddm.edp}, which is the main \ff script
file for the computation of the stationary state, and {\em FFEM\_BdG\_\$case\_ddm.edp}
which is the main \ff\ script file for the computation of the BdG
eigenvalues (\$case=1c\_2D\_3D for the one-component case and
\$case=2c\_2D\_3D for the two-component case).

To run these codes, first of all, the user must install \ff
with \texttt{PETSc} following the instructions in
\texttt{https://doc.freefem.org/introduction/installation.html}.
Then, the user can run the \ff code for the computation of the
GP stationary state by using either the command\\
\texttt{mpirun -np 4 FreeFem++-mpi FFEM$\_$GP$\_$\$case\_ddm.edp}\\
or \\
\texttt{ff-mpirun -np 4 FFEM$\_$GP$\_$\$case\_ddm.edp}. \\The BdG
eigenvalues can then be computed by typing (in terminal) either\\
\texttt{mpirun -np 4 FreeFem++-mpi FFEM$\_$BdG$\_$\$case\_ddm.edp} \\or\\
\texttt{ff-mpirun -np 4 FFEM$\_$BdG$\_$\$case\_ddm.edp}.\\
Parameter files for the examples presented in this paper are stored
in the \texttt{INIT} folder.

The obtained solutions are saved in the \texttt{dircase} directory.
Depending on the output format selected by the user, data files are
generated in specific folders for visualization with
Tecplot\footnote{\texttt{https://tecplot.com}},
Paraview\footnote{\texttt{https://www.paraview.org}}, and
Gnuplot\footnote{\texttt{http://www.gnuplot.info}}.~We also provide
ready-made layouts for visualization with Tecplot in the folder \texttt{Figures}.
The user can thus obtain the figures from this paper using newly
generated data. More details about the output structure are given in
Sect.  \ref{sec-outputs-bdg}.

The complete architecture of the \texttt{BdG\_1comp\_ddm} directory is the following (the architecture of the \texttt{BdG\_2comp\_ddm} directory is almost identical):
\begin{enumerate}
	\item {\em FFEM\_GP\_\$case\_ddm.edp}: the main script for computing GP stationary states.
	\item {\em FFEM\_BdG\_\$case\_ddm.edp}: the main script for computing the BdG spectrum.
	\item {\em FFEM\_LL\_\$case\_ddm.edp}: the main script for solving the eigenvalue
	problem of Eq.  \eqref{eq-bdg-ll} (\ie for finding the eigenvalue-eigenvector pairs
	$(\mu_{2},\phi_{2})$).
	\item {\em param\_num\_common.inc}: a parameter file containing main numerical parameters.
	\item \texttt{INIT}: directory storing the parameter files for the examples presented in
	Sects.  \ref{sec-valid1c} and  \ref{sec-valid2c}.
	\item \texttt{Figures}:  directory containing Tecplot layouts used to replot the figures
	shown in Sects.  \ref{sec-valid1c} and  \ref{sec-valid2c}.~The main code must be run with
	the associated example before opening the layout to replot the figure.
	\item \texttt{A\_macro}: directory containing macros used in the main scripts for GP and BdG
	problems.
	\item \texttt{A\_macro\_LL}: directory containing macros used in the main scripts for
	the study of respective linear limits (LL).
\end{enumerate}

\subsection{Macros and functions}

The different macros and functions used in the toolbox for the sequential code are stored in the \texttt{A\_macro} folders:
\begin{itemize}
	\item {\em Macro\_BdGsolve.edp}: macro for computing the BdG eigenvalues associated with matrices of
	Eqs.  \eqref{eq-num-BdG-weak} and  \eqref{eq-num-BdG2c-weak}.
	\item {\em Macro\_createdir.edp}: macro for creating the file structure of the \texttt{dircase} folder.
	\item {\em Macro\_globalpartition.edp}: macro for creating a partition of the global mesh, and sending
	the solution from the global mesh to the local one.
	\item {\em Macro\_GPsolve.edp}: macro for computing the GP stationary state with Newton's method
	(see Eqs.  \eqref{eq-num-GP-stat-Newton} and  \eqref{eq-num-GP2c-stat-Newton-1}-\eqref{eq-num-GP2c-stat-Newton-4}).
	\item {\em Macro\_LLsolve.edp}: macro for computing the eigenvalues of Eq.  \eqref{eq-bdg-ll}.
	\item {\em Macro\_meshAdapt.edp}: macro for adapting the mesh to the wave function.
	\item {\em Macro\_onedomainsol.edp}: macro for sending the solution from the local domain to the global one.
	\item {\em Macro\_operator.edp}: collection of useful macros and functions:
	gradients, energy \eqref{eq-NRJ}, chemical potential, %\eqref{eq-GP-mu},
Hermite polynomials, etc. Also contains a macro creating a spherical mesh for 3D problems.
	\item {\em Macro\_output.edp}: macros used for saving data in Tecplot and Paraview formats.
	\item {\em Macro\_plotEigenvector.edp}: macro for plotting the real and imaginary parts of a BdG eigenvector.
	\item {\em Macro\_plotphi.edp}: macro for plotting the complex wave function.
	The user can press "k"  to alternate between plots of the density, phase and
	real and imaginary parts of the wave function.
	\item {\em Macro\_problem.edp}: definitions of the weak formulations for the GP
	[cf.~Eqs.  \eqref{eq-num-GP-stat-Newton} or  \eqref{eq-num-BdG2c-weak}],
	the BdG problems [cf.~Eqs.  \eqref{eq-num-BdG-weak} or  \eqref{eq-num-GP2c-stat-Newton-1}-\eqref{eq-num-GP2c-stat-Newton-4}]
	and the linear limit problem [cf.~Eq.  \eqref{eq-bdg-ll}].
	\item {\em Macro\_readmu.edp}: macro to read the $\mu$ from \textbf{dircase/Gnuplot/GP\_results.dat},
	and compute the corresponding BdG eigenvalues.
	\item {\em Macro\_readmubeta.edp}: macro to read the values of $\mu$ or $\beta$ from \textbf{GP\_mucont\_results.dat} or \textbf{GP\_betacont\_results.dat} that are contained in \textbf{dircase/Gnuplot/} in order to compute the
	corresponding BdG spectrum.
	\item {\em Macro\_restart.edp}: macros used to save and load the wave function to or from \ff files.
	\item {\em Macro\_saveData.edp}: macro for saving the stationary wave function.
	\item {\em Macro\_saveEigenvalues.edp}: macro for saving the BdG eigenvalues and eigenvectors.
\end{itemize}

\subsection{Input parameters}

Parameters are separated in two files.~Numerical parameters used in
all computations are specified in {\em param\_num\_common.inc}.~Files
in the \texttt{INIT} directory specify physical parameters associated
with the state of interest, computation and numerical parameters specific
to this problem.~The files distributed with the toolbox provide a variety
of examples that can be used as a starting point when selecting parameters
for the study of new states. \\
\textbf{(1)} In the file \texttt{param\_num\_common.inc}, the parameters are:
\begin{itemize}
	\item \textbf{displayplot}: controls the output information to plot.~Possible
	values range from $0$ (no plots), to {$2$} (plots data at all iterations of
	Newton's method, and all eigenvectors computed by the BdG code).
	\item \textbf{iwait}: Boolean indicating if the code must wait for user's
	input when a plot is shown (\texttt{true}) or it can continue
	(\texttt{false}) with the next plot.
	\item \textbf{cutXY}, \textbf{cutXZ}, \textbf{cutYZ}: (only for 3D cases
	in the one-component case) Booleans indicating whether to plot cuts of
	the wave function along the different axis at $x=0$, $y=0$ or $z=0$.
	\item \textbf{Tecplot}: Boolean indicating whether to save data in the Tecplot format.
	\item \textbf{Tecplotddm}: for saving solution for Tecplot with DDM or not.
	\item \textbf{Paraview}: Boolean indicating whether to save data in the Paraview format (only in 2D and 3D). 
	\item \textbf{adaptinit}: if \texttt{true}, the initial solution is recomputed after the first mesh adaptation.
	\item \textbf{adaptmeshFF}: determines if mesh adaptation is used (\texttt{true}) or not (\texttt{false}).
	\item \textbf{useShift}: Boolean indicating whether to use a shift when computing the BdG eigenvalues (see, Sec.  \ref{sec-num-meth-bdg}). %(see Eq. \eqref{eq-bdg-shift}).
	\item \textbf{Nadapt}: if mesh adaptation is used, then the mesh is adaptated every \textbf{Nadapt} iterations during the continuation.
	\item \textbf{Nplot}: the wave function is plotted every \textbf{Nplot} iterations during the continuation.
	\item \textbf{Nsave}: the wave function is saved for Paraview or Tecplot every \textbf{Nsave} iterations during the continuation.
	\item \textbf{Nrst}: the wave function is saved for the BdG computation every \textbf{Nrst} iterations during the continuation.
	\item \textbf{tolerrF}: the tolerance value $\epsilon_{\scriptscriptstyle F}$ in Eq. \eqref{eq-bdg-err1}.
	\item \textbf{tolNewton}: the tolerance value $\epsilon_{\scriptscriptstyle q}$ in Eq. \eqref{eq-bdg-err1}.
	\item \textbf{shift}: the value of the shift $\sigma$ used when computing eigenvalues.
	\item \textbf{shiftLL}: the value of the shift $\sigma$ used when computing eigenvalues close to the linear limit.
	\item \textbf{shiftFLL}: the value of the shift $\sigma$ used when computing eigenvalues far from the linear limit.
	\item \textbf{adaptboundary}: to adapt (==0) or not (==1) the boundary of the mesh in 3D.
	\item \textbf{skipBdG}: the value to skip $\mu$ or $\beta_{12}$ computed with GP for BdG computation.
	\item \textbf{muL, mubetaL}: to switch between using \textbf{shiftFLL} or \textbf{shiftLL}, if $\mu$ or $\mu_1$ or $\mu_2$ or $\beta_{12}<$\textbf{mubetaL} we use \textbf{shiftLL} otherwise we use \textbf{shiftFLL}.
	\item \textbf{LL}: Boolean indicating whether we want to compute the linear limit, \ie eigenvalue problem for $(\mu_{2},\phi_{2})$ for the second component or no.
	\item \textbf{NNZ}: contain the non zero elements (nnz) for the BdG matrix.
	\item \textbf{dmuk}: counter for dmu adaptation.
	\item \textbf{FINAL}: Boolean  to run the final solution of \textbf{endmu} in the GP continuation or to stop the BdG computation.
	\item \textbf{newtonMax}: the maximum number of Newton iterations.
\end{itemize}
\vspace{1cm}
\textbf{(2)} In the file \texttt{\$case.inc}, stored in the \texttt{INIT} directory, the parameters are:
\begin{itemize}
	\item General parameters for the case:\\
	$\bullet$ \textbf{dimension}: the dimension of the problem (2 or 3).\\
	$\bullet$ \textbf{FEchoice}: the type of finite element used. Usually $P2$.\\
	$\bullet$ \textbf{nev}: the number of eigenvalues computed by the BdG code.
	\item Parameters used to restart a computation:\\
	$\bullet$ \textbf{restart}: Boolean indicating if the initial solution is a restart from a previous computation. 
	If \texttt{true}, the solution and mesh stored in \textbf{dirrestart} for the value of $\mu$ given by \textbf{murestart} will be used as  initial solution.\\
	$\bullet$ \textbf{murestart}: the initial value of $\mu$ in the case of a restart.\\
	$\bullet$ \textbf{dirrestart}: the folder where the initial solution is stored in the case of a restart.
	\item Parameters of the continuation:\\
	$\bullet$ \textbf{kpol}, \textbf{lpol}, \textbf{mpol}: integers defining the initial state in the linear limit.\\
	$\bullet$ \textbf{startmu}: the initial value of $\mu$.\\
	$\bullet$ \textbf{endmu}: the final value of $\mu$.\\
	$\bullet$ \textbf{dmu}: the increment in $\mu$ during the continuation.\\
	$\bullet$ \textbf{facmu}: when using the linear limit, the initial value of $\mu$ is given by $\textbf{facmu}\cdot \mu_{\ket{klm}}$.
	$\bullet$ \textbf{mubeta}: a macro that contains the name of the variable that we want to do the continuation over it: $\mu_1, \mu_2, \beta_{12}$ or $\beta_{21}$.
	\item Coefficients of the GP equation:\\
	$\bullet$ \textbf{beta}: the nonlinear coefficient (we set  $\beta=1$ in all test cases except for the linear limit cases where $\beta=1.03$).\\
	$\bullet$ \textbf{ax}, \textbf{ay}, \textbf{az}: the frequencies of the trapping potential along the three coordinate axes.\\
	$\bullet$ \textbf{Ctrap}: a function defining the trapping potential.   
	\item Parameters for the mesh generation:\\
	$\bullet$ \textbf{Dx}: the distance between points on the mesh border.\\
	$\bullet$ \textbf{scaledom}: a coefficient used to control the size of the domain: the mesh radius is given by $\textbf{Rdom} = \textbf{scaledom}\cdot r_{\TF}$, where $r_\TF$ is the Thomas-Fermi radius.\\
	$\bullet$ \textbf{createMesh}: a macro creating the initial mesh \texttt{Th}.
	\item Parameters for the mesh adaptation:\\
	$\bullet$ \textbf{errU}: the interpolation error level.\\
	$\bullet$ \textbf{hmin}: the minimum length of a mesh element edge in the new mesh.\\
	$\bullet$ \textbf{hmax}: the maximum length of a mesh element edge in the new mesh.\\
	$\bullet$ \textbf{adaptratio}: the ratio for a prescribed smoothing of the metric. No smoothing is done if the value is less than $1.1$.
	\item Parameters for the initial solution:\\
	$\bullet$ \textbf{initname}: the name given to the initial solution.\\
	$\bullet$ \textbf{initcond}: a macro defining the initial solution for the \texttt{phi} variable.
	\item Definitions of the boundary conditions:\\
	$\bullet$ \textbf{BCGP}: the boundary conditions used in the GP code for Eqs. \eqref{eq-num-GP-stat-Newton} and \eqref{eq-num-GP2c-stat-Newton-1}-\eqref{eq-num-GP2c-stat-Newton-4}.\\
	$\bullet$ \textbf{BCBdG}: the boundary conditions used in the BdG code for Eqs. \eqref{eq-num-BdG-weak} and \eqref{eq-num-BdG2c-weak}.\\
	$\bullet$ \textbf{BCLL}: the boundary conditions used in the LL code for Eqs. \eqref{eq-bdg-ll}.\\
	$\bullet$ \textbf{fcase}: the name given to the current computation.\\
	$\bullet$ \textbf{dircase}: the directory where the results are stored.
\end{itemize}
%\vspace{1cm}
\textbf{(3)} In a two component case, some new parameters are defined in the \texttt{\$case.inc} file:
\begin{itemize}
	\item Parameters used to restart a computation:\\
	$\bullet$ \textbf{mu1restart}, \textbf{mu2restart}: initial values of $\mu_1$ and $\mu_2$ in the case of a restart.\\
	$\bullet$ \textbf{beta12restart}, \textbf{beta21restart} initial values of $\beta_{12}$ and $\beta_{21}$ in the case of a restart.
	\item Parameters of the continuation:\\
	$\bullet$ \textbf{startmu1}, \textbf{startmu2}: initial values of $\mu_1$ and $\mu_2$.\\
	$\bullet$ \textbf{endmu1}, \textbf{endmu2}: final values of $\mu_1$ and $\mu_2$.\\
	$\bullet$ \textbf{dmu1}, \textbf{dmu2}: increments of $\mu_1$ and $\mu_2$ during the continuation.\\
	$\bullet$ \textbf{startbeta12}, \textbf{startbeta21}: initial values of $\beta_{12}$ and $\beta_{21}$.\\
	$\bullet$ \textbf{endbeta12}, \textbf{endbeta21}: final values of $\beta_{12}$ and $\beta_{21}$.\\
	$\bullet$ \textbf{dbeta12}, \textbf{dbeta21}: increments of $\beta_{12}$ and $\beta_{21}$ during the continuation.
	\item Coefficients of the GP equation:\\
	$\bullet$ \textbf{beta11}, \textbf{beta12}: nonlinear coefficients $\beta_{11}$ and $\beta_{22}$.
	\item Parameters for the initial solution:\\
	$\bullet$ \textbf{initname1}: the name given to the initial solution for the first component.\\
	$\bullet$ \textbf{initname2}: the name given to the initial solution for the second component.\\
	$\bullet$ \textbf{initcond}: a macro defining the initial solution for \texttt{[phi1,phi2]} variables.\\
\end{itemize}

\subsection{Outputs}\label{sec-outputs-bdg}

When a computation starts, the \texttt{OUTPUT$\_$\$case} directory is created.%
~It contains up to eight folders.~The \texttt{RUNPARAM\_GP}, \texttt{RUNPARAM\_BdG},
and \texttt{RUNPARAM\_LL} directories contain a copy of the code and data files, thus
allowing an easy identification of each case, and preparing an eventual rerun of the
same case at a later time.~The other folders contain different output format files of
the computed solution for its visualization using Tecplot, Paraview or Gnuplot.%
~The content of these subfolders depends on the case and on the computation parameters
(differences in the two component code are given in parenthesis):
\begin{enumerate}
	\item The \texttt{Gnuplot} folder contains two files:\\
	$\bullet$ Information about the stationary states are stored in the \texttt{GP\_results.dat} file (\texttt{GP\_mucont\_results.dat} or \texttt{GP\_betacont\_results.dat} file). The columns appear in the following order: the non-linear coefficient $\beta$ ($\beta_{12}$ and $\beta_{21}$),
	the imposed chemical potential $\mu$ ($\mu_1$ and $\mu_2$), the number of Newton iterations used for this value of $\mu$, the norms
	associated with $\epsilon_{\scriptscriptstyle F}$ and $\epsilon_{\scriptscriptstyle q}$ in Eq.  \eqref{eq-bdg-err1}, the computed value
	of the chemical potential
	%\eqref{eq-GP-mu}
	(computed values of $\mu_1$ and $\mu_2$),
	the number of atoms \eqref{eq-GP-N} (the number of atoms in the two components),
	the GP energy,
	%\eqref{eq-NRJ},
	the mesh size, the number of degrees of freedom, the CPU time  to compute the stationary state,
	and the value of the current $\delta\mu$ ($\delta\mu_1$, $\delta\mu_2$).\\
	$\bullet$ BdG eigenvalues are stored in the \texttt{BdG\_results.dat} file.%
	~The columns appear in the following order: the non-linear coefficient $\beta$ ($\beta_{12}$ and $\beta_{21}$), the imposed chemical potential $\mu$ ($\mu_1$ and $\mu_2$), the eigenvalue number between 0 and \textbf{nev}, the real and
	imaginary part of the eigenvalues, the Krein signature and its sign (the Krein signature and its sign for the two components).\\
	$\bullet$ BdG's numerical information is stored in the \texttt{BdG\_num\_results.dat} file. The columns appear in the following order: the non-linear coefficient $\beta$ ($\beta_{12}$ and $\beta_{21}$), the imposed chemical potential $\mu$ ($\mu_1$ and $\mu_2$),
	the non zero element for the BdG matrix, the number of degrees of freedom, the CPU time to compute the eigenvalues, and the cumulative CPU time.
	\item The \texttt{Paraview} folder contains the wave functions stored as {\em .vtk} or {\em .vtu} and {\em .pvd} files:\\
	$\bullet$ \texttt{phi\_init.vtu} and \texttt{phi\_final.vtu} are the initial and final solutions.\\
	$\bullet$ \texttt{phi\_mu\_\$mu.vtu} contains the stationary wave function for a given value of $\mu$.\\
	$\bullet$ \texttt{phi\_mu1\_\$mu1\_mu2\_\$mu2.vtu} contains the stationary wave function for given values of $\mu_1$ and $\mu_2$ in the first continuation.\\
	$\bullet$ \texttt{phi\_beta12\_\$beta12\_beta21\_\$beta21.vtu} \enlargethispage{\baselineskip}contains the stationary wave function for given values of $\beta_{12}$ and $\beta_{21}$ in the second continuation.
	\item The \texttt{Paraview\_Eigenvectors} folder contains the eigenvectors stored as:\\
	$\bullet$ \texttt{eVec\_mu\_\$mu\_\$nev.vtu} in the one-component code.\\
	$\bullet$ \texttt{eVec\_beta12\_\$beta12\_beta21\_\$beta21\_mu1\_\$mu1\_mu2\_\$nev.vtu} in the two-component code.
	\item The \texttt{RST} folder contains the stationary states stored as \ff files. The names are:\\
	$\bullet$ \texttt{RST-\$mu.rst} or \texttt{RST-\$mu1-\$mu2-\$beta12-\$beta21.rst} for the data.\\
	$\bullet$ \texttt{RSTTh-\$mu} or \texttt{RSTTh-\$mu1-\$mu2-\$beta12-\$beta21} for the mesh files. The file extensions are {\em .msh} (in 2D) or {\em .meshb} (in 3D).
	
	\item The \texttt{RST\_LL} folder contains the stationary states stored as \ff files. The names are:\\
	$\bullet$ \texttt{LL\_mu1-\$mu1\_ip-\$mu2.rst} for the data.\\
	$\bullet$ \texttt{LLTh\_mu1-\$mu1\_ip-\$mu2} for the mesh files. The file extensions are {\em .msh} (in 2D) or {\em .meshb} (in 3D).
	
	\item The \texttt{Tecplot} folder contains the wave functions stored as {\em .dat} Tecplot files:\\
	$\bullet$ \texttt{phi\_init.dat} and \texttt{phi\_final.dat} are the initial and final solutions.\\
	$\bullet$ \texttt{phi\_mu\_\$mu.dat} contains the stationary wave function for a given value of $\mu$.\\
	$\bullet$ \texttt{phi\_mu1\_\$mu1\_mu2\_\$mu2.dat} contains the stationary wave function for given values of $\mu_1$ and $\mu_2$ in the first continuation.\\
	$\bullet$ \texttt{phi\_beta12\_\$beta12\_beta21\_\$beta21.dat} contains the stationary wave function for given values of $\beta_{12}$ and $\beta_{21}$ in the second continuation.
	\item The \texttt{Tecplot\_Eigenvectors} folder contains the eigenvectors stored in the Tecplot format:\\
	$\bullet$ \texttt{eVec\_mu\_\$mu\_\$nev.dat} in the one-component code.\\
	$\bullet$ \texttt{eVec\_beta12\_\$beta12\_beta21\_\$beta21\_mu1\_\$mu1\_mu2\_\$nev.dat} in the two-component code.
	\item The \texttt{Tecplot\_Eigenvalues} folder contains the file \texttt{BdG\_results\_eig.dat} that contain all the eigenvalues stored in the Tecplot format.
\end{enumerate}

%\clearpage
\section{Summary and conclusions}\label{sec-conclusions}
%%%%%%%%%%%%%%%%%%%%%%%%%%%%%%%%%%%%%%%%%%%%%%%%%%%%%%%%%%%%%%%%

The experimental realization of single- and two-component BECs
in higher spatial dimensions has admittedly been an exciting
journey in understanding the fundamental properties
of matter at ultracold temperatures. In parallel, however, this
journey has posed computational challenges pertaining about not
only the existence of matter waves in GP equations (single and
two-component versions thereof) but more crucially, their spectral
stability analysis, \ie BdG spectrum. The study
of the BdG spectrum often results in solving a very large eigenvalue
problem, a task that is quite computationally demanding and requires
the use of parallelization. With the present work, we took up
this challenge, and presented as well as delivered a parallel finite-element
toolbox for computing the BdG spectrum of stationary solutions to
one- and two-component GP equations in 2D and 3D.

The toolbox was created with the open-source, finite-element software
\ff which is now interfaced with parallel libraries such as \texttt{PETSc}
and \texttt{SLEPc}. The ability of \ff to perform adaptive mesh refinements,
together with the use of parallel linear solvers such as domain decomposition
and algebraic multigrid methods in  \texttt{PETSc}, makes the present toolbox a versatile tool for
studying 2D and 3D configurations to GP equations within reasonable CPU
times. The computation of the BdG spectrum that is carried out in the
present toolbox consists of two steps. At first, stationary states are
identified by using Newton's method which now has access to  parallel
linear solvers from \texttt{PETSc}. Moreover, a natural parameter continuation
method is adopted to obtain branches of solutions to GP equations
over the chemical potential $\mu$ or the inter-component interaction parameters
$\beta_{12}$ and $\beta_{21}$. Upon tracing branches of solutions, the BdG
spectrum is computed afterwards by solving the associated eigenvalue problem
with \texttt{SLEPc}.

We successfully verified our toolbox's results against known theoretical and
numerical findings that have been published in the open literature. We
reported typical CPU times that render the toolbox to be used on ordinary
laptops and small workstations (of course, depending on the complexity of
the state of interest). The parameter files of the toolbox
correspond to the test cases we presented in this paper, and they can be used
by the user to reproduce the results. We further provide these files from
the scope of getting used as templates, if the user intends to compute a new BEC setup or case of interest. We hope that
the description and documentation of the toolbox will allow the user
in a convenient way to consider other types of trapping potentials
 \eg quartic $\pm$ quadratic trapping ones \citep{BEC-physV-2004-bretin},
and nonlinearities, such as the non-local ones appearing in dipolar
settings,  \eg see \citep{tang2022spectrally}.

There is clearly a broad array of future computational explorations
and developments stemming from this work that we briefly mention here.
First, we implemented a natural (or sequential) continuation approach
to trace branches of solutions in the present toolbox. It will be quite
interesting to consider other types of continuation approaches in \ff
including the pseudo-arclength continuation  \cite{doedel_tuckerman_book,seydel_book_2010},
asymptotic numerical method (ANM)  \cite{ventura_2020}, and deflation-based
techniques \citep{deflation_2018,charalampidis2020bifurcation,boulle2020deflation,panos-boulle-2022},
among many others. Another possibility concerns about the interfacing of
other libraries for eigenvalue computations, including the FEAST eigenvalue
solver \citep{polizzi2009density} which enjoys multiple levels of parallelization%
 \citep{feast4}. Finally, with the recent experimental developments on
spinor condensates \citep{dshall_comm_phys_2021,dshall_nature_2022} described
by more than two GP equations (see,  \eg  \cite{PhysRevA.105.053303} where
the authors considered a three-component GP system for studying monopoles
and Alice rings), it is thus timely to bring forth state-of-the-art computing
methodologies in order to elucidate the configuration space of solutions in
these experimentally accessible systems. Such computational studies and
software development in \ff are currently in progress and will be reported
in future contributions.

\section*{Acknowledgments}
% ======================================================================
The authors acknowledge financial support from the French ANR grant
ANR-18-CE46-0013 QUTE-HPC. The work of EGC has been partially supported
by the U.S. National Science Foundation under Grant No. DMS-2204782. He
expresses his gratitude to the Centre national de la recherche scientifique
(CNRS) for awarding him a visiting professorship during the summer of 2023
at the Laboratoire de Math\'ematiques Rapha\"el Salem (LMRS) at University of Rouen
Normandie.
Part
of this work used computational resources provided by the Institut du
d{\'e}veloppement et des ressources en informatique scientifique (IDRIS)
and Centre R{\'e}gional Informatique et d'Applications Num{\'e}riques de Normandie
(CRIANN). 
The authors are grateful to Prof.~P.~Kevrekidis for stimulating
discussions, and Dr.~N.~Boull\'e for providing his numerical results that helped for numerical validations in Sec. \ref{sec-3D-twoc}.
They also thank Dr.~V.~Kalt for his insights and earlier collaboration.

%\pagebreak
%\clearpage

%\section*{Bibliography}

%\bibliographystyle{model1-num-names}
%\bibliography{\bibpath/bib_bdg,\bibpath/danaila_publis,\bibpath/bib-fem-2019,\bibpath/bib-bec-2019}

\begin{thebibliography}{82}
	\expandafter\ifx\csname natexlab\endcsname\relax\def\natexlab#1{#1}\fi
	\providecommand{\url}[1]{\texttt{#1}}
	\providecommand{\href}[2]{#2}
	\providecommand{\path}[1]{#1}
	\providecommand{\DOIprefix}{doi:}
	\providecommand{\ArXivprefix}{arXiv:}
	\providecommand{\URLprefix}{URL: }
	\providecommand{\Pubmedprefix}{pmid:}
	\providecommand{\doi}[1]{\href{http://dx.doi.org/#1}{\path{#1}}}
	\providecommand{\Pubmed}[1]{\href{pmid:#1}{\path{#1}}}
	\providecommand{\bibinfo}[2]{#2}
	\ifx\xfnm\relax \def\xfnm[#1]{\unskip,\space#1}\fi
	%Type = Article
	\bibitem[{Anderson et~al.(1995)Anderson, Ensher, Matthews, Wieman, and
		Cornell}]{anderson1995observation}
	\bibinfo{author}{M.~H. Anderson}, \bibinfo{author}{J.~R. Ensher},
	\bibinfo{author}{M.~R. Matthews}, \bibinfo{author}{C.~E. Wieman},
	\bibinfo{author}{E.~A. Cornell},
	\newblock \bibinfo{title}{Observation of {B}ose-{E}instein condensation in a
		dilute atomic vapor},
	\newblock \bibinfo{journal}{Science} \bibinfo{volume}{269}
	(\bibinfo{year}{1995}) \bibinfo{pages}{198--201}.
	%Type = Article
	\bibitem[{Davis et~al.(1995)Davis, Mewes, Andrews, van Druten, Durfee, Kurn,
		and Ketterle}]{davis1995bose}
	\bibinfo{author}{K.~B. Davis}, \bibinfo{author}{M.~O. Mewes},
	\bibinfo{author}{M.~R. Andrews}, \bibinfo{author}{N.~J. van Druten},
	\bibinfo{author}{D.~S. Durfee}, \bibinfo{author}{D.~M. Kurn},
	\bibinfo{author}{W.~Ketterle},
	\newblock \bibinfo{title}{{B}ose-{E}instein condensation in a gas of sodium
		atoms},
	\newblock \bibinfo{journal}{Phys. Rev. Lett.} \bibinfo{volume}{75}
	(\bibinfo{year}{1995}) \bibinfo{pages}{3969--3973}.
	%Type = Book
	\bibitem[{Pethick and Smith(2011)}]{pethick_book2011}
	\bibinfo{author}{C.~Pethick}, \bibinfo{author}{H.~Smith},
	\bibinfo{title}{{B}ose-{E}instein condensation in Dilute Gases},
	\bibinfo{publisher}{Cambridge University Press}, \bibinfo{year}{2011}.
	%Type = Book
	\bibitem[{Pitaevskii and Stringari(2015)}]{pitaevskii2015bose}
	\bibinfo{author}{L.~P. Pitaevskii}, \bibinfo{author}{S.~Stringari},
	\bibinfo{title}{{B}ose-{E}instein condensation and {S}uperfluidity},
	\bibinfo{publisher}{Oxford University Press}, \bibinfo{year}{2015}.
	%Type = Article
	\bibitem[{Fetter and Svidzinsky(2001)}]{fetter_prl2001}
	\bibinfo{author}{A.~L. Fetter}, \bibinfo{author}{A.~A. Svidzinsky},
	\newblock \bibinfo{title}{Vortices in a trapped dilute bose-einstein
		condensate},
	\newblock \bibinfo{journal}{Journal of Physics: Condensed Matter}
	\bibinfo{volume}{13} (\bibinfo{year}{2001}) \bibinfo{pages}{R135}.
	%Type = Article
	\bibitem[{P.~Engels and Cornell(2004)}]{engels_2004}
	\bibinfo{author}{V.~S. P.~Engels, I.~Coddington}, \bibinfo{author}{E.~A.
		Cornell},
	\newblock \bibinfo{title}{Vortex lattice dynamics in a dilute-gas bec},
	\newblock \bibinfo{journal}{J. Low Temp. Phys} \bibinfo{volume}{134}
	(\bibinfo{year}{2004}) \bibinfo{pages}{683--688}.
	%Type = Article
	\bibitem[{Kevrekidis et~al.(2004)Kevrekidis, Carretero-Gonz\'alez,
		Frantzeskakis, and Kevrekidis}]{pgk_mod_2004}
	\bibinfo{author}{P.~G. Kevrekidis}, \bibinfo{author}{R.~Carretero-Gonz\'alez},
	\bibinfo{author}{D.~J. Frantzeskakis}, \bibinfo{author}{I.~G. Kevrekidis},
	\newblock \bibinfo{title}{Vortices in bose-einstein condensates: Some recent
		developments},
	\newblock \bibinfo{journal}{Mod. Phys. Lett. B} \bibinfo{volume}{18}
	(\bibinfo{year}{2004}) \bibinfo{pages}{1481--1505}.
	%Type = Article
	\bibitem[{Fetter(2009)}]{fetter_mod_2009}
	\bibinfo{author}{A.~L. Fetter},
	\newblock \bibinfo{title}{Rotating trapped bose-einstein condensates},
	\newblock \bibinfo{journal}{Rev. Mod. Phys.} \bibinfo{volume}{81}
	(\bibinfo{year}{2009}) \bibinfo{pages}{647--691}.
	%Type = Article
	\bibitem[{Matthews et~al.(1999)Matthews, Anderson, Haljan, Hall, Wieman, and
		Cornell}]{PhysRevLett.83.2498}
	\bibinfo{author}{M.~R. Matthews}, \bibinfo{author}{B.~P. Anderson},
	\bibinfo{author}{P.~C. Haljan}, \bibinfo{author}{D.~S. Hall},
	\bibinfo{author}{C.~E. Wieman}, \bibinfo{author}{E.~A. Cornell},
	\newblock \bibinfo{title}{Vortices in a bose-einstein condensate},
	\newblock \bibinfo{journal}{Phys. Rev. Lett.} \bibinfo{volume}{83}
	(\bibinfo{year}{1999}) \bibinfo{pages}{2498--2501}.
	%Type = Article
	\bibitem[{Leanhardt et~al.(2002)Leanhardt, G\"orlitz, Chikkatur, Kielpinski,
		Shin, Pritchard, and Ketterle}]{BEC-physV-2002-imprint}
	\bibinfo{author}{A.~E. Leanhardt}, \bibinfo{author}{A.~G\"orlitz},
	\bibinfo{author}{A.~P. Chikkatur}, \bibinfo{author}{D.~Kielpinski},
	\bibinfo{author}{Y.~Shin}, \bibinfo{author}{D.~E. Pritchard},
	\bibinfo{author}{W.~Ketterle},
	\newblock \bibinfo{title}{Imprinting vortices in a {B}ose-{E}instein condensate
		using topological phases},
	\newblock \bibinfo{journal}{Phys. Rev. Lett.} \bibinfo{volume}{89}
	(\bibinfo{year}{2002}) \bibinfo{pages}{190403}.
	%Type = Article
	\bibitem[{Becker et~al.(2008)Becker, Stellmer, Soltan-Panahi, D{\"o}rscher,
		Baumert, Richter, Kronj{\"a}ger, Bongs, and
		Sengstock}]{becker2008oscillations}
	\bibinfo{author}{C.~Becker}, \bibinfo{author}{S.~Stellmer},
	\bibinfo{author}{P.~Soltan-Panahi}, \bibinfo{author}{S.~D{\"o}rscher},
	\bibinfo{author}{M.~Baumert}, \bibinfo{author}{E.-M. Richter},
	\bibinfo{author}{J.~Kronj{\"a}ger}, \bibinfo{author}{K.~Bongs},
	\bibinfo{author}{K.~Sengstock},
	\newblock \bibinfo{title}{Oscillations and interactions of dark and
		dark--bright solitons in {B}ose-{E}instein condensates},
	\newblock \bibinfo{journal}{Nature Physics} \bibinfo{volume}{4}
	(\bibinfo{year}{2008}) \bibinfo{pages}{496--501}.
	%Type = Article
	\bibitem[{Madison et~al.(2000)Madison, Chevy, Wohlleben, and
		Dalibard}]{madison_prl_2000}
	\bibinfo{author}{K.~W. Madison}, \bibinfo{author}{F.~Chevy},
	\bibinfo{author}{W.~Wohlleben}, \bibinfo{author}{J.~Dalibard},
	\newblock \bibinfo{title}{Vortex formation in a stirred bose-einstein
		condensate},
	\newblock \bibinfo{journal}{Phys. Rev. Lett.} \bibinfo{volume}{84}
	(\bibinfo{year}{2000}) \bibinfo{pages}{806--809}.
	%Type = Article
	\bibitem[{Haljan et~al.(2001)Haljan, Coddington, Engels, and
		Cornell}]{BEC-physV-2001-haljan}
	\bibinfo{author}{P.~C. Haljan}, \bibinfo{author}{I.~Coddington},
	\bibinfo{author}{P.~Engels}, \bibinfo{author}{E.~A. Cornell},
	\newblock \bibinfo{title}{Driving {B}ose-{E}instein condensate vorticity with a
		rotating normal cloud},
	\newblock \bibinfo{journal}{Phys. Rev. Lett.} \bibinfo{volume}{87}
	(\bibinfo{year}{2001}) \bibinfo{pages}{210403--210407}.
	%Type = Article
	\bibitem[{Yan et~al.(2011)Yan, Chang, Hamner, Kevrekidis, Engels, Achilleos,
		Frantzeskakis, Carretero-Gonz\'alez, and Schmelcher}]{yan2011multiple}
	\bibinfo{author}{D.~Yan}, \bibinfo{author}{J.~J. Chang},
	\bibinfo{author}{C.~Hamner}, \bibinfo{author}{P.~G. Kevrekidis},
	\bibinfo{author}{P.~Engels}, \bibinfo{author}{V.~Achilleos},
	\bibinfo{author}{D.~J. Frantzeskakis},
	\bibinfo{author}{R.~Carretero-Gonz\'alez}, \bibinfo{author}{P.~Schmelcher},
	\newblock \bibinfo{title}{Multiple dark-bright solitons in atomic
		{B}ose-{E}instein condensates},
	\newblock \bibinfo{journal}{Phys. Rev. A} \bibinfo{volume}{84}
	(\bibinfo{year}{2011}) \bibinfo{pages}{053630}.
	%Type = Article
	\bibitem[{Theocharis et~al.(2010)Theocharis, Weller, Ronzheimer, Gross,
		Oberthaler, Kevrekidis, and Frantzeskakis}]{theocharis2010multiple}
	\bibinfo{author}{G.~Theocharis}, \bibinfo{author}{A.~Weller},
	\bibinfo{author}{J.~P. Ronzheimer}, \bibinfo{author}{C.~Gross},
	\bibinfo{author}{M.~K. Oberthaler}, \bibinfo{author}{P.~G. Kevrekidis},
	\bibinfo{author}{D.~J. Frantzeskakis},
	\newblock \bibinfo{title}{Multiple atomic dark solitons in cigar-shaped
		{B}ose-{E}instein condensates},
	\newblock \bibinfo{journal}{Phys. Rev. A} \bibinfo{volume}{81}
	(\bibinfo{year}{2010}) \bibinfo{pages}{063604}.
	%Type = Article
	\bibitem[{Scherer et~al.(2007)Scherer, Weiler, Neely, and
		Anderson}]{anderson_prl_2007}
	\bibinfo{author}{D.~R. Scherer}, \bibinfo{author}{C.~N. Weiler},
	\bibinfo{author}{T.~W. Neely}, \bibinfo{author}{B.~P. Anderson},
	\newblock \bibinfo{title}{Vortex formation by merging of multiple trapped
		bose-einstein condensates},
	\newblock \bibinfo{journal}{Phys. Rev. Lett.} \bibinfo{volume}{98}
	(\bibinfo{year}{2007}) \bibinfo{pages}{110402}.
	%Type = Article
	\bibitem[{Weiler et~al.(2008)Weiler, Neely, Scherer, Bradley, Davis, and
		Anderson}]{anderson_nature_2008}
	\bibinfo{author}{C.~N. Weiler}, \bibinfo{author}{T.~W. Neely},
	\bibinfo{author}{D.~R. Scherer}, \bibinfo{author}{A.~S. Bradley},
	\bibinfo{author}{M.~J. Davis}, \bibinfo{author}{B.~P. Anderson},
	\newblock \bibinfo{title}{Spontaneous vortices in the formation of
		bose-einstein condensates},
	\newblock \bibinfo{journal}{Nature} \bibinfo{volume}{455}
	(\bibinfo{year}{2008}) \bibinfo{pages}{948--951}.
	%Type = Article
	\bibitem[{Aftalion and {Danaila}(2003)}]{dan-2003-aft}
	\bibinfo{author}{A.~Aftalion}, \bibinfo{author}{I.~{Danaila}},
	\newblock \bibinfo{title}{Three-dimensional vortex configurations in a rotating
		{B}ose-{E}instein condensate},
	\newblock \bibinfo{journal}{Physical Review A} \bibinfo{volume}{68}
	(\bibinfo{year}{2003}) \bibinfo{pages}{023603}.
	%Type = Book
	\bibitem[{P.~G.~Kevrekidis and Carretero-Gonz\'alez(2015)}]{pgk_siam_book}
	\bibinfo{author}{D.~J.~F. P.~G.~Kevrekidis},
	\bibinfo{author}{R.~Carretero-Gonz\'alez}, \bibinfo{title}{The Defocusing
		Nonlinear Schrödinger Equation: From Dark Solitons to Vortices and Vortex
		Rings}, \bibinfo{publisher}{Society for Industrial and Applied Mathematics,
		Philadelphia}, \bibinfo{year}{2015}.
	%Type = Article
	\bibitem[{Malomed(2019)}]{malomed_physicaD_2019}
	\bibinfo{author}{B.~A. Malomed},
	\newblock \bibinfo{title}{(invited) vortex solitons: Old results and new
		perspectives},
	\newblock \bibinfo{journal}{Physica D: Nonlinear Phenomena}
	\bibinfo{volume}{399} (\bibinfo{year}{2019}) \bibinfo{pages}{108--137}.
	%Type = Article
	\bibitem[{Crasovan et~al.(2004)Crasovan, P{\'e}rez-Garc{\`{\i}}a, {Danaila},
		Mihalache, and Torner}]{dan-2004-cras}
	\bibinfo{author}{L.-C. Crasovan}, \bibinfo{author}{V.~M.
		P{\'e}rez-Garc{\`{\i}}a}, \bibinfo{author}{I.~{Danaila}},
	\bibinfo{author}{D.~Mihalache}, \bibinfo{author}{L.~Torner},
	\newblock \bibinfo{title}{Three--dimensional parallel vortex rings in
		{B}ose--{E}instein condensates},
	\newblock \bibinfo{journal}{Physical Review A} \bibinfo{volume}{70}
	(\bibinfo{year}{2004}) \bibinfo{pages}{033605(1--5)}.
	%Type = Article
	\bibitem[{Bisset et~al.(2015)Bisset, Wang, Ticknor, Carretero-Gonz\'alez,
		Frantzeskakis, Collins, and Kevrekidis}]{bisset2015robust}
	\bibinfo{author}{R.~N. Bisset}, \bibinfo{author}{W.~Wang},
	\bibinfo{author}{C.~Ticknor}, \bibinfo{author}{R.~Carretero-Gonz\'alez},
	\bibinfo{author}{D.~J. Frantzeskakis}, \bibinfo{author}{L.~A. Collins},
	\bibinfo{author}{P.~G. Kevrekidis},
	\newblock \bibinfo{title}{Robust vortex lines, vortex rings, and hopfions in
		three-dimensional {B}ose-{E}instein condensates},
	\newblock \bibinfo{journal}{Phys. Rev. A} \bibinfo{volume}{92}
	(\bibinfo{year}{2015}) \bibinfo{pages}{063611}.
	%Type = Article
	\bibitem[{Wang et~al.(2017)Wang, Bisset, Ticknor, Carretero-Gonz\'alez,
		Frantzeskakis, Collins, and Kevrekidis}]{wang2017single}
	\bibinfo{author}{W.~Wang}, \bibinfo{author}{R.~N. Bisset},
	\bibinfo{author}{C.~Ticknor}, \bibinfo{author}{R.~Carretero-Gonz\'alez},
	\bibinfo{author}{D.~J. Frantzeskakis}, \bibinfo{author}{L.~A. Collins},
	\bibinfo{author}{P.~G. Kevrekidis},
	\newblock \bibinfo{title}{Single and multiple vortex rings in three-dimensional
		{B}ose-{E}instein condensates: Existence, stability, and dynamics},
	\newblock \bibinfo{journal}{Phys. Rev. A} \bibinfo{volume}{95}
	(\bibinfo{year}{2017}) \bibinfo{pages}{043638}.
	%Type = Article
	\bibitem[{V.~Kalt and Danaila(2023)}]{dan-2023-CPC-postproc}
	\bibinfo{author}{F.~H. V.~Kalt, G.~SADAKA}, \bibinfo{author}{I.~Danaila},
	\newblock \bibinfo{title}{Identification of vortices in quantum fluids: Finite
		element},
	\newblock \bibinfo{journal}{Computer Physics Communications}
	\bibinfo{volume}{284} (\bibinfo{year}{2023}) \bibinfo{pages}{108606}.
	%Type = Article
	\bibitem[{Charalampidis et~al.(2015)Charalampidis, Kevrekidis, Frantzeskakis,
		and Malomed}]{PhysRevE.91.012924}
	\bibinfo{author}{E.~G. Charalampidis}, \bibinfo{author}{P.~G. Kevrekidis},
	\bibinfo{author}{D.~J. Frantzeskakis}, \bibinfo{author}{B.~A. Malomed},
	\newblock \bibinfo{title}{Dark-bright solitons in coupled nonlinear
		schr\"odinger equations with unequal dispersion coefficients},
	\newblock \bibinfo{journal}{Phys. Rev. E} \bibinfo{volume}{91}
	(\bibinfo{year}{2015}) \bibinfo{pages}{012924}.
	%Type = Article
	\bibitem[{Law et~al.(2010)Law, Kevrekidis, and Tuckerman}]{kody_prl_2011}
	\bibinfo{author}{K.~J.~H. Law}, \bibinfo{author}{P.~G. Kevrekidis},
	\bibinfo{author}{L.~S. Tuckerman},
	\newblock \bibinfo{title}{Stable vortex--bright-soliton structures in
		two-component bose-einstein condensates},
	\newblock \bibinfo{journal}{Phys. Rev. Lett.} \bibinfo{volume}{105}
	(\bibinfo{year}{2010}) \bibinfo{pages}{160405}.
	%Type = Article
	\bibitem[{Charalampidis et~al.(2016)Charalampidis, Kevrekidis, Frantzeskakis,
		and Malomed}]{PhysRevE.94.022207}
	\bibinfo{author}{E.~G. Charalampidis}, \bibinfo{author}{P.~G. Kevrekidis},
	\bibinfo{author}{D.~J. Frantzeskakis}, \bibinfo{author}{B.~A. Malomed},
	\newblock \bibinfo{title}{Vortex-soliton complexes in coupled nonlinear
		schr\"odinger equations with unequal dispersion coefficients},
	\newblock \bibinfo{journal}{Phys. Rev. E} \bibinfo{volume}{94}
	(\bibinfo{year}{2016}) \bibinfo{pages}{022207}.
	%Type = Article
	\bibitem[{{Danaila} et~al.(2016){Danaila}, Khamehchi, Gokhroo, Engels, and
		Kevrekidis}]{dan-2016-PRA}
	\bibinfo{author}{I.~{Danaila}}, \bibinfo{author}{M.~A. Khamehchi},
	\bibinfo{author}{V.~Gokhroo}, \bibinfo{author}{P.~Engels},
	\bibinfo{author}{P.~G. Kevrekidis},
	\newblock \bibinfo{title}{Vector dark-antidark solitary waves in multicomponent
		{B}ose-{E}instein condensates},
	\newblock \bibinfo{journal}{Physical Review A} \bibinfo{volume}{94}
	(\bibinfo{year}{2016}) \bibinfo{pages}{053617}.
	%Type = Article
	\bibitem[{Wang and Kevrekidis(2017)}]{wang2017two}
	\bibinfo{author}{W.~Wang}, \bibinfo{author}{P.~G. Kevrekidis},
	\newblock \bibinfo{title}{Two-component dark-bright solitons in
		three-dimensional atomic {B}ose-{E}instein condensates},
	\newblock \bibinfo{journal}{Phys. Rev. E} \bibinfo{volume}{95}
	(\bibinfo{year}{2017}) \bibinfo{pages}{032201}.
	%Type = Article
	\bibitem[{Ruostekoski and Anglin(2001)}]{anglin_prl_2001}
	\bibinfo{author}{J.~Ruostekoski}, \bibinfo{author}{J.~R. Anglin},
	\newblock \bibinfo{title}{Creating vortex rings and three-dimensional skyrmions
		in bose-einstein condensates},
	\newblock \bibinfo{journal}{Phys. Rev. Lett.} \bibinfo{volume}{86}
	(\bibinfo{year}{2001}) \bibinfo{pages}{3934--3937}.
	%Type = Article
	\bibitem[{Battye et~al.(2002)Battye, Cooper, and
		Sutcliffe}]{sutcliffe_prl_2002}
	\bibinfo{author}{R.~A. Battye}, \bibinfo{author}{N.~R. Cooper},
	\bibinfo{author}{P.~M. Sutcliffe},
	\newblock \bibinfo{title}{Stable skyrmions in two-component bose-einstein
		condensates},
	\newblock \bibinfo{journal}{Phys. Rev. Lett.} \bibinfo{volume}{88}
	(\bibinfo{year}{2002}) \bibinfo{pages}{080401}.
	%Type = Article
	\bibitem[{Ruostekoski and Anglin(2003)}]{anglin_prl_2006}
	\bibinfo{author}{J.~Ruostekoski}, \bibinfo{author}{J.~R. Anglin},
	\newblock \bibinfo{title}{Monopole core instability and alice rings in spinor
		bose-einstein condensates},
	\newblock \bibinfo{journal}{Phys. Rev. Lett.} \bibinfo{volume}{91}
	(\bibinfo{year}{2003}) \bibinfo{pages}{190402}.
	%Type = Article
	\bibitem[{Mithun et~al.(2022)Mithun, Carretero-Gonz\'alez, Charalampidis, Hall,
		and Kevrekidis}]{PhysRevA.105.053303}
	\bibinfo{author}{T.~Mithun}, \bibinfo{author}{R.~Carretero-Gonz\'alez},
	\bibinfo{author}{E.~G. Charalampidis}, \bibinfo{author}{D.~S. Hall},
	\bibinfo{author}{P.~G. Kevrekidis},
	\newblock \bibinfo{title}{Existence, stability, and dynamics of monopole and
		alice ring solutions in antiferromagnetic spinor condensates},
	\newblock \bibinfo{journal}{Phys. Rev. A} \bibinfo{volume}{105}
	(\bibinfo{year}{2022}) \bibinfo{pages}{053303}.
	%Type = Article
	\bibitem[{Charalampidis et~al.(2020)Charalampidis, Boull\'e, Farrell, and
		Kevrekidis}]{charalampidis2020bifurcation}
	\bibinfo{author}{E.~Charalampidis}, \bibinfo{author}{N.~Boull\'e},
	\bibinfo{author}{P.~Farrell}, \bibinfo{author}{P.~Kevrekidis},
	\newblock \bibinfo{title}{Bifurcation analysis of stationary solutions of
		two-dimensional coupled {G}ross-{P}itaevskii equations using deflated
		continuation},
	\newblock \bibinfo{journal}{Communications in Nonlinear Science and Numerical
		Simulation} \bibinfo{volume}{87} (\bibinfo{year}{2020})
	\bibinfo{pages}{105255}.
	%Type = Article
	\bibitem[{Boull\'e et~al.(2023)Boull\'e, Newell, Farrell, and
		Kevrekidis}]{panos-boulle-2022}
	\bibinfo{author}{N.~Boull\'e}, \bibinfo{author}{I.~Newell},
	\bibinfo{author}{P.~E. Farrell}, \bibinfo{author}{P.~G. Kevrekidis},
	\newblock \bibinfo{title}{Two-component three-dimensional atomic bose-einstein
		condensates supporting complex stable patterns},
	\newblock \bibinfo{journal}{Phys. Rev. A} \bibinfo{volume}{107}
	(\bibinfo{year}{2023}) \bibinfo{pages}{012813}.
	%Type = Book
	\bibitem[{Kapitula and Promislow(2013)}]{keith_book}
	\bibinfo{author}{T.~Kapitula}, \bibinfo{author}{K.~Promislow},
	\bibinfo{title}{Spectral and Dynamical Stability of Nonlinear Waves},
	\bibinfo{publisher}{Springer}, \bibinfo{year}{2013}.
	%Type = Article
	\bibitem[{Bogolyubov(1947)}]{Bogolyubov-1947}
	\bibinfo{author}{N.~N. Bogolyubov},
	\newblock \bibinfo{title}{{On the theory of superfluidity}},
	\newblock \bibinfo{journal}{J. Phys. (USSR)} \bibinfo{volume}{11}
	(\bibinfo{year}{1947}) \bibinfo{pages}{23--32}.
	%Type = Book
	\bibitem[{De~Gennes(1966)}]{deGennes-1966}
	\bibinfo{author}{P.~G. De~Gennes}, \bibinfo{title}{Superconductivity of Metals
		and Alloys}, \bibinfo{publisher}{CRC Press}, \bibinfo{year}{1966}.
	%Type = Article
	\bibitem[{Dion and Canc{\`e}s(2007)}]{BEC-CPC-2007-dion-cances}
	\bibinfo{author}{C.~M. Dion}, \bibinfo{author}{E.~Canc{\`e}s},
	\newblock \bibinfo{title}{Ground state of the time-independent
		{G}ross-{P}itaevskii equation},
	\newblock \bibinfo{journal}{Comput. Phys. Comm.} \bibinfo{volume}{177}
	(\bibinfo{year}{2007}) \bibinfo{pages}{787--798}.
	%Type = Article
	\bibitem[{Caliari and Rainer(2013)}]{BEC-CPC-2013-Caliari}
	\bibinfo{author}{M.~Caliari}, \bibinfo{author}{S.~Rainer},
	\newblock \bibinfo{title}{{GSGPEs}: A {M}atlab code for computing the ground
		state of systems of {G}ross-{P}itaevskii equations},
	\newblock \bibinfo{journal}{Comput. Phys. Comm.} \bibinfo{volume}{184}
	(\bibinfo{year}{2013}) \bibinfo{pages}{812 -- 823}.
	%Type = Article
	\bibitem[{Antoine and Duboscq(2014)}]{BEC-CPC-2014-antoine-duboscq}
	\bibinfo{author}{X.~Antoine}, \bibinfo{author}{R.~Duboscq},
	\newblock \bibinfo{title}{{GPELab}, a {M}atlab toolbox to solve
		{G}ross-{P}itaevskii equations {I}: Computation of stationary solutions},
	\newblock \bibinfo{journal}{Comput. Phys. Comm.} \bibinfo{volume}{185}
	(\bibinfo{year}{2014}) \bibinfo{pages}{2969--2991}.
	%Type = Article
	\bibitem[{Marojevi{\'c} et~al.(2016)Marojevi{\'c}, G{\"o}kl{\"u}, and
		L{\"a}mmerzahl}]{BEC-CPC-2016-FEM}
	\bibinfo{author}{Z.~Marojevi{\'c}}, \bibinfo{author}{E.~G{\"o}kl{\"u}},
	\bibinfo{author}{C.~L{\"a}mmerzahl},
	\newblock \bibinfo{title}{{ATUS-PRO}: A {FEM}-based solver for the
		time-dependent and stationary {G}ross-{P}itaevskii equation},
	\newblock \bibinfo{journal}{Computer Physics Communications}
	\bibinfo{volume}{202} (\bibinfo{year}{2016}) \bibinfo{pages}{216 -- 232}.
	%Type = Article
	\bibitem[{Vergez et~al.(2016)Vergez, {Danaila}, Auliac, and
		Hecht}]{dan-2016-CPC}
	\bibinfo{author}{G.~Vergez}, \bibinfo{author}{I.~{Danaila}},
	\bibinfo{author}{S.~Auliac}, \bibinfo{author}{F.~Hecht},
	\newblock \bibinfo{title}{A finite-element toolbox for the stationary
		{G}ross-{P}itaevskii equation with rotation},
	\newblock \bibinfo{journal}{Comput. Phys. Comm.} \bibinfo{volume}{209}
	(\bibinfo{year}{2016}) \bibinfo{pages}{144--162}.
	%Type = Article
	\bibitem[{Uecker et~al.(2014)Uecker, Wetzel, and
		Rademacher}]{uecker_wetzel_rademacher_2014}
	\bibinfo{author}{H.~Uecker}, \bibinfo{author}{D.~Wetzel},
	\bibinfo{author}{J.~Rademacher},
	\newblock \bibinfo{title}{{pde2path - A Matlab Package for Continuation and
			Bifurcation in 2D Elliptic Systems}},
	\newblock \bibinfo{journal}{Num. Math.: Theory, Methods and Applications}
	\bibinfo{volume}{7} (\bibinfo{year}{2014}) \bibinfo{pages}{58–106}.
	%Type = Article
	\bibitem[{Muruganandam and Adhikari(2009)}]{BEC-CPC-2009-Muruganandam}
	\bibinfo{author}{P.~Muruganandam}, \bibinfo{author}{S.~Adhikari},
	\newblock \bibinfo{title}{Fortran programs for the time-dependent
		{G}ross-{P}itaevskii equation in a fully anisotropic trap},
	\newblock \bibinfo{journal}{Comput. Phys. Comm.} \bibinfo{volume}{180}
	(\bibinfo{year}{2009}) \bibinfo{pages}{1888--1912}.
	%Type = Article
	\bibitem[{Vudragovi\'c et~al.(2012)Vudragovi\'c, Vidanovi\'c, Balaz,
		Muruganandam, and Adhikari}]{BEC-CPC-2012-Vudragovic}
	\bibinfo{author}{D.~Vudragovi\'c}, \bibinfo{author}{I.~Vidanovi\'c},
	\bibinfo{author}{A.~Balaz}, \bibinfo{author}{P.~Muruganandam},
	\bibinfo{author}{S.~K. Adhikari},
	\newblock \bibinfo{title}{C programs for solving the time-dependent
		{G}ross-{P}itaevskii equation in a fully anisotropic trap},
	\newblock \bibinfo{journal}{Comput. Phys. Comm.} \bibinfo{volume}{183}
	(\bibinfo{year}{2012}) \bibinfo{pages}{2021 -- 2025}.
	%Type = Article
	\bibitem[{Kong et~al.(2014)Kong, Hong, and Zhang}]{BEC-CPC-2014-simplectic}
	\bibinfo{author}{L.~Kong}, \bibinfo{author}{J.~Hong},
	\bibinfo{author}{J.~Zhang},
	\newblock \bibinfo{title}{{LOD}-ms for {G}ross-{P}itaevskii equation in
		{B}ose-{E}instein condensates},
	\newblock \bibinfo{journal}{Communications in Computational Physics}
	\bibinfo{volume}{14} (\bibinfo{year}{2014}) \bibinfo{pages}{219--241}.
	%Type = Article
	\bibitem[{Hohenester(2014)}]{BEC-CPC-2014-Hohenester}
	\bibinfo{author}{U.~Hohenester},
	\newblock \bibinfo{title}{{OCTBEC} a {M}atlab toolbox for optimal quantum
		control of {B}ose-{E}instein condensates},
	\newblock \bibinfo{journal}{Comput. Phys. Comm.} \bibinfo{volume}{185}
	(\bibinfo{year}{2014}) \bibinfo{pages}{194--216}.
	%Type = Article
	\bibitem[{{Kishor Kumar} et~al.(2019){Kishor Kumar}, Lon{\v{c}}ar,
		Muruganandam, Adhikari, and Bala{\v{z}}}]{BEC-CPC-2019-rotating}
	\bibinfo{author}{R.~{Kishor Kumar}}, \bibinfo{author}{V.~Lon{\v{c}}ar},
	\bibinfo{author}{P.~Muruganandam}, \bibinfo{author}{S.~K. Adhikari},
	\bibinfo{author}{A.~Bala{\v{z}}},
	\newblock \bibinfo{title}{{C} and {F}ortran {O}pen{MP} programs for rotating
		{B}ose-{E}instein condensates},
	\newblock \bibinfo{journal}{Computer Physics Communications}
	\bibinfo{volume}{240} (\bibinfo{year}{2019}) \bibinfo{pages}{74 -- 82}.
	%Type = Book
	\bibitem[{Allgower and Georg(1990)}]{Allgower-1990}
	\bibinfo{author}{E.~L. Allgower}, \bibinfo{author}{K.~Georg},
	\bibinfo{title}{Spectral and Dynamical Stability of Nonlinear Waves},
	\bibinfo{publisher}{Springer-Verlag}, \bibinfo{year}{1990}.
	%Type = Book
	\bibitem[{Kelley(2003)}]{kelley_book_2003}
	\bibinfo{author}{C.~T. Kelley}, \bibinfo{title}{Solving Nonlinear Equations
		with Newton's Method}, \bibinfo{publisher}{Society of Industrial and Applied
		Mathematics, Philadelphia}, \bibinfo{year}{2003}.
	%Type = Article
	\bibitem[{Carretero-Gonz\'alez et~al.(2016)Carretero-Gonz\'alez, Kevrekidis,
		and Kolokolnikov}]{carreterogonzalez2016vortex}
	\bibinfo{author}{R.~Carretero-Gonz\'alez}, \bibinfo{author}{P.~Kevrekidis},
	\bibinfo{author}{T.~Kolokolnikov},
	\newblock \bibinfo{title}{Vortex nucleation in a dissipative variant of the
		nonlinear {S}chr\"odinger equation under rotation},
	\newblock \bibinfo{journal}{Physica D: Nonlinear Phenomena}
	\bibinfo{volume}{317} (\bibinfo{year}{2016}) \bibinfo{pages}{1--14}.
	%Type = Article
	\bibitem[{Boull\'e et~al.(2020)Boull\'e, Charalampidis, Farrell, and
		Kevrekidis}]{boulle2020deflation}
	\bibinfo{author}{N.~Boull\'e}, \bibinfo{author}{E.~G. Charalampidis},
	\bibinfo{author}{P.~E. Farrell}, \bibinfo{author}{P.~G. Kevrekidis},
	\newblock \bibinfo{title}{Deflation-based identification of nonlinear
		excitations of the three-dimensional {G}ross-{P}itaevskii equation},
	\newblock \bibinfo{journal}{Phys. Rev. A} \bibinfo{volume}{102}
	(\bibinfo{year}{2020}) \bibinfo{pages}{053307}.
	%Type = Article
	\bibitem[{Roy et~al.(2020)Roy, Pal, Gautam, Angom, and
		Muruganandam}]{arko2020fact}
	\bibinfo{author}{A.~Roy}, \bibinfo{author}{S.~Pal},
	\bibinfo{author}{S.~Gautam}, \bibinfo{author}{D.~Angom},
	\bibinfo{author}{P.~Muruganandam},
	\newblock \bibinfo{title}{{FACt}: {FORTRAN} toolbox for calculating
		fluctuations in atomic condensates},
	\newblock \bibinfo{journal}{Computer Physics Communications}
	\bibinfo{volume}{256} (\bibinfo{year}{2020}) \bibinfo{pages}{107288}.
	%Type = Article
	\bibitem[{SADAKA et~al.(2024)SADAKA, Kalt, Hecht, and Danaila}]{sadaka_2023}
	\bibinfo{author}{G.~SADAKA}, \bibinfo{author}{V.~Kalt},
	\bibinfo{author}{F.~Hecht}, \bibinfo{author}{I.~Danaila},
	\newblock \bibinfo{title}{A finite element toolbox for the bogoliubov-de gennes
		stability analysis of bose-einstein condensates},
	\newblock \bibinfo{journal}{Computer Physics Communications}
	\bibinfo{volume}{294} (\bibinfo{year}{2024}) \bibinfo{pages}{108948}.
	%Type = Article
	\bibitem[{Hecht(2012)}]{hecht-2012-JNM}
	\bibinfo{author}{F.~Hecht},
	\newblock \bibinfo{title}{New developments in {F}reefem++},
	\newblock \bibinfo{journal}{Journal of Numerical Mathematics}
	\bibinfo{volume}{20} (\bibinfo{year}{2012}) \bibinfo{pages}{251--266}.
	%Type = Book
	\bibitem[{Balay and et. al.(2022)}]{petsc}
	\bibinfo{author}{S.~Balay}, \bibinfo{author}{et. al.},
	\bibinfo{title}{PETSc/TAO Users Manual}, \bibinfo{publisher}{Argonne National
		Laboratory-21/39, \url{https://petsc.org/release/docs/manual/manual.pdf}},
	\bibinfo{year}{2022}.
	%Type = Book
	\bibitem[{Jolivet(2023)}]{pierre_interfacing}
	\bibinfo{author}{P.~Jolivet}, \bibinfo{title}{Introduction to FreeFEM with an
		emphasis on parallel computing},
	\bibinfo{publisher}{\url{https://joliv.et/FreeFem-tutorial/main.pdf}},
	\bibinfo{year}{2023}.
	%Type = Book
	\bibitem[{Dolean et~al.(2016)Dolean, Jolivet, and Nataf}]{pierre_book_2015}
	\bibinfo{author}{V.~Dolean}, \bibinfo{author}{P.~Jolivet},
	\bibinfo{author}{F.~Nataf}, \bibinfo{title}{An Introduction to Domain
		Decomposition Methods: Algorithms, Theory, and Parallel Implementation},
	\bibinfo{publisher}{Society for Industrial and Applied Mathematics,
		Philadelphia}, \bibinfo{year}{2016}.
	%Type = Book
	\bibitem[{Tournier et~al.(2019)Tournier, Jolivet, and
		Nataf}]{FFD:Tournier:2019}
	\bibinfo{author}{P.-H. Tournier}, \bibinfo{author}{P.~Jolivet},
	\bibinfo{author}{F.~Nataf}, \bibinfo{title}{{FFDDM}: FreeFem Domain
		Decomposition Method},
	\bibinfo{publisher}{\url{https://doc.freefem.org/documentation/ffddm/index.html}},
	\bibinfo{year}{2019}.
	%Type = Article
	\bibitem[{Hernandez et~al.(2005)Hernandez, Roman, and
		Vidal}]{hernandez2005slepc}
	\bibinfo{author}{V.~Hernandez}, \bibinfo{author}{J.~E. Roman},
	\bibinfo{author}{V.~Vidal},
	\newblock \bibinfo{title}{{SLEPc}: A scalable and flexible toolkit for the
		solution of eigenvalue problems},
	\newblock \bibinfo{journal}{ACM Trans. Math. Softw.} \bibinfo{volume}{31}
	(\bibinfo{year}{2005}) \bibinfo{pages}{351–362}.
	%Type = Article
	\bibitem[{Frantzeskakis(2010)}]{frantzeskakis2010dark}
	\bibinfo{author}{D.~J. Frantzeskakis},
	\newblock \bibinfo{title}{Dark solitons in atomic {B}ose-{E}instein
		condensates: from theory to experiments},
	\newblock \bibinfo{journal}{Journal of Physics A: Mathematical and Theoretical}
	\bibinfo{volume}{43} (\bibinfo{year}{2010}) \bibinfo{pages}{213001}.
	%Type = Article
	\bibitem[{Bao and Cai(2013)}]{bao2013mathematical}
	\bibinfo{author}{W.~Bao}, \bibinfo{author}{Y.~Cai},
	\newblock \bibinfo{title}{Mathematical theory and numerical methods for
		{B}ose-{E}instein condensation},
	\newblock \bibinfo{journal}{Kinetic and Related Models} \bibinfo{volume}{6}
	(\bibinfo{year}{2013}) \bibinfo{pages}{1--135}.
	%Type = Book
	\bibitem[{Adams and Fournier(2003)}]{adams_sobolev_book}
	\bibinfo{author}{R.~A. Adams}, \bibinfo{author}{J.~J.~F. Fournier},
	\bibinfo{title}{Sobolev Spaces}, \bibinfo{publisher}{Academic Press},
	\bibinfo{year}{2003}.
	%Type = Inproceedings
	\bibitem[{Borouchaki et~al.(1996)Borouchaki, Castro-Diaz, George, Hecht, and
		Mohammadi}]{hecht-1996-missi}
	\bibinfo{author}{H.~Borouchaki}, \bibinfo{author}{M.~J. Castro-Diaz},
	\bibinfo{author}{P.~L. George}, \bibinfo{author}{F.~Hecht},
	\bibinfo{author}{B.~Mohammadi},
	\newblock \bibinfo{title}{Anisotropic adaptive mesh generation in two
		dimensions for {CFD}},
	\newblock in: \bibinfo{booktitle}{5th Inter. Conf. on Numerical Grid Generation
		in Computational Field Simulations}, \bibinfo{organization}{Mississipi State
		Univ.}, \bibinfo{year}{1996}.
	%Type = Book
	\bibitem[{Frey and George(1999)}]{frey-george-1999}
	\bibinfo{author}{P.~J. Frey}, \bibinfo{author}{P.~L. George},
	\bibinfo{title}{Maillages}, \bibinfo{publisher}{Herm{\`e}s, Paris},
	\bibinfo{year}{1999}.
	%Type = Book
	\bibitem[{Mohammadi and Pironneau(2000)}]{moham-piron-2000}
	\bibinfo{author}{B.~Mohammadi}, \bibinfo{author}{O.~Pironneau},
	\bibinfo{title}{Applied Shape Design for Fluids}, \bibinfo{publisher}{Oxford
		Univ. Press}, \bibinfo{year}{2000}.
	%Type = Article
	\bibitem[{Dapogny et~al.(2014)Dapogny, Dobrzynski, and Frey}]{dapogny2014three}
	\bibinfo{author}{C.~Dapogny}, \bibinfo{author}{C.~Dobrzynski},
	\bibinfo{author}{P.~Frey},
	\newblock \bibinfo{title}{Three-dimensional adaptive domain remeshing, implicit
		domain meshing, and applications to free and moving boundary problems},
	\newblock \bibinfo{journal}{Journal of Computational Physics}
	\bibinfo{volume}{262} (\bibinfo{year}{2014}) \bibinfo{pages}{358--378}.
	%Type = Article
	\bibitem[{{Danaila} and Hecht(2010)}]{dan-2010-JCP}
	\bibinfo{author}{I.~{Danaila}}, \bibinfo{author}{F.~Hecht},
	\newblock \bibinfo{title}{A finite element method with mesh adaptivity for
		computing vortex states in fast-rotating {B}ose-{E}instein condensates},
	\newblock \bibinfo{journal}{J. Comput. Physics} \bibinfo{volume}{229}
	(\bibinfo{year}{2010}) \bibinfo{pages}{6946--6960}.
	%Type = Article
	\bibitem[{Kevrekidis and Pelinovsky(2010)}]{kevrekidis2010distribution}
	\bibinfo{author}{P.~G. Kevrekidis}, \bibinfo{author}{D.~E. Pelinovsky},
	\newblock \bibinfo{title}{Distribution of eigenfrequencies for oscillations of
		the ground state in the {T}homas-{F}ermi limit},
	\newblock \bibinfo{journal}{Phys. Rev. A} \bibinfo{volume}{81}
	(\bibinfo{year}{2010}) \bibinfo{pages}{023627}.
	%Type = Article
	\bibitem[{Charalampidis et~al.(2018)Charalampidis, Kevrekidis, and
		Farrell}]{deflation_2018}
	\bibinfo{author}{E.~Charalampidis}, \bibinfo{author}{P.~Kevrekidis},
	\bibinfo{author}{P.~Farrell},
	\newblock \bibinfo{title}{Computing stationary solutions of the two-dimensional
		gross–pitaevskii equation with deflated continuation},
	\newblock \bibinfo{journal}{Communications in Nonlinear Science and Numerical
		Simulation} \bibinfo{volume}{54} (\bibinfo{year}{2018})
	\bibinfo{pages}{482--499}.
	%Type = Article
	\bibitem[{Middelkamp et~al.(2010)Middelkamp, Kevrekidis, Frantzeskakis,
		Carretero-Gonz\'alez, and Schmelcher}]{middelkamp2010bifurcations}
	\bibinfo{author}{S.~Middelkamp}, \bibinfo{author}{P.~G. Kevrekidis},
	\bibinfo{author}{D.~J. Frantzeskakis},
	\bibinfo{author}{R.~Carretero-Gonz\'alez}, \bibinfo{author}{P.~Schmelcher},
	\newblock \bibinfo{title}{Bifurcations, stability, and dynamics of multiple
		matter-wave vortex states},
	\newblock \bibinfo{journal}{Phys. Rev. A} \bibinfo{volume}{82}
	(\bibinfo{year}{2010}) \bibinfo{pages}{013646}.
	%Type = Article
	\bibitem[{Bisset et~al.(2015)Bisset, Wang, Ticknor, Carretero-Gonz\'alez,
		Frantzeskakis, Collins, and Kevrekidis}]{bisset2015bifurcation}
	\bibinfo{author}{R.~N. Bisset}, \bibinfo{author}{W.~Wang},
	\bibinfo{author}{C.~Ticknor}, \bibinfo{author}{R.~Carretero-Gonz\'alez},
	\bibinfo{author}{D.~J. Frantzeskakis}, \bibinfo{author}{L.~A. Collins},
	\bibinfo{author}{P.~G. Kevrekidis},
	\newblock \bibinfo{title}{Bifurcation and stability of single and multiple
		vortex rings in three-dimensional {B}ose-{E}instein condensates},
	\newblock \bibinfo{journal}{Phys. Rev. A} \bibinfo{volume}{92}
	(\bibinfo{year}{2015}) \bibinfo{pages}{043601}.
	%Type = Article
	\bibitem[{Bretin et~al.(2004)Bretin, Stock, Seurin, and
		Dalibard}]{BEC-physV-2004-bretin}
	\bibinfo{author}{V.~Bretin}, \bibinfo{author}{S.~Stock},
	\bibinfo{author}{Y.~Seurin}, \bibinfo{author}{J.~Dalibard},
	\newblock \bibinfo{title}{Fast rotation of a {B}ose-{E}instein condensate},
	\newblock \bibinfo{journal}{Phys. Rev. Lett.} \bibinfo{volume}{92}
	(\bibinfo{year}{2004}) \bibinfo{pages}{050403}.
	%Type = Article
	\bibitem[{Tang et~al.(2022)Tang, Xie, Zhang, and Zhang}]{tang2022spectrally}
	\bibinfo{author}{Q.~Tang}, \bibinfo{author}{M.~Xie},
	\bibinfo{author}{Y.~Zhang}, \bibinfo{author}{Y.~Zhang},
	\newblock \bibinfo{title}{A spectrally accurate numerical method for computing
		the {B}ogoliubov-de {G}ennes excitations of dipolar {B}ose-{E}instein
		condensates},
	\newblock \bibinfo{journal}{SIAM Journal on Scientific Computing}
	\bibinfo{volume}{44} (\bibinfo{year}{2022}) \bibinfo{pages}{B100--B121}.
	%Type = Book
	\bibitem[{Doedel and Tuckerman(2000)}]{doedel_tuckerman_book}
	\bibinfo{author}{E.~Doedel}, \bibinfo{author}{L.~S. Tuckerman},
	\bibinfo{title}{Numerical Methods for Bifurcation Problems and Large-Scale
		Dynamical Systems}, \bibinfo{publisher}{(The IMA Volumes in Mathematics and
		its Applications), Springer}, \bibinfo{year}{2000}.
	%Type = Book
	\bibitem[{Seydel(2010)}]{seydel_book_2010}
	\bibinfo{author}{R.~Seydel}, \bibinfo{title}{Practical Bifurcation and
		Stability Analysis}, volume~\bibinfo{volume}{5},
	\bibinfo{publisher}{Interdisciplinary Applied Mathematics, Springer-Verlag
		(New York)}, \bibinfo{year}{2010}.
	%Type = Article
	\bibitem[{Ventura et~al.(2020)Ventura, Ferry, and Zahrouni}]{ventura_2020}
	\bibinfo{author}{P.~Ventura}, \bibinfo{author}{M.~P. Ferry},
	\bibinfo{author}{H.~Zahrouni},
	\newblock \bibinfo{title}{A secure version of asymptotic numerical method via
		convergence acceleration},
	\newblock \bibinfo{journal}{Compt. Rend. Mecan.} \bibinfo{volume}{348}
	(\bibinfo{year}{2020}) \bibinfo{pages}{361--374}.
	%Type = Article
	\bibitem[{Polizzi(2009)}]{polizzi2009density}
	\bibinfo{author}{E.~Polizzi},
	\newblock \bibinfo{title}{Density-matrix-based algorithm for solving eigenvalue
		problems},
	\newblock \bibinfo{journal}{Phys. Rev. B} \bibinfo{volume}{79}
	(\bibinfo{year}{2009}) \bibinfo{pages}{115112}.
	%Type = Book
	\bibitem[{Polizzi(2023)}]{feast4}
	\bibinfo{author}{E.~Polizzi}, \bibinfo{title}{FEAST Eigenvalue Solver},
	\bibinfo{publisher}{\url{https://www.feast-solver.org/index.htm}},
	\bibinfo{year}{2023}.
	%Type = Article
	\bibitem[{Xiao et~al.(2021)Xiao, Borgh, Weiss, Blinova, Ruostekoski, and
		Hall}]{dshall_comm_phys_2021}
	\bibinfo{author}{Y.~Xiao}, \bibinfo{author}{M.~O. Borgh},
	\bibinfo{author}{L.~S. Weiss}, \bibinfo{author}{A.~A. Blinova},
	\bibinfo{author}{J.~Ruostekoski}, \bibinfo{author}{D.~S. Hall},
	\newblock \bibinfo{title}{Controlled creation and decay of singly-quantized
		vortices in a polar magnetic phase},
	\newblock \bibinfo{journal}{Communications Physics} \bibinfo{volume}{4}
	(\bibinfo{year}{2021}).
	%Type = Article
	\bibitem[{Xiao et~al.(2022)Xiao, Borgh, Blinova, Ollikainen, Ruostekoski, and
		Hall}]{dshall_nature_2022}
	\bibinfo{author}{Y.~Xiao}, \bibinfo{author}{M.~O. Borgh},
	\bibinfo{author}{A.~A. Blinova}, \bibinfo{author}{T.~Ollikainen},
	\bibinfo{author}{J.~Ruostekoski}, \bibinfo{author}{D.~S. Hall},
	\newblock \bibinfo{title}{Topological superfluid defects with discrete point
		group symmetries},
	\newblock \bibinfo{journal}{Nature Communications} \bibinfo{volume}{13}
	(\bibinfo{year}{2022}).
	
\end{thebibliography}

%
\end{document}